\documentclass[aps,prx,amsmath, amssymb, english, twocolumn,reprint,floatfix, superscriptaddress]{revtex4-1}

\usepackage{float}
\usepackage{graphicx}
\usepackage{subcaption}
\usepackage{amsmath}
\usepackage{physics}
\usepackage{cancel}

\usepackage{hyperref}
\hypersetup{
colorlinks,
citecolor=blue,
filecolor=blue,
linkcolor=blue,
urlcolor=blue
}

\date{}
\begin{document}
\title{Simulation of Thermal Relaxation in Spin Chemistry Systems on a Quantum Computer Using Inherent Qubit Decoherence}

\author{Brian Rost} 
\affiliation{IBM Research - Almaden, 95120 San Jose, CA}
\affiliation{Department of Physics, Georgetown University, 20057 Washington, DC}

\author{Barbara Jones} 
\affiliation{IBM Research - Almaden, 95120 San Jose, CA}

\author{Mariya Vyushkova} 
\affiliation{Center for Research Computing, University of Notre Dame, 46556 Notre Dame, IN}

\author{Aaila Ali} 
\affiliation{IBM Research - Almaden, 95120 San Jose, CA}
\affiliation{Department of Physics, DePaul University, 60614 Chicago, IL}
\affiliation{Department of Computer Engineering, Illinois Institute of Technology, 60616 Chicago, IL}

\author{Charlotte Cullip}
\affiliation{IBM Research - Almaden, 95120 San Jose, CA}
\affiliation{Departments of Chemistry and Computer Science, Occidental College, 90041 Los Angeles, CA}

\author{Alexander Vyushkov} 
\affiliation{Center for Research Computing, University of Notre Dame, 46556 Notre Dame, IN}

\author{Jarek Nabrzyski}
\affiliation{Center for Research Computing, University of Notre Dame, 46556 Notre Dame, IN}
\date{\today}

\begin{abstract}
Current and near term quantum computers (i.e. NISQ devices) are limited in their computational power in part due to qubit decoherence. Here we seek to take advantage of qubit decoherence as a resource in simulating the behavior of real world quantum systems, which are always subject to decoherence, with no additional computational overhead. As a first step toward this goal we simulate the thermal relaxation of quantum beats in radical ion pairs (RPs) on a quantum computer as a proof of concept of the method. We present three methods for implementing the thermal relaxation, one which explicitly applies the relaxation Kraus operators, one which combines results from two separate circuits in a classical post-processing step, and one which relies on leveraging the inherent decoherence of the qubits themselves. We use our methods to simulate two real world systems and find excellent agreement between our results, experimental data, and the theoretical prediction.
% Because current quantum devices are not yet robust enough to simulate the dynamics of this system, we classically precompute the isolated dynamics, encode this onto a pair of qubits and use the quantum computer only for the thermal relaxation. As quantum hardware improves both the dynamics and relaxation can be simulated simultaneously. We also provide a method to explicitly simulate the thermal relaxation by implementing decoherence channels using ancillary qubits. 
\end{abstract}
\maketitle
\section{Introduction}

As spin $1/2$ particles are ``nature's qubits" (simple two-state systems), electron spins map onto quantum devices in a straightforward, natural way. Because of that, it is interesting to focus on quantum applications of chemistry problems involving electron spin dynamics, such as spin chemistry simulations. Spin chemistry is an interdisciplinary subfield of physics and chemistry dealing with magnetic and spin effects in chemical reactions, which connect quantum phenomena like superposition and entanglement directly to macroscopic, measurable chemical parameters such as reaction yields, and make it possible to dynamically visualize those phenomena in chemistry experiments.
\par 
Historically, the fields of spin chemistry and quantum information science are closely related. On the one hand, spin chemistry expertise has contributed significantly to development of quantum computation, especially to spin-based qubit implementations\cite{Matsuoka_2016,Atzori_2019,Forbes_2019,Hore_2020,Nelson_2017,Wu_2018,Nelson_2020,Salikhov_2006,Volkov_2011}. At the same time, novel and promising approaches to spin chemistry problems (most prominently, avian magnetoreception mechanism), originating from the quantum information science field, have recently been introduced\cite{Kominis_2009,Cai_2010,Gauger_2011,Kominis_2011,Tiersch_2012_Deco,Hogben_2012,Pauls_2013,Kritsotakis_2014,Zhang_2015,Guo_2017,Vitalis_2017,Mouloudakis_2017,Kominis_2020,Fay_2020}.

\par Among spin chemistry problems, we have identified a simple yet experimentally important model simulation problem - quantum beats in radiation-generated radical pairs\cite{Bagryansky2007,molin2004,yu1999quantum,brocklehurst2002magnetic}. In a typical experiment, a burst of ionizing radiation is sent through a solution, forming radical pairs in a spin-correlated singlet state, due to spin conservation in the ionization process. This state then evolves due to hyperfine couplings (HFC), unequal Larmor precession rates (caused by unequal $g$-factors) in the presence of a magnetic field, and paramagnetic relaxation\cite{Bagryansky2007,molin2004,yu1999quantum,brocklehurst2002magnetic,bagryansky2005spin}. The radical pairs undergo spin-selective recombination reactions, resulting in a fluorescent product when recombining in the singlet state, but not when recombining in a triplet state. The dynamics of the spin state of the radical pairs, often called ``quantum beats", offers insights into the properties of the pairs' constituent molecules, allowing for determination of e.g. $\!g$-factors and HFC constants. Recently, a new theoretical approach treating spin-selective recombination of radical pairs as a quantum measurement that collapses the wavefunction into one of the total spin eigenstates (either singlet or triplet), has been suggested and actively discussed \cite{Kominis_2009,Kominis_2011,Kritsotakis_2014,Jones_2010,Shushin_2010,il2010should,Ivanov_2010,Purtov_2010,Jones_2011,Kominis_2011_Com,Jones_2011_Com,Tiersch_2012_Op,Dellis_2012,Bagryansky_2013}. Therefore, the radical pair simulation problem can be described as time evolution of qubit pairs initialized in an entangled state \cite{Hore_2020,Wu_2018,Nelson_2020}, where spin-selective recombination plays a role of quantum measurement, which makes the radical pair mechanism a very interesting and promising application for quantum simulation.

%For this paper we focus on radical pairs 
%2,2,6,6-tetramethyl-piperidine$+$.(TMP)/para-terphenyl-d14$-$.(PTP) 
%which undergo spin-selective recombination fluorescence. That is, they fluoresces when recombining in the singlet state, but not when recombining in a triplet state. A theoretical approach treating spin-selective recombination of radical pairs as a quantum measurement that collapses the wavefunction into one of the eigenstates (either singlet or triplet), has recently been suggested and actively discussed \cite{Kominis_2009,Kominis_2011,Kritsotakis_2014,Jones_2010,Shushin_2010,il2010should,Ivanov_2010,Purtov_2010,Jones_2011,Kominis_2011_Com,Jones_2011_Com,Tiersch_2012_Op,Dellis_2012,Bagryansky_2013}. This makes the RPM a very interesting and promising application for quantum simulation.
\par The intensity of the observed fluorescence is approximately given by\cite{Bagryansky2007,bagryansky2005spin}
\begin{equation}\label{eq:intense}
  I(t)=F(t)\left(\theta S(t)+\frac{1}{4}(1-\theta)\right)
\end{equation}
%$\theta\approx0.108$\cite{Bagryansky2007}
where $F(t)$ is the time-dependent radical pair recombination rate, $\theta$ is the experimentally determined fraction of recombination events occurring between RPs produced from the same precursor and $S(t)$ is the probability of a RP being in the singlet state, which generally depends on magnetic field strength $B$. The spin dynamics are extracted from the fluorescence profile using the time-resolved magnetic field effect (TR-MFE) method. This is done by taking a ratio of $I(t)$ at high-field and low-field to cancel $F(t)$, which has no $B$ dependence. A quantity, $Q$, subject to a high-field is denoted as $Q_B$ and subject to a low-field as $Q_0$. Using Eq. \ref{eq:intense} we have for the TR MFE
\begin{equation}\label{eq:MFE}
M(t)=\frac{I_B(t)}{I_0(t)}=\cancel{\frac{F(t)}{F(t)}}\left(\frac{4\theta S_B(t)+(1-\theta)}{4\theta S_0(t)+(1-\theta)}\right).
\end{equation}
In this paper we present two methods for computing various quantities of interest, subject to thermal relaxation, on currently available quantum computers. A quantity, $Q$, calculated for an isolated system is denoted as $\tilde Q$ when subject to thermal relaxation.
\par In section \ref{sec:theory} we cover the relevant theory of spin chemistry and thermal relaxation, as well as introducing the two real world systems we use to test our algorithms. In section \ref{sec:algorithm} we present our algorithms for simulating thermal relaxation in these systems and discuss their scope and scalability. In section \ref{sec:results} we present data taken from applying our algorithms to these systems on real quantum computers provided by IBMQ\cite{ibm}.

\section{Theory}\label{sec:theory}
\subsection{Calculation of \texorpdfstring{$S(t)$}{S(t)}}\label{sec:ST}
We consider systems of radical pairs in non-viscous solutions with no exchange between the pair's constitutes. Such systems are described by Hamiltonians of the form\cite{yu1999quantum,Mariya1,Bagryansky2007,Brocklehurst_1976}
\begin{equation}\label{eq:ham}
H=\sum_{i=1}^{2}\left[\mu_B g_i\vb{B}+\sum_{j=1}^{N_i}a_{ij}\vb{I}_{ij}\right]\cdot\vb{S}_i.
\end{equation}
Here $\mu_B$ is the Bohr magneton, $\vb{S}_i \ (g_i)$ is the spin ($g$-factor) of the unpaired electron in molecule $i$, $\vb{I}_{ij}$ refers to the $j^{\text{th}}$ nuclear spin in molecule $i$, $a_{ij}$ is the HFC constant between $\vb{S}_i$ and $\vb{I}_{ij}$, and $\vb{B}$ is the magnetic field.

\par From this Hamiltonian, we seek an expression for $S(t)$, the probability of a pair being in the singlet state at time $t$, given it was initially created in the singlet state at time $t=0$. Alternatively, $S(t)$ may be thought of as the fraction of pairs, which were initially in the singlet state, found in the singlet state at time $t$. While the initial electronic state of the pair preserves the spin multiplicity of its precursor\cite{yu1999quantum}, the initial nuclear state has no constraints. Therefore, following Brocklehurst\cite{brocklehurst2002magnetic}, we take the initial nuclear state of a newly formed RP to be the maximally mixed state (i.e. all nuclear states are equally likely). This gives an initial state of
\begin{equation}
  \rho(0)=\frac{1}{N_{\vb{I}}}\ket{S}\!\bra{S}\otimes I_{N_{\vb{I}}}\qq{with} \ket{S}=\frac{\ket{01}-\ket{10}}{\sqrt{2}}.
\end{equation} 
where $I_n$ is the $n\times n$ identity matrix and $N_{\vb{I}}$ is the size of the Hilbert space of the nuclear spins.
\par We then compute $S(t)$ by evolving $\rho(0)$ by time $t$ under the $H$ from Eq. \ref{eq:ham}, tracing out the nuclear degrees of freedom (DOF) and taking the expectation value of the resulting reduced electronic density matrix w.r.t. $\ket{S}$
\begin{equation}\label{eq:sb}
  S(t)=\bra{S}\Tr_{\vb{I}}\left\{e^{-iHt}\rho(0)e^{iHt}\right\}\ket{S}.
\end{equation}
\par There is no analytic expression for Eq. \ref{eq:sb} in general, although it is known for a variety of simple cases\cite{Bagryansky2007,brocklehurst2002magnetic,Bagryansky_2013}. Evaluating Eq. \ref{eq:sb} numerically is straightforward, however it becomes exponentially expensive in the number of nuclear DOFs.

\subsection{Thermal Relaxation}
\par The main mechanisms of relaxation in these systems are generalized amplitude damping (i.e. longitudinal paramagnetic relaxation or $T_1$ decay) and dephasing (i.e. transverse paramagnetic relaxation or $T_2$ decay)\cite{Brocklehurst_1976,bagryansky1997spin,brocklehurst2002magnetic,steiner1989magnetic,Bagryansky2007,yu1999quantum}. It can be shown that only the effective $T_1$ and $T_2$ for the pair are relevant to $\tilde S(t)$ so for simplicity we take $T_1$ and $T_2$ to be equal for both radicals. The following remains unchanged if $T_1$ and $T_2$ are different for each radical. In general, we have $T^{-1}=T_a^{-1}+T_c^{-1}$ \cite{bagryansky1997spin} where $a$ and $c$ stand for anion and cation respectively.

\par We derive an expression for the evolution of the singlet state probability, $\tilde{S}(t)$, subject to these decoherence channels, in terms of the singlet state probability for the isolated system, $S(t)$. In general, a proper description of the dynamics of a system undergoing both relaxation and coherent time evolution is very complicated because the relaxation channels and coherent time evolution do not commute. However, this is not always the case, as we demonstrate in this work for spin chemistry systems at room temperature.
%because the evolution of the electronic subsystem and the infinite temperature decoherence channels commute.
\par A quantum channel, $\mathcal{E}$, can be characterized by its Kraus operators, $K_i$, and performs the operation\cite{Preskill}
\begin{equation}\label{eq:channel}
  \mathcal{E}(\rho)=\sum_i K_i\rho K_i^\dagger.
\end{equation}

% Define commuting quantum channels as any two quantum channels, $\mathcal{E}_1$ and $\mathcal{E}_2$ that satisfy 
% \begin{equation}\label{eq:commute}
%   \left[\mathcal{E}_1,\mathcal{E}_2\right]=0\implies\mathcal{E}_1\left(\mathcal{E}_2(\cdot)\right)=\mathcal{E}_2\left(\mathcal{E}_1(\cdot)\right).
% \end{equation}
% \par A simple first step is to show that the generalized amplitude damping and dephaisng channels commute. 

The Kraus operators for the generalized amplitude damping channel, $\mathcal{E}_x$, are\cite{fujiwara2004ampdamp}:

\begin{equation}\label{eq:krausX}
\begin{split}
& K_0^x=
\sqrt{p_n}\left(
\begin{array}{cc}
1 & 0 \\
0 & \sqrt{\bar p_x} \\
\end{array}
\right)=\sqrt{p_n}\left(\frac{I\!+\!Z}{2}+\sqrt{\bar p_x}\frac{I\!-\!Z}{2}\right)
\\
&K_1^x=
\sqrt{p_n}\left(
\begin{array}{cc}
0 & \sqrt{p_x} \\
0 & 0 \\
\end{array}
\right)=\sqrt{p_n}\left(\frac{X\!+\!iY}{2}\right)
\\
&K_2^x=
\sqrt{\bar p_n}\left(
\begin{array}{cc}
\sqrt{\bar p_x} & 0 \\
0 & 1\\
\end{array}
\right)=\sqrt{\bar p_n}\left(\frac{I\!-\!Z}{2}+\sqrt{\bar p_x}\frac{I\!+\!Z}{2}\right)
\\
&K_3^x=
\sqrt{\bar p_n}\left(
\begin{array}{cc}
0 & 0 \\
\sqrt{p_x} & 0 \\
\end{array}
\right)=\sqrt{\bar p_n}\left(\frac{X\!-\!iY}{2}\right)
\end{split}
\end{equation}
Where $p_n=\left(1 + e^{-\beta}\right)^{-1}$ is the Fermi function which gives the equilibrium population of $\ket{0}$ as a function of inverse temperature, $\beta$. $p_n$ can be viewed as a temperature parameter controlling the ratio of damping of $\ket{1}$ and $\ket{0}$. $p_x$ is the damping parameter, and $\bar \alpha \equiv 1-\alpha$. $T\!=\!0 (\beta\!=\!\infty)\to p_n=1$ and we recover the standard amplitude damping channel, only $\ket{1}\mapsto\ket{0}$ occurs, with probability $p_x$. $T\!=\!\infty (\beta\!=\!0)\to p_n=1/2$ so $\ket{1}\mapsto\ket{0}$ and $\ket{0}\mapsto\ket{1}$ both occur with probability $p_x/2$. 

\par The dephasing channel, $\mathcal{E}_z$, is given by the Kraus operators
\begin{equation}\label{eq:krausZ}
K_0^z=\sqrt{\bar p_z}I
\qquad
K_1^z=\sqrt{p_z}Z
\end{equation}

\noindent which intuitively reads, ``With probability $p_z$ apply $Z$ otherwise do nothing".

It is shown in Appendix \ref{appendixC} that $\mathcal{E}_x$ and $\mathcal{E}_z$ commute. Define commuting quantum channels as any two quantum channels, $\mathcal{E}_1$ and $\mathcal{E}_2$ that satisfy 
\begin{equation}\label{eq:commute}
  \left[\mathcal{E}_1,\mathcal{E}_2\right]=0\implies\mathcal{E}_1\left(\mathcal{E}_2(\cdot)\right)=\mathcal{E}_2\left(\mathcal{E}_1(\cdot)\right).
\end{equation}
Denote $\mathcal{E}\left(\rho\right)\equiv\mathcal{E}_x\left(\mathcal{E}_z(\rho)\right)=\mathcal{E}_z\left(\mathcal{E}_x(\rho)\right)$. It is also shown that for Hamiltonians of the form given in Eq. \ref{eq:ham}, the time evolution of the electronic subsystem commutes with $\mathcal{E}^\infty$, $\mathcal{E}$ at infinite temperature. 

This is useful because, to a very good approximation, room temperature is infinite temperature for spin chemistry systems. The energies associated with the Hamiltonian in Eq. \ref{eq:ham} for even the larger $B$ fields used in spin chemistry experiments are orders of magnitude smaller than $k_B T$ at room temperature.

Note that the time evolution of the electronic subsystem is not a unitary operation and is itself given by a quantum channel, $\mathcal{E}_U$, such that
\begin{equation}
  \mathcal{E}_U(\rho)=\Tr_{\vb{I}}\left\{e^{-iHt}\rho \ e^{iHt}\right\}
\end{equation}

\par We are thus able to compute $\tilde{S}(t)$ by applying $\mathcal{E}^\infty$ and $\mathcal{E}_U$ only once, instead of needing to use more complicated methods, e.g. time ordered products or Trotterization, precisely because these channel commute. To find the correct $p_x$ and $p_z$ to apply at time $t$, we apply $\mathcal{E}^0$ ($\mathcal{E}$ at $T=0$) to a generic $\rho$ and equate it to the known iterative infinitesimal result\cite{Preskill}.
\begin{equation}
  \mathcal{E}^0\left(\rho\right):
  \begin{cases}
  \rho_{11}\mapsto\bar p_x\rho_{11}=e^{-t/T_1}\rho_{11}\\
  \rho_{10}\mapsto \sqrt{\bar p_x} \left(1-2 p_z\right)\rho _{10}=e^{-t/T_2}\rho _{10}
  \end{cases}
\end{equation}
which gives 
\begin{equation}
p_x=1-e^{\frac{-t}{T_1}}\qq{and} p_z=\frac{1}{2} \left(1-e^{-t\left(\frac{1}{2T_1}+\frac{1}{T_2}\right)}\right).
\end{equation}

\noindent Finally, $\tilde{S}(t)$ is computed as 
\begin{equation}\label{eq:stilde}
\begin{split}
  \bra{S}\mathcal{E}^\infty(\mathcal{E}_U(\rho(0)))\ket{S}=\bra{S}\Tr_{\vb{I}}\left\{\mathcal{E}^\infty(\rho(t))\right\}\ket{S}=\\
  =\frac{1}{4}\left(1+e^{-\flatfrac{t}{T_1}} + e^{-\flatfrac{t}{T_2}}\left(4S(t)-2 \right)\right)
\end{split}
\end{equation}
which agrees with Ref. \onlinecite{Bagryansky2007}. Although this result is already known, the algorithm given in the Sec. \ref{sec:Ex} relies on the fact that $\mathcal{E}^\infty$ and $\mathcal{E}_U$ commute. To the best of our knowledge, this has not previously been shown.

\section{Algorithms}\label{sec:algorithm}

\par We present two approaches for simulating the dynamics of these types of spin systems undergoing thermal relaxation on a quantum computer, and discuss the advantages and disadvantages of each. Both methods require implementing time evolution under the action of the Hamiltonian in Eq. \ref{eq:ham}. A wide variety of algorithms already exist for this step\cite{low2017optimal,berry2015simulating,whitfield2011simulation,babbush2018low}. The novel part in the approaches is the implementation of thermal relaxation.
\par Firstly, we explicitly simulate the generalized amplitude damping (Eq. \ref{eq:krausX}) and dephasing channels (Eq. \ref{eq:krausZ}). This has the advantage of being highly controllable since we are able to implement decoherence with any parameters we wish. However, it has the disadvantage of requiring a larger and deeper circuit which makes it less reliable, and scalable on NISQ devices.
\par Secondly, we leverage the decoherence of the qubits themselves to model the decoherence of the real spin system. This has the advantage that it requires no additional gates or qubits, so the circuit remains small and simple and should scale well to larger systems. However, it has the disadvantage that we are not in control of the decoherence parameters. A further complication is that the protocol requires waiting for the qubits to relax, which can amplify errors due to miscalibration. An in-depth discussion follows. 

\subsection{Explicit Decay, Kraus Method}\label{sec:Ex}
\par
\begin{figure}[htp]
	\includegraphics[width = \columnwidth]{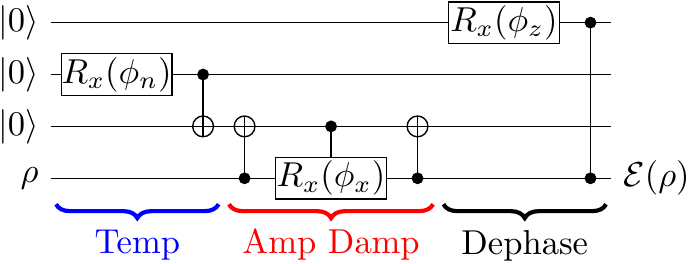}
	\caption{Circuit to perform generalized amplitude damping, $\mathcal{E}_x$, and dephasing, $\mathcal{E}_z$, on a qubit. $\phi_{x,z}=2 \sin ^{-1}\left(\sqrt{p_{x,z}}\right)$ and $\phi_{n}=2 \cos ^{-1}\left(\sqrt{p_{n}}\right)$.}
	\label{circ:ADDPgen}
\end{figure}

 We can implement the Kraus operators given in Eq. \ref{eq:krausX} and \ref{eq:krausZ} using the quantum circuit shown in Fig. \ref{circ:ADDPgen}, following the approaches given in \onlinecite{Barreiro2011, rost2020}. The five gates operating only on the bottom three qubits implement the generalized amplitude damping channel, $\mathcal{E}_x$, while the two gates involving the top qubit implement the dephasing channel, $\mathcal{E}_z$.
\par It is easy to make sense of this circuit by thinking of the standard ampltidue damping channel as: ``If the qubit is in $\ket{1}$, apply an $X$ gate with probability $p_x$.", the temperature correction part (added to give the generalized amplitude damping channel) as: ``with probability $p_n$ apply $X$ to the amplitude damping control line, i.e. damp $\ket{0}$ instead of $\ket{1}$," and similarly, the dephasing channel as: ``With probability $p_z$ apply a $Z$ gate."

The circuit in Fig. \ref{circ:ADDPgen} can be optimized greatly. Both the ``Temp" and ``Dephase" parts of the circuit in Fig. \ref{circ:ADDPgen} can be thought of as probabilistically implementing a gate ($X$ and $Z$ respectively). This means the rotation gate and controlled operation can replaced by a probabilistic gate. 
\par 1) We can run the circuit with and without each gate ($X$ or $Z$), with the gate operating where the target of the (now deleted) controlled operation was, then taking a weighted average over the results. The weights should be the probabilities of the (now deleted) rotation gate leaving the ancilla in $\ket{1}$, i.e. $w_{XZ}=\bar p_n p_z$, $w_{Z}= p_n p_z$, $w_{X}=\bar p_n \bar p_z$, $w_{I}=p_n \bar p_z$. Since the number of possible combinations grows exponentially as more qubits are added, this approach will not scale well. However, it is easy to implement using currently available quantum software and the systems under consideration here only have decoherence on the two electrons - making this our method of choice in this work.
\par 2) Alternatively, we can run the circuit with an $R_x(\theta_n)$ and $R_z(\theta_z)$, with the gate operating where the target of the (now deleted) controlled operation was, where $\theta_i$ is drawn from a normal distribution of mean $0$ and variance $\log \left((1-2 p)^{-2}\right)$ for each shot. This has the advantage that it avoids the need to compute exponentially many weights, but with the disadvantage that each shot needs a different circuit. This may or may not be an issue with future quantum compilers.
\par
\begin{figure}[htp]
	\includegraphics[width = \columnwidth]{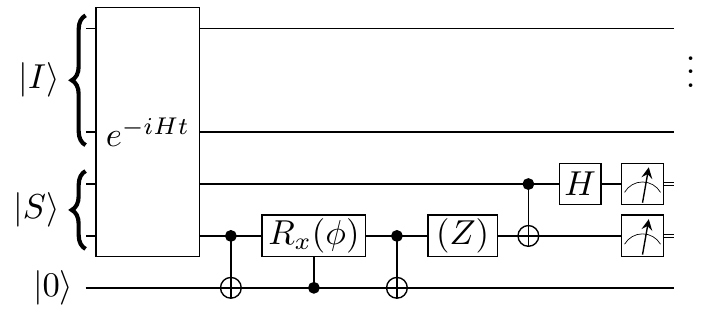}
	\caption{Circuit to simulate the dynamics of a spin system subject to thermal relaxation using the Kraus method. $\phi=2\sin^{-1}{\sqrt{p_x}}$ $(Z)$ indicates that the shots are to be split between implementing the circuit including the $Z$ gate and omitting the $Z$ gate. Results are then combined as a weighted average with weights $w_z=p_z$, $w_{\bar z}=\bar p_z$. The final $cX$ and $H$ gates transform from the Bell basis and map $\ket{S}\mapsto\ket{11}$ for measurement.}
	\label{circ:Kraus}
\end{figure}

 Another simplification can be made by noting that $\mathcal{E}^\infty$ acting on both qubits ($\mathcal{E}_2^\infty$) produces the same behavior as $\mathcal{E}^\infty$ acting on a single qubit ($\mathcal{E}_1^\infty$) twice. This is a consequence of the fact that $\mathcal{E}_2^\infty$, $\mathcal{E}_U$ and $\ket{S}$ are all invariant under $\ket{ab}\mapsto\ket{ba}$. This means we can eliminate half of the cost of implementing $\mathcal{E}_2^\infty$, because acting with $\mathcal{E}_1^\infty$ twice is the same as acting with $\mathcal{E}_1^\infty$ once but with $T_{1/2}\mapsto T_{1,2}/2$. Furthermore,
since $\mathcal{E}_1^\infty$, $\mathcal{E}_U$ and $\ket{S}$ are all also invariant under $\ket{\uparrow}\leftrightarrow\ket{\downarrow}$, we see that 
\begin{equation}\label{eq:mapSimp}
  \tilde{S}(t)=\bra{S}\mathcal{E}_2^\infty\!\!\left(\mathcal{E}_U(\ket{S}\!\!\bra{S})\right)\!\ket{S}=\bra{S}\mathcal{E}_1^0\!\left(\mathcal{E}_U(\ket{S}\!\!\bra{S})\right)\!\ket{S}
\end{equation}
because any losses in the population of $\ket{\uparrow\downarrow}$ (from $\ket{\uparrow\downarrow}\mapsto\ket{\downarrow\downarrow}$) are exactly offset by the gain in the population of $\ket{\downarrow\uparrow}$ (from $\ket{\uparrow\uparrow}\mapsto\ket{\downarrow\uparrow}$), and the temperature doesn't affect the coherences. This allows us to simply delete the ``Temp" part of the circuit in Fig. \ref{circ:ADDPgen}. Note that Eq. \ref{eq:mapSimp} does \textit{not} hold if $\mathcal{E}_1^0$ is replaced by $\mathcal{E}_2^0$ because losses in the population of $\ket{\uparrow\downarrow}$ are offset by gains in $\ket{\downarrow\uparrow}$ \textit{and} $\ket{\downarrow\downarrow}$.
Putting this all together, we get the circuit shown in Fig. \ref{circ:Kraus}.

%%%%%%%%%%%%%%%%%%%%%%%%%%%%%%%%%%%%%%%%%%%%%
\subsection{Leveraging Inherent Qubit Noise}\label{sec:inher}
%%%%%%%%%%%%%%%%%%%%%%%%%%%%%%%%%%%%%%%%%%%%%

An entirely different approach to simulating the thermal relaxation of a system is to try to take advantage of the inherent thermal relaxation of the qubits in the quantum computer. At the most basic level, the protocol is to simply run the simulation and rely on the qubits' natural decoherence behavior to implement the dissipative channels for free. This approach is appealing since simulating dissipative channels will become increasingly expensive in time and space as system sizes grow. It is also an interesting application of NISQ devices, whose noise often hinders quantum simulation. Here we hope to turn that around and use the noise as an integral part of the simulation. Such a protocol will be most beneficial for NISQ devices when gate errors are sufficiently low and qubit decoherence is the limiting factor in a simulation.
\par Despite the apparent simplicity, there are a variety of challenges inherent to such an approach. One immediate challenge that arises is that the qubits are an effective zero-temperature system while the spin chemistry systems are effectively infinite temperature. Furthermore, a more practical issue is that during the long time-scales for this protocol systematic errors tend to be amplified. Specifically, the errors from effective $Z\otimes I$ and $I\otimes Z$ terms arising due to qubit frequency miscalibration\cite{mckay2017efficient} were particularly problematic in our simulations.

%are particularly noticeable on the IBM quantum computers. We overcome both of these issues by introducing echo pulses ($X$ gates) on the qubits representing the electronic degrees of freedom.

% \par We overcome both of these issues by introducing echo pulses ($X$ gates) on the qubits representing the electronic degrees of freedom. Whenever an $X$ gate occurs, the relative phase accumulation due to the $ZI$ and $IZ$ errors is reversed and further phase accumulation cancels with, rather than adding to, the previously accumulated phase. Furthermore, since the zero temperature amplitude damping experienced by the qubits sends $\ket{1}\mapsto\ket{0}$ and an $X$ gate flips $\ket{0}\leftrightarrow\ket{1}$, the application of an $X$ gate reverses the direction of the amplitude damping. By performing these echo gates at well chosen intervals, we are able to effectively mitigate the frequency errors as well as mimic the infinite temperature decay to arbitrary accuracy.

\par The qubits in IBMQ's machines are always precessing under the action of the Hamiltonian term $\Omega \sigma_z/2$\cite{Qiskit}. I.e. a qubit in state $\ket{\psi}$ at $t=0$ will evolve to $e^{-i \Omega \sigma_z t/2}\ket{\psi}=R_z(\Omega)\ket{\psi}$ after time $t$. This precession is corrected for in software, but miscalibrations of $\Omega$ will result in imperfect corrections. If the true frequency is $\Omega$ but the backend is calibrated to $\omega=\Omega-\Delta \omega$, then after a time $t$ the corrected state will be $R_z(\Delta \omega t)\ket{\psi}$ instead of $\ket{\psi}$.
\par Because $\Delta \omega t$ tends to be small for $t$ values associated with typical circuits, this error tends to be negligible. However, our circuits run to long enough times where this becomes a significant error. We can mitigate this error by introducing $X$ gates (aka echo pulses) into our circuit, which has the effect of flipping the sign of the phase, so further phase accumulation will cancel instead of compound\cite{jurcevic2020demonstration,sundaresan2020reducing}. It is easy to verify that using an even number of echo pulses, $n$, at $(k+1/2)t/n, \ k\in\{0,1,\ldots, n-1\}$ corrects this error
$$R_z\!\left(\frac{\Delta \omega t}{2n}\right)\cdot \prod_{k=1}^n\left[XR_z\!\left(\frac{\Delta \omega t}{n}\right)\right]\cdot R_z\!\left(\frac{\Delta \omega t}{2n}\right)=I$$

\par
\begin{figure}[htp]
	\includegraphics[width = \columnwidth]{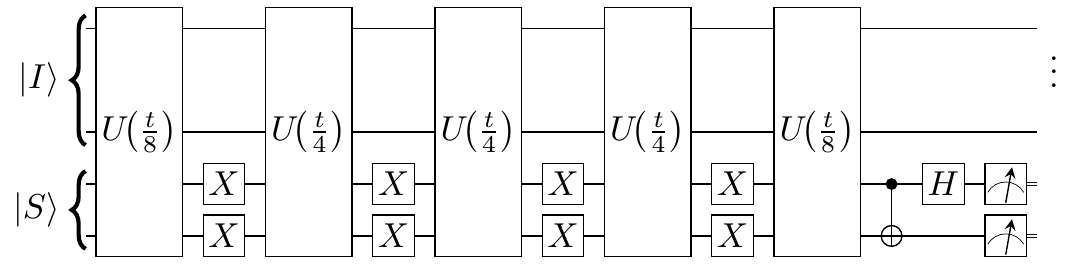}
	\caption{Circuit to simulate the dynamics of a spin system subject to thermal relaxation using the Inherent method where $U(t)=e^{-iHt}$. The final $cX$ and $H$ gates transform from the Bell basis and map $\ket{S}\mapsto\ket{11}$ for measurement.}
	\label{circ:Inher}
\end{figure}

 Additionally, these $X$ gates also have an effect on the decay characteristics of the state. As the qubits are at $T\approx0$ they are always decaying toward $\ket{0}$. The $X$ gates flip $\ket{0}\leftrightarrow\ket{1}$ so the amplitude damping proceeds ``in reverse" while the dephasing is unaffected. One would expect this to be a crude approximation to the infinite temperature amplitude damping where a qubit is decaying toward an equal mixed state of $\ket{0}$ and $\ket{1}$. Indeed as $n\to\infty$ we recover the exact $\mathcal{E}^\infty$. Denote the probability of measuring $\ket{S}$ after running the protocol using $n$ (even) echo gates as $S^e(t)$. Then we find

\begin{equation}\label{eq:tempDis}
\tilde S(t)=S^e(t)-\left(\frac{\left(e^{-\frac{t}{T_1}}-1\right) \sinh ^2\left(\frac{t}{4 n T_1}\right)}{\cosh \left(\frac{t}{2 n T_1}\right)}\right)^2
\end{equation}
\noindent and $S^e(t)$ converges to $\tilde S(t)$ for all $t$ as $n$ increases. However, increasing $n$ also incurs errors due to imperfect $X$ gates. In this work, we find using $n=4$ gives good results and the corresponding circuit is shown in Fig. \ref{circ:Inher}.

\par Another potential issue is that the $T_1$ and $T_2$ of the qubits may be significantly longer than typical simulation times. Typically the timescales for decoherence in superconducting qubits are $10$s of microseconds\cite{Qiskit} and potentially much longer for other technology, like trapped ions\cite{wang2017single}.
\par This issue can be overcome by slowing the execution of the circuit. This can potentially by done by, for example, pulse stretching\cite{kandala2019error}, using large numbers of small Trotter steps and/or adding wait cycles into the circuit. Pulse stretching is especially appealing because it is continuously applying $\mathcal{E}_U$ while $\mathcal{E}^0$ is naturally occurring, overcoming the issue associated with the fact that $\left[\mathcal{E}_U,\mathcal{E}^0\right]\neq0$. It also uses fewer gates than applying a large number of small trotter steps. Adding in wait cycles between trotter steps is the easiest to implement but also incurs the largest error due to $\left[\mathcal{E}_U,\mathcal{E}^0\right]\neq0$.
\par Additionally, the ratio $T_1/T_2$ may be quite different between the radical pair and the qubits. One can always use the Kraus method above to implement dephasing to lower $T_2$ as far as desired. We show another general approach in the following subsection that can be used to fix the decoherence parameter mismatch between qubits and radical pairs.

\subsection{Correction Circuits}\label{sec:corrCirc}
In principle, the time evolution and decoherence can be run separately and the results combined in a classical post-processing step to recover the full dynamics. We do this as follows:\\
\par Step 1) Implement the time evolution with no decoherence, i.e. use the circuit in Fig. \ref{circ:Kraus} without the decoherence part. Let the measured populations of $\{\ket{S},\ket{T_0},\ket{T_{\pm}}\}$ for this step be $\{S,T_0,T_{\pm}\}$ respectively.\\
\par Step 2) In parallel, apply decoherence to the singlet state with no coherent time evolution, i.e. use circuit in Fig. \ref{circ:Kraus} with $t=0$ or Fig. \ref{circ:Inher} with $U(t)$ replaced by identity gates. Let the measured populations of $\{\ket{S},\ket{T_0},\ket{T_{\pm}}\}$ for this step be $\{S',T_0',T'_{\pm}\}$ respectively.\\
\par Step 3) Combine the results from steps 1 and 2 to recover the thermally relaxed dynamics, $\tilde S$
\begin{equation}\label{eq:corS}
  \tilde S = SS'+T_0T_0'+T_{+}T'_{+}+T_{-}T'_{-}
\end{equation}
Because of the symmetry in $\mathcal{E}^\infty$, the probability of $\ket{S}\mapsto\ket{T}$, where $\ket{T}$ is any triplet state, is equal to the probability $\ket{T}\mapsto\ket{S}$. In this way we can view $\{S,T_0,T_{\pm}\}$ as the undamped populations and $\{S',T_0',T'_{\pm}\}$ as the probability of those states transitioning to $\ket{S}$.
\par This method can be used to add decoherence to a simulation implemented without decoherence. Additionally, this method can be used to ``undo" any decoherence occurring during a simulation, which is often unavoidable on NISQ devices. Given simulation results subject to $T_1$ and $T_2$ decay i.e. $\{\tilde S,\tilde T_0,\tilde T_{\pm}\}$ and $\{S',T_0',T'_{\pm}\}$, this provides a procedure to recover the undamped dynamics. This can be done by using Eq. \ref{eq:corS} together with 
\begin{equation}\label{eq:CorrUndo}
\begin{split}
  \tilde T_0 &= T_0S'+ST_0'+T_{+}T'_{+}+T_{-}T'_{-}\\
  \tilde T_{\pm}&= T_{+}+T'_{+}-4T_{+}T'_{+}
\end{split}
\end{equation}
to solve for the undamped populations $\{S,T_0,T_{\pm}\}$.
\par Furthermore, the ability of correction circuits to both apply and ``undo" decoherence allows for more flexible decoherence protocols. In particular when using the inherent method, if the qubits have a $T_1/T_2$ value which is different from that of the radical pairs, the decoherence dynamics of the qubits will not mimic that of the system. In this situation, we can apply the correction circuit procedure twice to fix these issues and recover the desired decay dynamics of the radical pair system. The idea is to first use the correction circuit method to ``undo" the decoherence from the qubits and then use the method a second time in conjunction with the Kraus method to apply the desired decoherence.
%\par Furthermore, and of particular interest here, given a set of qubits whose $T_1/T_2$ value is different from that of the radical pairs, this provides a procedure to transform the results from the inherent decay of the qubits to the decay of the radical pair system. The general strategy is to use the correction circuit method to ``undo" the decoherence from the qubits and then use the method a second time in conjunction with the Kraus method to apply the desired decoherence. This can be useful when implementing the decoherence directly using the Kraus.

To do this, we perform the time evolution with decoherence using the inherent method yielding $\{\tilde S,\tilde T_0,\tilde T_{\pm}\}$. We also perform the decoherence using the inherent method without time evolution, giving us $\{S',T_0',T'_{\pm}\}$. Next, we decohere the singlet state using the Kraus method (circuit in Fig. \ref{circ:Kraus} without the time evolution), inputting the desired $T_1$ and $T_2$ parameters, giving $\{S'',T_0'',T''_{\pm}\}$. We then use $\{\tilde S,\tilde T_0,\tilde T_{\pm}\}$ and $\{S',T_0',T'_{\pm}\}$ to recover $\{S,T_0,T_{\pm}\}$ via Eqs. \ref{eq:corS}, \ref{eq:CorrUndo} and then apply Eq. \ref{eq:corS} with these values and $\{S'',T_0'',T''_{\pm}\}$ to produce the correct $\tilde S(t)$ with the radical pair and qubits having arbitrary $T_1$ and $T_2$ values. In certain situation, e.g. $T_1\to\infty$ as outlined in the next section, this protocol may simplify. 

\section{Applications}\label{sec:results}

\begin{figure}[htp]
	\includegraphics[width = \columnwidth]{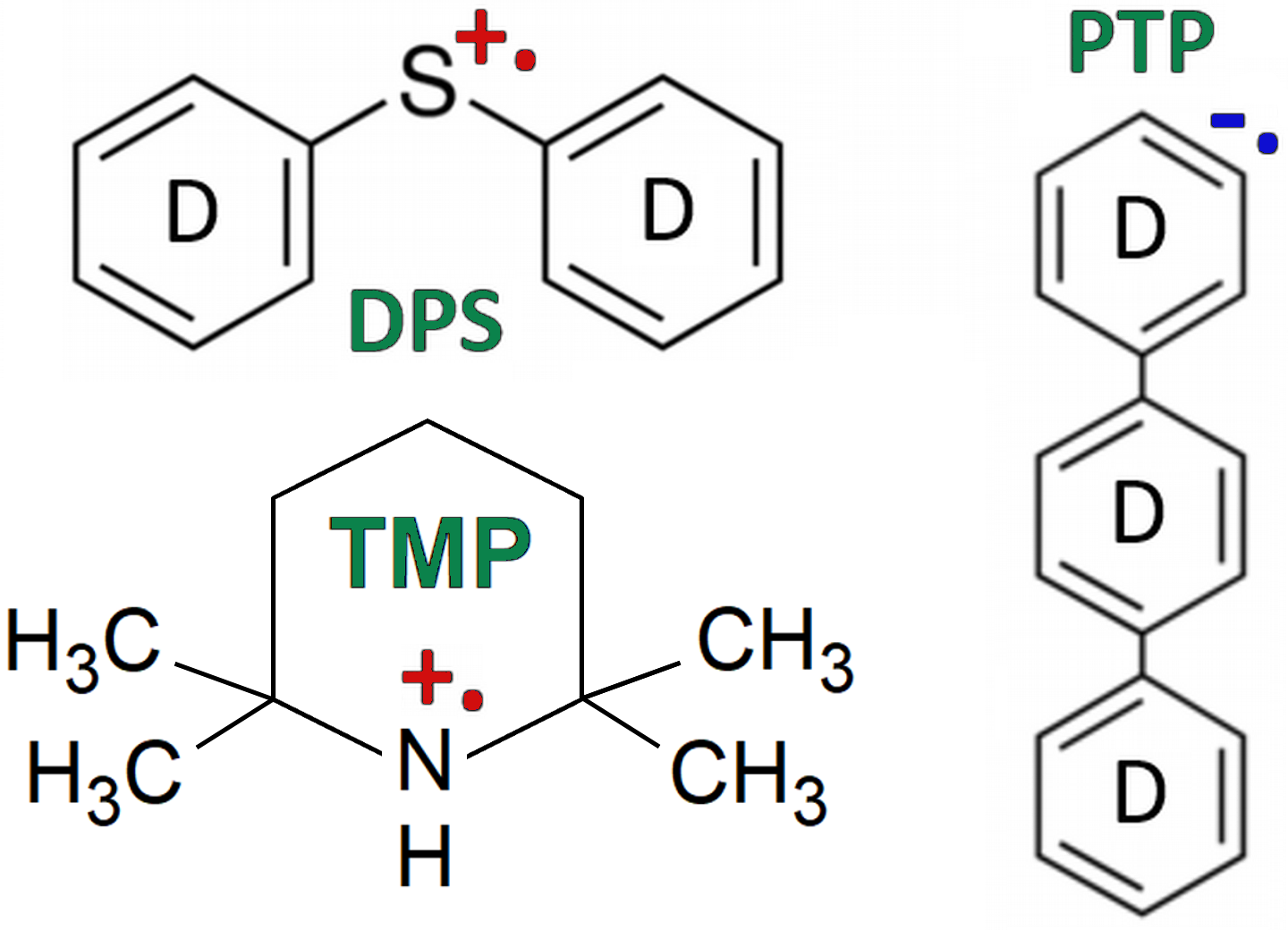}
	\caption{Chemical structures of radical ions DPS, TMP and PTP which are used in the remainder of the paper.}
	\label{fig:chem}
\end{figure}

In this section we benchmark our methods by using them to simulate two different radical pair systems that have been previously studied in spin chemistry experiments. The first pair consists of a perdeuterated para-terphenyl radical anion (PTP) and a perdeuterated diphenyl sulphide radical cation (DPS) and the second pair consists of a 2,2,6,6-tetramethylpiperidine radical cation (TMP) and PTP. The chemical structures of these species are shown in Fig. \ref{fig:chem}.

\subsection{Quantum beats in diphenyl sulfide-d\texorpdfstring{$_{10}^{+\cdot}$}{+.} (DPS) / para-terphenyl-d\texorpdfstring{$_{10}^{-\cdot}$}{-.} (PTP)}\label{sec:DPS}

\par We begin with an application to a very simple system, the radical pair perdeuterated diphenyl sulfide$^{+\cdot}$ (DPS) and perdeuterated para-terphenyl$^{-\cdot}$ (PTP) in a dilute alkane solution. In this system, hyperfine couplings observed in the experiment are largely negligible due to the much smaller magnetic moment of deuteron, and due to fast ion-molecule electron exchange among the diphenyl sulfide molecules in solution\cite{bagryansky1997spin}.
% \begin{figure}[htp]
% 	\includegraphics[width = \columnwidth]{Mol2}
% 	\caption{Structures of DPS and PTP radical ions}
% 	\label{fig:DPSPTP}
% \end{figure}
\noindent However, while largely negligible, the large number of small hyperfine couplings can be approximated semi-classically by treating the effects of these couplings as a Gaussian dephasing term\cite{Molin1993}. In this case, the expression Eq. \ref{eq:stilde} for $\tilde S(t)$ becomes\cite{bagryansky1997spin}:

\begin{equation}\label{eq:MGaussian}
\tilde S(t)=
\frac{1}{4}\left(1+e^{-\flatfrac{t}{T_1}} + e^{-\flatfrac{t}{T_2}-\sigma^2 t^2}\left(4S(t)-2 \right)\right)
\end{equation}
Where\cite{Molin1993} $\sigma=\sqrt{\dfrac{1}{3}\sum_n a_n^2I_n(I_n+1)}$ is the second moment of the Gaussian distribution of the hyperfine components. This is interesting to note, because superconducting qubits are also potentially subject to such broadband dephasing processes from things like flux noise, charge noise and critical-current noise\cite{krantz2019quantum}. Indeed we did observe Gaussian type dephasing noise in some of our runs, but until these noise sources are more predictable and well documented, it will be difficult to leverage this noise as a resource.

\par If we neglect the HFC in PTP, we can directly simulate the Hamiltonian time evolution on currently available devices - due to the simple form the Hamiltonian takes:

\begin{equation}\label{eq:hamDPS}
  H=\mu_B\vb{B}\cdot\left(g_1\vb{S}_1+g_2\vb{S}_2\right)\propto \alpha S_1^z+\beta S_2^z.
\end{equation}

where $g_1=2.0028$, $g_2=2.0082$, $\vb{B}=B\hat{\vb{z}}$ and $B= 17$mT for low-field and $960$mT for high-field\cite{bagryansky1997spin}. Note that this simple Hamiltonian gives $$U(t)=e^{-iHt}=%e^{-i\omega_1Z_1t/2}e^{-i\omega_2Z_2t/2}=
R_z(\omega_1 t)\otimes R_z(\omega_2 t)$$ where $\omega_i$ is proportional to $g_i$. In this case, with no hyperfine couplings present in both radicals, singlet-triplet ($\ket{S}\leftrightarrow\ket{T_0}$) oscillations in external magnetic field result from the difference in Larmor precession rates of the two electron spins caused by difference in their g-factors 
%(Fig. \ref{fig:Lamor}) 
with no transitions to $\ket{T_{\pm}}$. The singlet state population in this case is described by the expression\cite{Bagryansky2007}

% \begin{equation}
% S_B(t)=\frac{1}{2}\left[\vphantom{\frac{1}{2}}1+\cos\left((\omega_1-\omega_2)t\right)\right]
% \end{equation}
\begin{equation}
S_B(t)=\cos^2\left(\frac{\omega_1-\omega_2}{2} \ t\right)
\end{equation}

% \begin{figure}[htp]
% 	\includegraphics[width = \columnwidth]{Larmor}
% 	\caption{Vector diagram representing singlet-to-triplet oscillations in a radical pair in a magnetic field with no HFCs on either radical. Figure adapted from Ref.~\onlinecite{Bagryansky2007}}
% 	\label{fig:Lamor}
% \end{figure}

\par
\begin{figure}[htp]
	\includegraphics[width = \columnwidth]{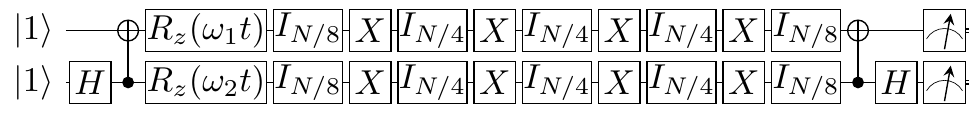}
	\caption{Circuit to simulate the dynamics and thermal relaxation of the DPS/PTP radical pair. $I$ is the identity gate, $N$ is given by \ref{eq:numID} and $\omega_i\propto g_i$. $\tilde S(t)$ is given as the probability of measuring $\ket{11}$. This circuit uses four echo pulses ($XX$ gates) to mitigate errors from frequency miscalibration and to approximate the infinite temperature dynamics. The deviation from the true infinite temperature dynamics is given in Eq. \ref{eq:tempDis}. Over the time range simulated, the maximum deviation between the theoretical $\tilde S(t)$ and the ideal result of this circuit is less than $3\times 10^{-5}$.}
	\label{circ:DPSxx}
\end{figure}

 Since the Hamiltonian is composed of only $S^z$ operators, it's easy to see that $\left[\mathcal{E}_U,\mathcal{E}^0\right]=0$. This allows us to do all of the time evolution in one step and then the decoherence in the next. For this system, we employ the inherent method (Sec. \ref{sec:inher}). For this simple case, the time evolution is accomplished with $R_z$ gates and the decoherence step is just the echo pulses with pauses in between. We implement the pauses in IBMQ by using identity gates, which act as wait cycles. We choose the number of identity gates to use as
\begin{equation}\label{eq:numID}
N=\frac{t \ T_{qu}}{T_{rp} \ t_{id}}
\end{equation} 
where $t$ is the simulated time, $T_{qu}$ is the average of $T_1$ and $T_2$ over both qubits, $T_{rp}$ is the decay time of the radical pairs and $t_{id}$ is the duration of the identity gate. $T_{qu}$ and $t_{id}$ are provided by the backend in Qiskit. The circuit implementing this protocol is shown in Fig. \ref{circ:DPSxx}.

\par As a proof of concept of the Inherent method (Sec. \ref{sec:inher}), we begin by simulating the circuit shown in Fig. \ref{circ:DPSxx}, with thermal noise added to the qubits using Qiskit AER's ``thermal relaxation error" noise model. Qiskit takes as input the qubits temperature ($0$), time, $T_1$ and $T_2$ and then simulates the circuit assuming the qubits thermally relax following amplitude damping and dephasing. Additionally, we generate the semi-classical broadband dephasing noise ($\sigma$ from Eq. \ref{eq:MGaussian}) by modifying Qiskit AER's ``thermal relaxation error" module to assess the effect of neglecting the HFC in PTP. Note that one can easily implement this semi-classical dephasing on a quantum computer using the Kraus method outlined in Sec. \ref{sec:Ex} at essentially no computational cost. We run both circuits at 5,000 shots, which matches our runs on the actual quantum computers.

\par
\begin{figure*}[htp]
\begin{subfigure}[t]{0.66\columnwidth}
	\includegraphics[width = \columnwidth]{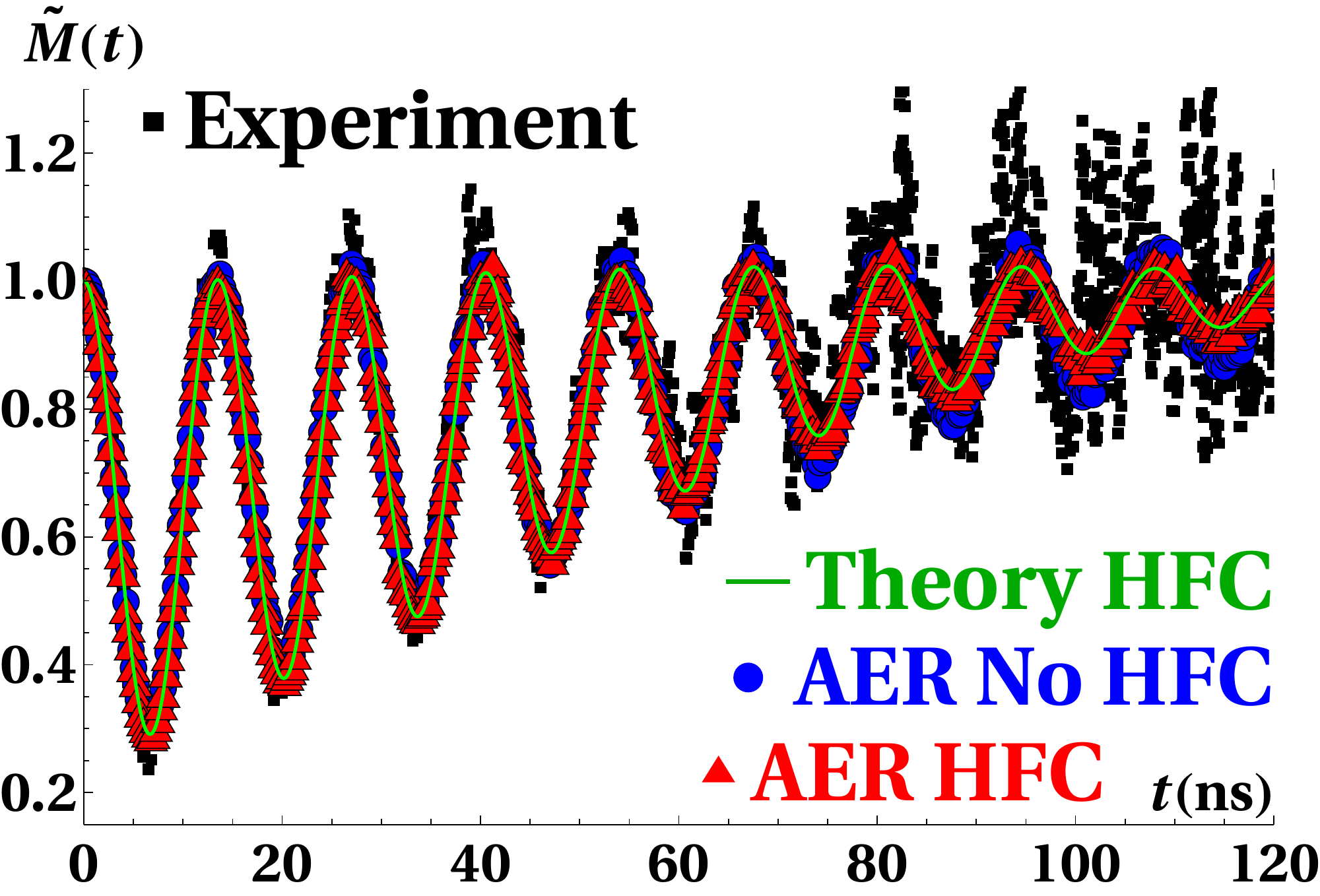}
	\caption{Results for $\tilde M(t)$ from Qiskit AER with thermal relaxation, both including and not including HFC, compared with theory and experimental data. Both with and without HFC, the AER data agree very well with the theory and experiment; we neglect the HFC in PTP moving forward.}
	\label{fig:MFEDPSAER}
	\end{subfigure}%
	~
\begin{subfigure}[t]{0.66\columnwidth}
	\includegraphics[width = \columnwidth]{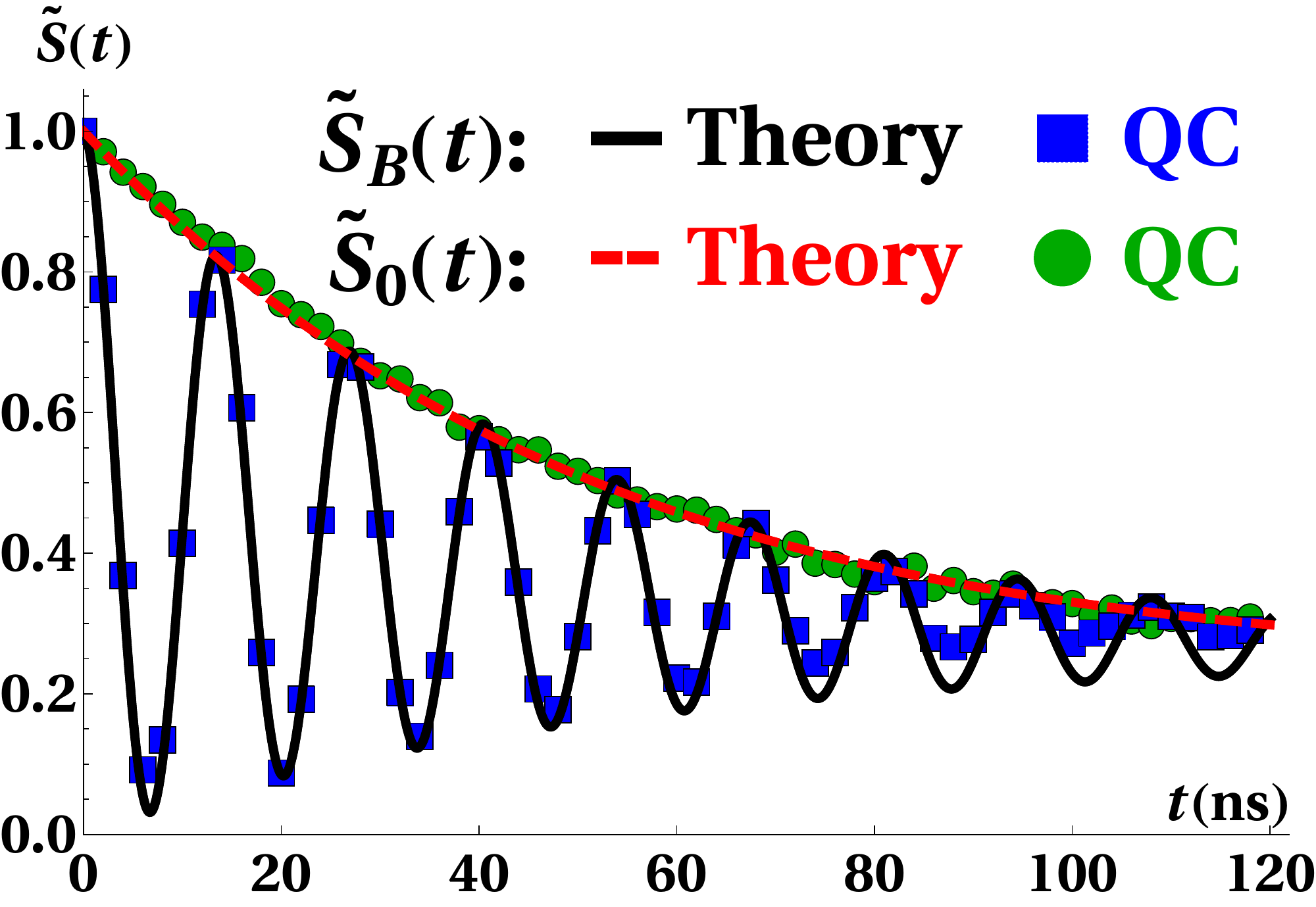}
	\caption{Results for $\tilde S(t)$, for both high- and low-fields, from the quantum computer compared to theory (without HFC). $\tilde S_0(t)$ has a MSE of 0.010\% and $\tilde S_B(t)$ has a MSE of 0.064\% compared to theory.}
	\label{fig:DPSHighLow}
	\end{subfigure}%
	~
\begin{subfigure}[t]{0.66\columnwidth}
	\includegraphics[width = \columnwidth]{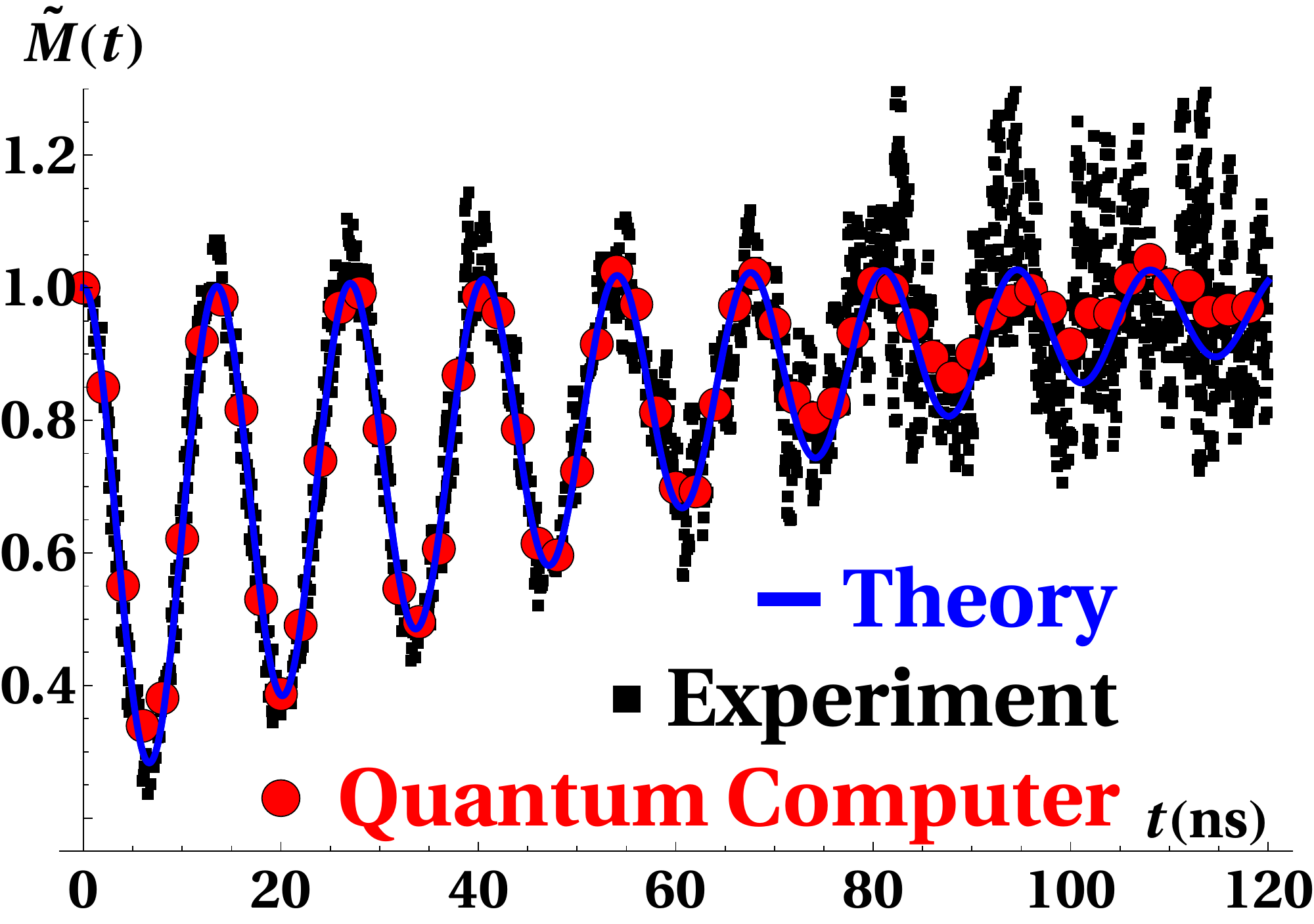}
	\caption{Results for $\tilde M(t)$ from combining the data in Fig. \ref{fig:DPSHighLow} as per Eq. \ref{eq:MFE} compared with theory (without HFC) and experimental data. $\tilde M(t)$ has a MSE of 0.073\% compared to theory.}
	\label{fig:MFEDPSQC}
	\end{subfigure}
\caption{Data for the TR MFE ($\tilde M(t)$) and $\tilde S(t)$ at high- and low-fields from running the Inherent method for the DPS/PTP radical pair (circuit in Fig. \ref{circ:DPSxx}) vs. theoretical calculations and experimental data. Runs were conducted by simulating our circuit in Qiskit AER with thermal noise as well as on IBM's Toronto quantum computer, both using 5,000 shots. For runs including the HFC of PTP we take $T_1=T_2=60$ns and take $T_1=T_2=50$ns for runs without HFC. $\tilde M(t)$ is constructed from $\tilde S(t)$ as per Eq. \ref{eq:MFE} with $\theta=0.425$\cite{usov1997highly,bagryansky1997spin}}
	\label{fig:DPSResults}
\end{figure*}

 These results are shown in Fig. \ref{fig:MFEDPSAER} and are in excellent agreement with both theory and experiment. We find that for PTP, the semi-classical hyperfine coupling approximation gives a correction much smaller than the other sources of uncertainty (e.g. experimental uncertainty in $T_1$/$T_2$ values\cite{bagryansky1997spin}) and so we will neglect it moving forward.

% \begin{figure}[htp]
% 	\includegraphics[width = \columnwidth]{MFEDPSAER}
% 	\caption{Results of simulating our protocol using Qiskit AER's ``thermal relaxation error", both including and not including the additional quadratic dephasing term, compared with the theoretical calculation and experimental data. We take $\theta=0.425$ and use $T_1=T_2=60$ns for HFC, and $T_1=T_2=50$ns without. Both with and without HFC, the AER data agree very well with the theory and experiment; we neglect the HFC in PTP moving forward.}
% 	\label{fig:MFEDPSAER}
% \end{figure}

\par Having confirmed the validity of neglecting the HFC in PTP, we proceed to run the circuit shown in Fig. \ref{circ:DPSxx} on IBMQ's Toronto quantum computer for both high- and low-field. These results are shown in Fig. \ref{fig:DPSHighLow} and show excellent agreement with the theory. We do note that our high-field run appears to have been subjected to some broadband dephasing noise, which becomes apparent at long times, in addition to the expected dephasing and amplitude damping.

% \begin{figure}[htp]
% 	\includegraphics[width = \columnwidth]{DPSHighLowData}
% 	\caption{Comparison of theoretical thermal relaxation vs. running the Inherent method (circuit shown in Fig. \ref{circ:DPSxx}) on the IBMQ Toronto quantum device.}
% 	\label{fig:DPSHighLow}
% \end{figure}

\par The data from Fig. \ref{fig:DPSHighLow} is then combined to calculate $\tilde M(t)$ as per Eq. \ref{eq:MFE}, using $\theta=0.425$\cite{bagryansky1997spin}. The result is shown in Fig. \ref{fig:MFEDPSQC} alongside the experimental results\cite{bagryansky1997spin} as well as our theoretical calculation. The simulation data are in excellent agreement with both the experimental and theory data, aside from the minor effects of broadband dephasing at long times.

% \begin{figure}[htp]
% 	\includegraphics[width = \columnwidth]{MFEDPSQC}
% 	\caption{Comparison of theory (without PTP HFC), experimental data and results of combining the data in Fig. \ref{fig:DPSHighLow} (from Q.C.) for the DPS/PTP radical pair.}
% 	\label{fig:MFEDPSQC}
% \end{figure}

\subsection{Quantum beats in 2,2,6,6-tetramethylpiperidine\texorpdfstring{$^{+\cdot}$}{+.}/para-terphenyl-d\texorpdfstring{$_{14}^{-\cdot}$}{-.}}\label{sec:TMP}

\par Now looking at another system, we consider the radical pair 2,2,6,6-tetramethylpiperidine$^{+\cdot}$ (TMP) and para-terphenyl-d$_{14}^{-\cdot}$ (PTP). As before, this radical pair can be formed by the passage of a burst of radiation through the solution of the two compounds, resulting in ionization of solvent molecules and subsequent capture of electrons by PTP molecules and holes by TMP molecules.
\par TMP is a radical cation, and its unpaired electron spin is localized on the nitrogen atom. There are two magnetic nuclei interacting with the electron spin, nitrogen with a nuclear moment of $1$, and amine hydrogen with a nuclear moment of $\frac{1}{2}$. 
\par In the PTP radical anion the unpaired electron spin density is delocalized, and hyperfine couplings are very small due to small magnetic moments of the deuterium nuclei and are neglected in this treatment, as justified in Sec. \ref{sec:DPS}. However, the HFC in TMP cannot be neglected because the electron density is localized and the couplings are much stronger.

% \begin{figure}[htp]
% 	\includegraphics[width = 0.25\columnwidth]{TMP}
% 	\caption{The radical TMP}
% 	\label{fig:TMP}
% \end{figure}

The Hamiltonian describing this system is
\begin{equation}
H=\mu_B\vb{B}\cdot\left(g_1\vb{S}_1+g_2\vb{S}_2\right)+a_H\vb{I}_H\cdot\vb{S}_1+a_N\vb{I}_N\cdot\vb{S}_1
\end{equation}
Here $\vb{S}_1$ and $\vb{S}_2$ refer to the electron spins in TMP and PTP respectively, $\vb{I}_i$ refers to the nuclear spin of species $i$ and $\vb{B}$ is the magnetic field. This is similar to the DPS/PTP Hamiltonian (Eq. \ref{eq:hamDPS}), with the addition of hyperfine couplings to the nitrogen and hydrogen nuclei on the TMP (last two terms in the Hamiltonian). The hydrogen and nitrogen HFC constants determined experimentally are $a_H = - 1.87$mT, and $a_N =1.8$mT \cite{bagryansky2005spin}. 

\par One large difference between the TMP and DPS radical pair with PTP is that the $T_1$ decay time is dependent on the magnetic field for TMP/PTP. In particular, we have $T_1=T_2$ for the low-field but $T_1\gg T_2$ for the high field. Therefore, to use the inherent method for high field, we will need to invoke a correction circuit as described in section \ref{sec:corrCirc}. This will be described in more detail below.

\par Simulating the coherent time evolution under this Hamiltonian is no longer trivial like for DPS/PTP. As of this writing, simulating this system out to times of interest is beyond the capabilities of current quantum computers. Because using the quantum computer to simulate $S(t)$ directly is not yet feasible, we demonstrate our protocols by classically precomputing $S(t)$, as described in \ref{sec:ST}, and encoding it onto a pair of qubits. 

\par
\begin{figure}[htp]
\begin{subfigure}[t]{0.49\columnwidth}
	\includegraphics[width = \columnwidth]{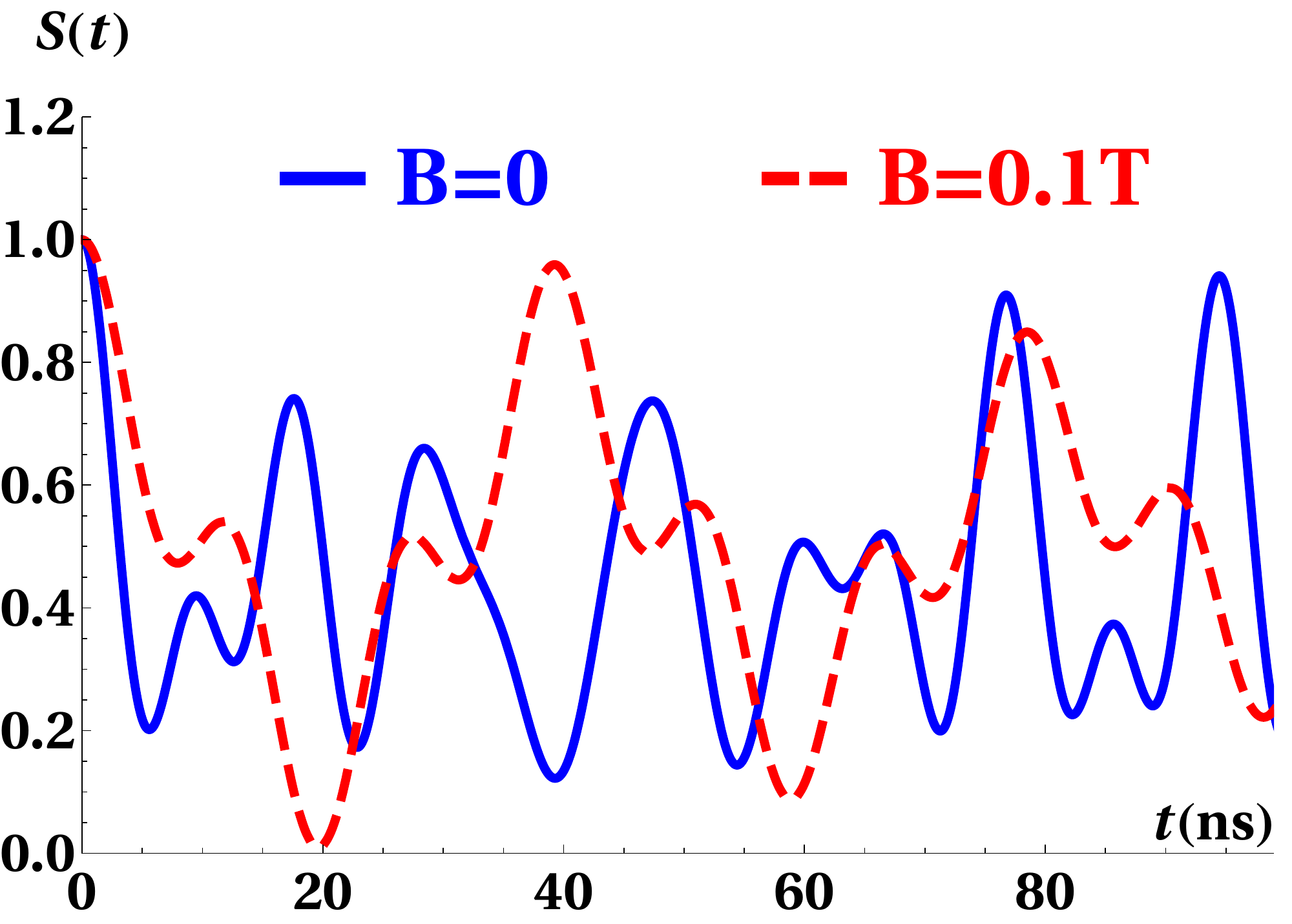}
	\caption{Plot of $S(t)$ for $B=0$ and large $B=0.1T$ corresponding to the fields used in the experiment.}
	\label{fig:ST}
	\end{subfigure}%
	~
\begin{subfigure}[t]{0.49\columnwidth}
	\includegraphics[width = \columnwidth]{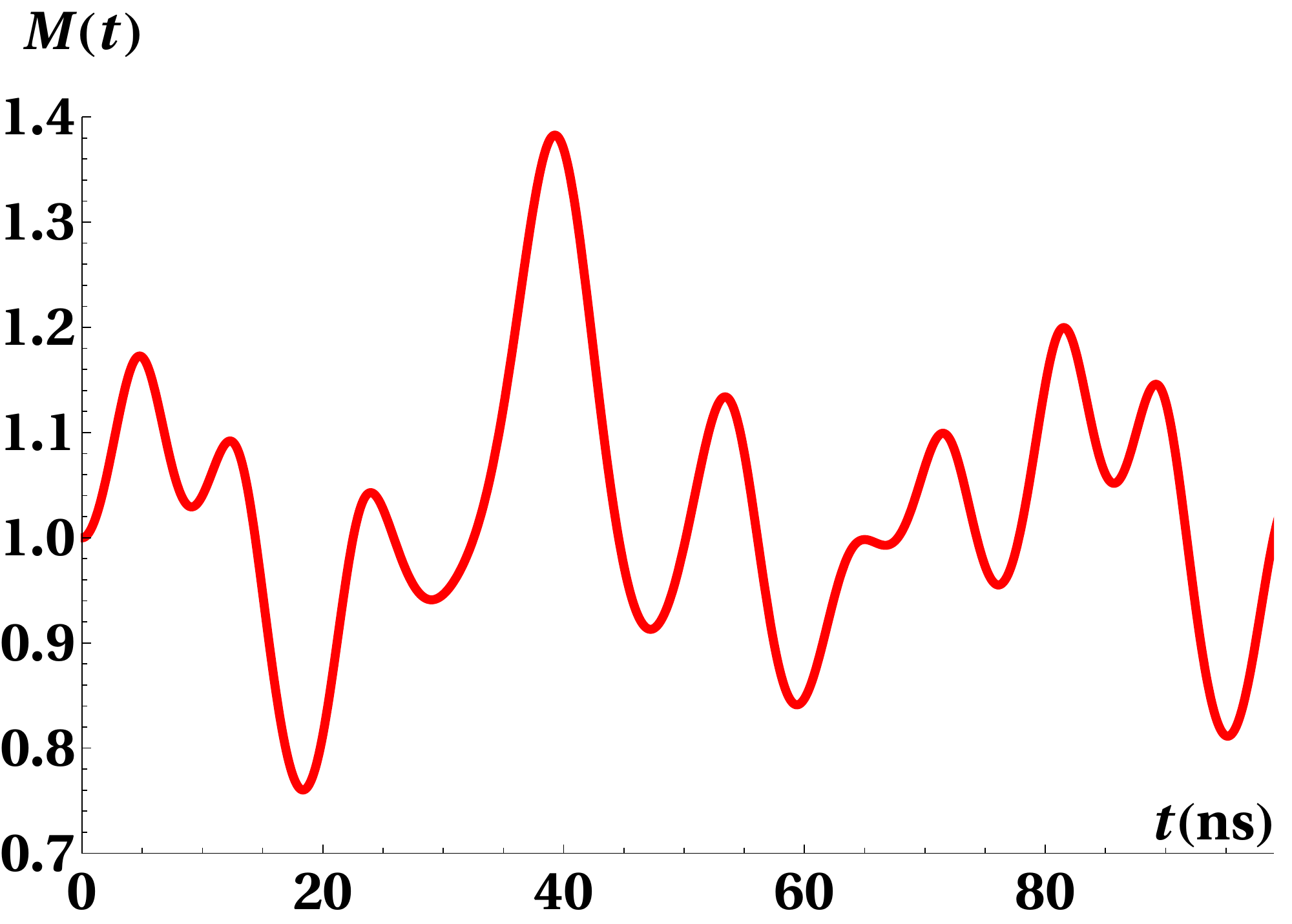}
	\caption{Plot of theoretical prediction for $M(t)$ without any thermal relaxation.}
	\label{fig:MFE0}
	\end{subfigure}
	\caption{Theoretical calculation of the quantities of interest without any thermal relaxation.}
	\label{fig:noDecay}
\end{figure}

 Fig. \ref{fig:ST} shows the classical precomputation of $S(t)$, as computed by Eq. \ref{eq:sb} for $B=0$ and $B=0.1$T, the values used in the experiment.

\begin{figure}[htp]
	\includegraphics[width = \columnwidth]{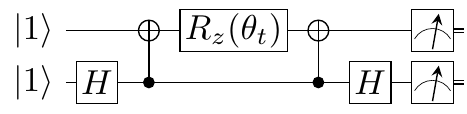}
	\caption{Raw circuit returning $S(t)$, the probability of the TMP/PTP radical pair being in a singlet state at time $t$, as the probability of measuring $\ket{11}$. Note that $S(t)$ is explicitly encoded in $\theta_t=2\cos^{-1}\left(\sqrt{S(t)}\right)$.}
	\label{circ:CircBare}
\end{figure}

These are then combined as per Eq. \ref{eq:MFE} to yield $M(t)$, shown in Fig. \ref{fig:MFE0}. Encoding $S(t)$ onto the qubits is simple and is accomplished with the circuit shown in Fig. \ref{circ:CircBare}.

\par
\begin{figure}[htp]
\begin{subfigure}[b]{\columnwidth}
	\includegraphics[width = \columnwidth]{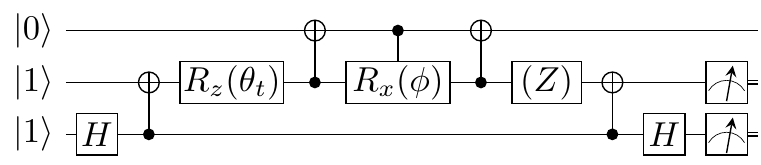}
	\caption{Circuit for $B=0$. Both amplitude damping and dephasing must be considered. $\theta_t=2\cos^{-1}\left(\sqrt{ S_0(t)}\right)$, $\phi=2\cos^{-1}\left(e^{-t/( 2\tau)}\right)$ and $w_z=\frac{1}{2}\left(1-e^{-t/( 2\tau)}\right)$}
	\label{circ:ADDPtmp}
	\end{subfigure}
\begin{subfigure}[b]{\columnwidth}
	\includegraphics[width = \columnwidth]{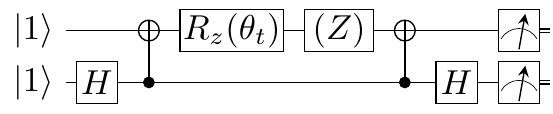}
	\caption{Circuit for large $B$. Only dephasing needs to be considered. $\theta_t=2\cos^{-1}\left(\sqrt{ S_B(t)}\right)$ and $w_z=\frac{1}{2}\left(1-e^{-t/T_2} \right)$}
	\label{circ:DP}
	\end{subfigure}
	\caption{Circuits to simulate the thermal relaxation of our system. The singlet state corresponds to measuring $\ket{11}$. $(Z)$ indicates that the shots are to be split between implementing the circuit including the $Z$ gate and omitting the $Z$ gate. Results are then combined as $S=w_zP_{11}^Z+(1-w_z)P_{11}^{\text{no }Z}$}
	\label{circ:circs}
\end{figure}

 With $S(t)$ encoded on the qubits, we may directly apply the Kraus method. Note that although this encoding of $S(t)$ onto the qubits does not reproduce the correct electronic state in full (since this encoding does not induce transitions to $\ket{T_{\pm}}$), the resulting $\tilde S(t)$ is the same. For this situation, we use the circuits in Fig. \ref{circ:circs} to implement the Kraus method for low- and high-field dynamics.

The results of running these circuits on an actual quantum computer (IBMQ Toronto at 5,000 shots) are shown in Fig. \ref{fig:TMPResults}. In general, they agree well with the theory and experimental data across the entire range of simulated time.

% \begin{figure}[htb!]
% 	\includegraphics[width = \columnwidth]{HighLowDataKraus}
% 	\caption{Comparison of theoretical thermal relaxation vs the Kraus method from running the circuit shown in Fig. \ref{circ:circs} on the IBMQ Almaden quantum device}
% 	\label{fig:HighLowKraus}
% \end{figure}

\par
\begin{figure}[htp]
	\includegraphics[width = \columnwidth]{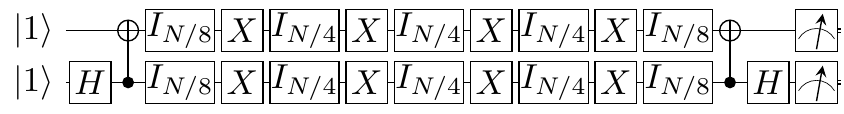}
	\caption{Correction circuit (Sec. \ref{sec:corrCirc}) for use with Inherent method (Sec. \ref{sec:inher}). $I_n$ means apply $n$ identity gates.}
	\label{circ:CircBIDs}
\end{figure}

 Because $T_1\gg T_2$ for the high-field case, an implementation of the Inherent method must use a correction circuit (shown in Fig. \ref{circ:CircBIDs}) as outlined in section \ref{sec:corrCirc}. 

However, since our encoding of $S(t)$ onto the qubits doesn't induce transitions to $\ket{T_{\pm}}$, this greatly simplifies the use of the correction circuit. This happens because any population in $\ket{T_{\pm}}$ can only come from $T_1$ decay. Since we wish to simulate $T_1\to\infty$, we can simply take our $T_1=T_2$ result and add the population of $\ket{T_{\pm}}$ from the correction circuit, in effect ``repopulating" the singlet state with any probability that leaked into $\ket{T_{-}}$ from the $T_1$ decay. 

% The population of $\ket{T_{-}}$ as measured by the correction circuit, $2P(\ket{00})$, is used to construct $\tilde S_B(t)$ with $T_1=\infty$ from $\tilde S_B(t)$ with $T_1=T_2$ as shown in Fig. \ref{fig:HighData}.

% \begin{figure}[htb!]
% 	\includegraphics[width = \columnwidth]{CorrDataB2}
% 	\caption{Comparison of theoretical thermal relaxation vs the correction circuit method from running the circuit shown in Fig. \ref{circ:CircBIDs} on the IBMQ Toronto quantum device}
% 	\label{fig:SPPT}
% \end{figure}

\begin{figure}[htp]
	\includegraphics[width = \columnwidth]{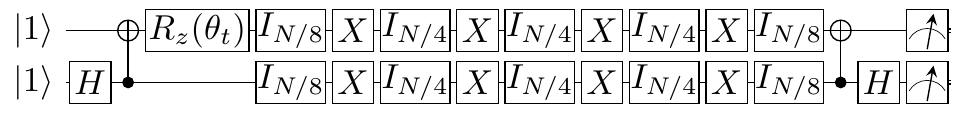}
	\caption{Circuit to implement the dynamics and thermal relaxation using the inherent method. $I_n$ means apply $n$ identity gates.}
	\label{circ:CircIDs}
\end{figure}

Denote the probability of measuring the singlet state (i.e. $\ket{11}$) from the circuit shown in Figs. \ref{circ:CircIDs} as $P(S)$ and the probability of measuring $\ket{T_{-}}$ (i.e. $2\ket{00}$) from the correction circuit shown in Fig. \ref{circ:CircBIDs} as $P(T)$. Then we have $\tilde S(t)=P(S)+P(T)$. Both $P(S)$ and $\tilde S(t)$ are shown in Fig. \ref{fig:HighData}. While this is a nice simplification for our case, it will not work when doing true Hamiltonian simulation and the more involved method described in section \ref{sec:corrCirc} must be used.
\par %The circuit used to simulate the $T_1=T_2$ dynamics for the Inherent method are shown in Fig. \ref{circ:CircIDs}. 
The results from running the circuits which are shown in Figs. \ref{circ:CircBIDs} (Simulation) and \ref{circ:CircIDs} (Correction) are shown in Figs. \ref{fig:LowData} and \ref{fig:HighData}. The final results for the TR-MFE obtained by classically combining the low- and high- field results are shown in Fig. \ref{fig:MFE1}. For this system we used a combination of all three methods, and find excellent agreement with theory. The Kraus method gives a MSE of 0.015\% and the Inherent method + correction circuit gives a MSE of 0.0059\% compared to theory. Our results are less noisy than the experimental data which, even after removing the anomalous data from $t=0$ to $t=12$, have a MSE of 0.109\%.

% \par Having $S(t)$ loaded on our qubits, we proceed to use both the Kraus (Figs. \ref{fig:HighLowKraus},\ref{fig:MFE1}) and Inherent (Figs. \ref{fig:HighLow},\ref{fig:MFE1}) methods to compute $\tilde S(t)$ and $\tilde M(t)$. For the large $B$ case, since $T_1\to\infty$, using the inherent method means we combine the results from running the simulation and correction circuit (Fig. \ref{fig:SPPT}) on qubits with $T_1\approx T_2$ in a classical post-processing step.

% \begin{figure}[htb!]
% 	\includegraphics[width = \columnwidth]{HighLowKData}
% 	\caption{Comparison of theoretical thermal relaxation vs the Inherent method from running the circuit shown in Fig. \ref{circ:CircIDs} on the IBMQ Toronto quantum device}
% 	\label{fig:HighLow}
% \end{figure}

% \begin{figure}
% 	\includegraphics[width = \columnwidth]{MFEDataGood}
% 	\caption{Reconstructed $\tilde{M}(t)$ from the results shown in Figs. \ref{fig:HighLow} and \ref{fig:HighLowKraus}. The results are highly accurate with a mean squared error of 0.015\% for the Kraus method and 0.0059\% for the inherent method, agreeing with the experimental data across all times within the experimental range of validity.}
% 	\label{fig:MFE1}
% \end{figure}

\begin{figure*}[htp]
\begin{subfigure}[t]{0.66\columnwidth}
	\includegraphics[width = \columnwidth]{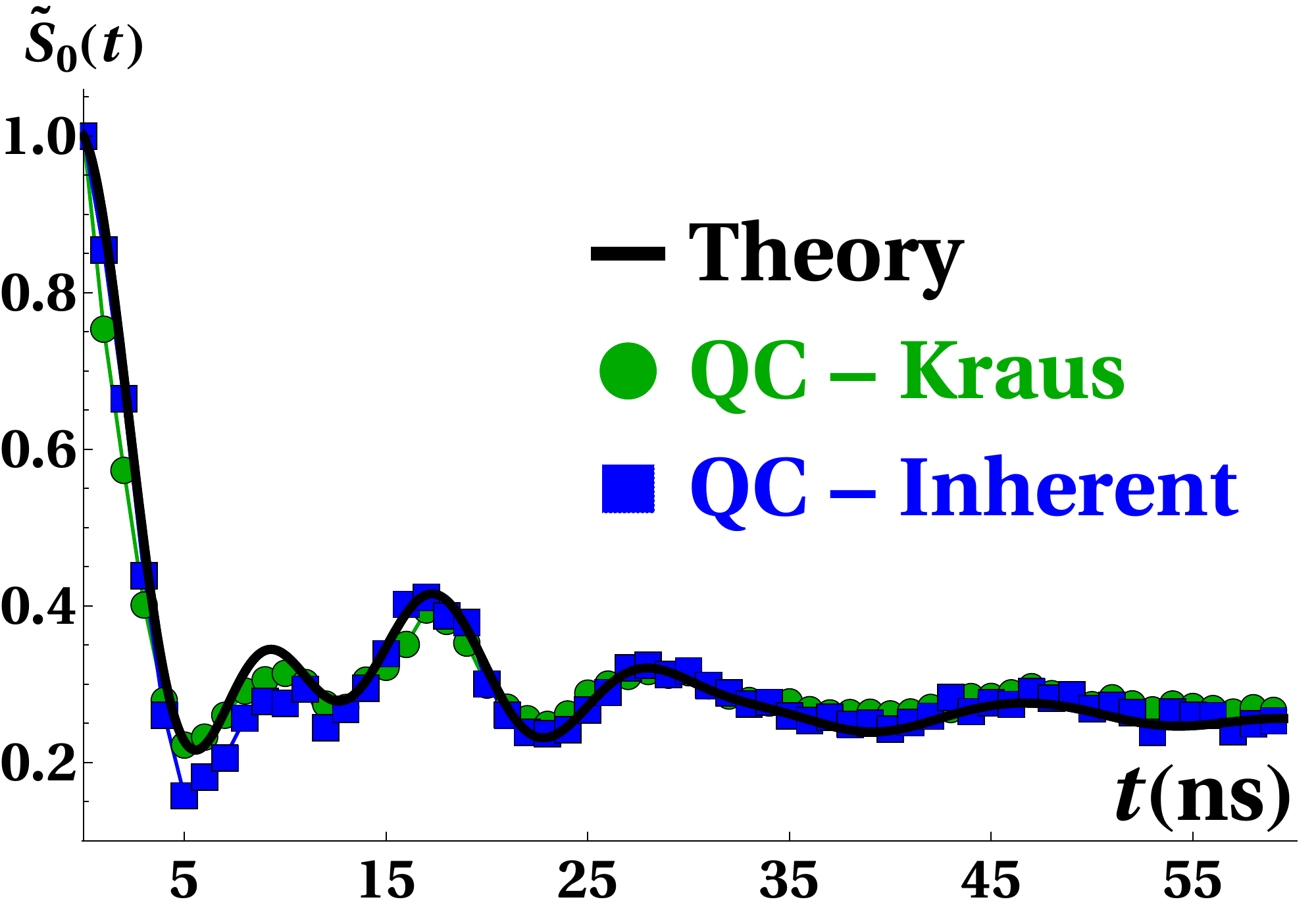}
	\caption{Results for $\tilde S_0(t)$ from the Kraus method and Inherent method compared with theory. $\tilde S_0(t)$ Kraus, $\tilde S_0(t)$ Inherent have MSEs of 0.097\%, 0.057\% respectively compared to theory.}
	\label{fig:LowData}
	\end{subfigure}%
	~
\begin{subfigure}[t]{0.66\columnwidth}
	\includegraphics[width = \columnwidth]{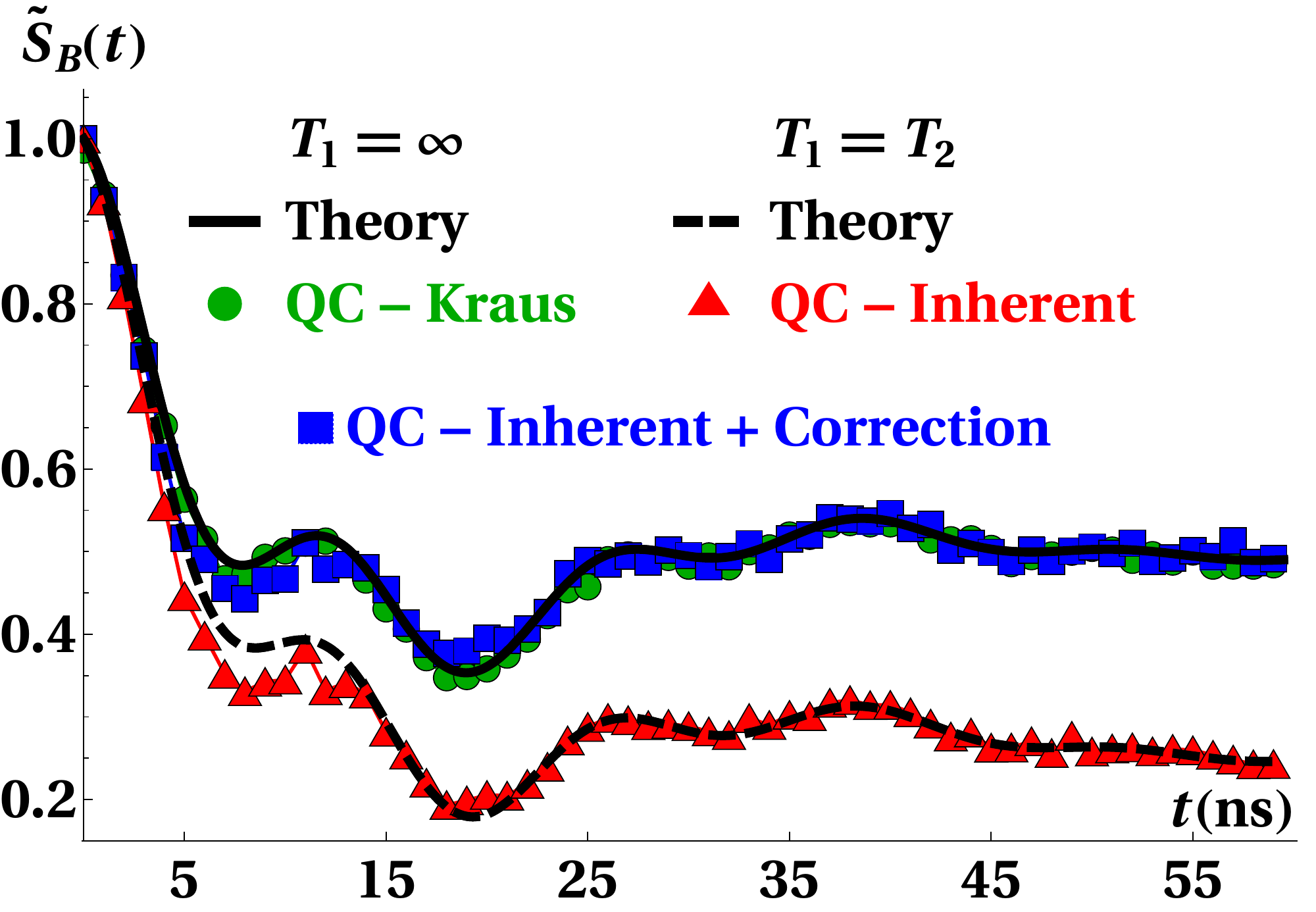}
	\caption{Results for $\tilde S_B(t)$ from the Kraus method and Inherent + Correction Circuit method compared with theory. $\tilde S_B(t)$ Kraus, $\tilde S_B(t): \ T_1=\infty$ Inherent, $\tilde S_B(t): \ T_1=T_2$ Inherent give MSEs of 0.011\%, 0.037\%, 0.045\% respectively compared to theory.}
	\label{fig:HighData}
	\end{subfigure}%
	~
\begin{subfigure}[t]{0.66\columnwidth}
	\includegraphics[width = \columnwidth]{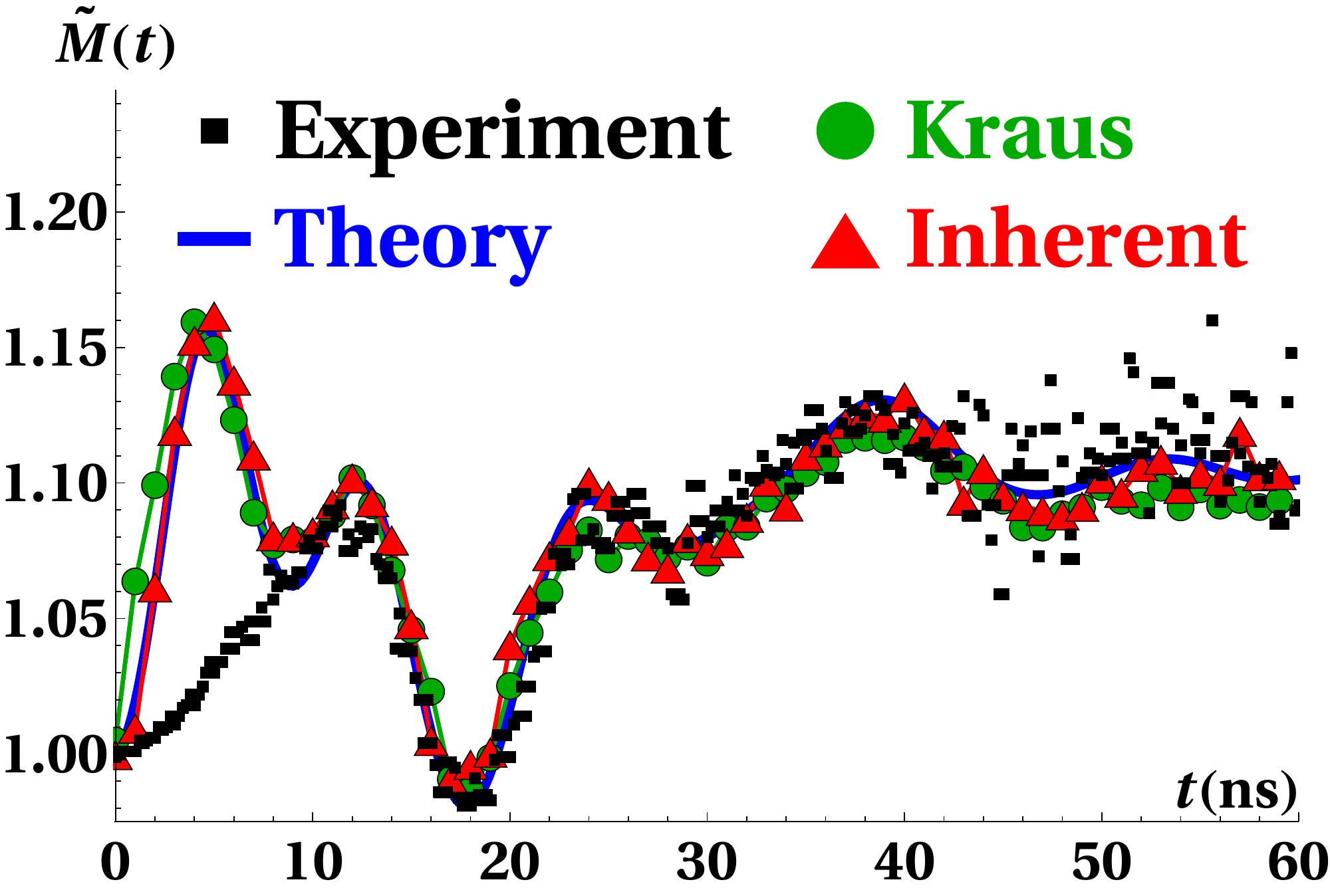}
	\caption{Reconstructed $\tilde{M}(t)$ from the results shown in Figs. \ref{fig:LowData} and \ref{fig:HighData}. Kraus, Inherent method gives a MSE of 0.015\%, 0.0059\% respectively compared to theory. The discrepancy between theory and experiment for $t \lesssim 12$\,ns is due to additional non-magnetosensitive fluorescence arising from the excitation of solvent molecules\cite{bagryansky2005spin}.}
	\label{fig:MFE1}
	\end{subfigure}
\caption{Data for the TR MFE ($\tilde M(t)$) and $\tilde S(t)$ at high- and low-fields from running both the Kraus and Inherent + Correction Circuit methods for the TMP/PTP radical pair (circuits in Figs. \ref{circ:circs}, \ref{circ:CircIDs}, \ref{circ:CircBIDs}) vs. theoretical calculations and experimental data. Runs for the Kraus method were conducted on IBM's Almaden quantum computer, the other circuits were conducted on IBM's Toronto quantum computer, all runs used 5,000 shots. $\tilde M(t)$ is constructed from $\tilde S(t)$ as per Eq. \ref{eq:MFE} with $\theta=0.108$. The results are highly accurate for both methods, which is not terribly surprising given the simplicity of these circuits due to bypassing the Hamiltonian simulation step.}
	\label{fig:TMPResults}
\end{figure*}

\section{Conclusion}

\par We have shown that thermal relaxation may be added to a spin chemistry simulation effectively on a quantum computer, with minimal added resources. This can be done by explicitly simulating the thermal relaxation Kraus operators (Sec. \ref{sec:Ex}), by leveraging the inherent thermal relaxation of the qubits themselves (Sec. \ref{sec:inher}), or by separately running the simulation without thermal relaxation and a correction circuit undergoing relaxation (using either previous method) and combining the results in a classical post-processing step (Sec. \ref{sec:corrCirc}).
\par The Kraus method provides the most control over the thermal relaxation, providing full control over the relaxation parameters such as temperature, $T_1$ and $T_2$. However, this method is only suitable when the simulation can be performed in a time much shorter than the qubits' natural decoherence time. Otherwise, the qubits' natural decoherence will interfere with the artificial Kraus decoherence, leading to inaccurate results. This issue can be mitigated by using the correction circuit method outlined in Sec. \ref{sec:corrCirc}. Furthermore, implementing the Kraus method requires ancilla qubits and controlled rotations which makes it unlikely to generalize well to larger systems on NISQ devices. An exception is when only dephasing is required which may be done at any scale, practically for free, by using $R_z$ gates with their angles drawn from a normal distribution each shot. Furthermore, when the Kraus map doesn't commute with the system's unitary time evolution (e.g. $T\neq\infty$), the map must be Trotterized and implemented each time step, further adding to the cost.
\par Alternatively, leveraging the inherent thermal relaxation of the qubits suffers none of these problems - remaining virtually free to implement for any system size and utilizing the qubits natural decoherence as a resource. However, it is much more restrictive in its applicability since the relaxation parameters of the qubits is fixed. We have shown some examples of how to get around these constraints by invoking a correction circuit to map the natural qubit relaxation dynamics onto the relaxation dynamics of a system of interest.
\par Furthermore, we show how this correction circuit can be used to invert the relaxation of a simulation to recover the un-relaxed dynamics as well as to add relaxation to the results of an un-relaxed simulation. More study is needed to determine how widely applicable such methods are to simulations outside of those described by Hamiltonians of the form given in Eq. \ref{eq:ham}.
\par We demonstrate the effectiveness of these techniques on two simple, real-world systems of interest in spin chemistry. Due to the simplicity of the DPS/PTP radical pair system, we are able to implement the full Hamiltonian simulation. Using the Inherent method, we were able to reproduce the theoretical result for the TR MFE with a mean squared error of $0.073$\% (Fig. \ref{fig:DPSResults}), despite the presence of broadband dephasing seen at longer times. This shows that such methods are an effective tool to model real world systems undergoing thermal relaxation that is usable on currently available quantum devices. Furthermore, these results are significantly less noisy than the experimental data (MSE $0.97$\%), which suffers from a noise problem outlined in Appendix \ref{appendix}.
\par We also demonstrate the effectiveness of these techniques for a more complicated system (the TMP/PTP radical pair) having two hyperfine coupling terms and a $B$-field dependent $T_1$ relaxation rate. Unfortunately, currently available quantum computers aren't sufficiently robust enough to implement Hamiltonian simulation of this system - largely due to CNOT error rates. As quantum computers continue to improve gate fidelity, we hope in the near future this will be possible. In order to work around this, we classically precompute $S(t)$, encode this onto a pair of qubits and use our protocols to recover $\tilde S(t)$. This classical precomputation step is exponentially hard in the size of the nuclear degrees of freedom, making it desirable to develop optimal methods for Hamiltonian time evolution on a quantum computer.
\par The results for TMP/PTP using both the Kraus and inherent + correction circuit method are shown in Fig. \ref{fig:TMPResults}. We find excellent agreement with theory, with both of our datasets again being less noisy than the experimental data. When directly simulating time evolution becomes reliable on NISQ devices, the encouraging results here suggest our methods could provide fast, easy and cheap ways to do primary investigations into spin chemistry problems to direct or supplement experiments. Because these methods do not suffer from the experimental noise problem outlined in Appendix \ref{appendix}, they may also provide a way to probe experimentally inaccessible regimes.
\par Further research is needed to explore the applicability of the three methods presented here to Hamiltonians other than those of the form in Eq. \ref{eq:ham}. It is likely that these methods will be applicable to many systems relaxing in an effectively infinite temperature environment.

\section{Acknowledgments}
We thank Dan Rugar, John Mamin, Andrew Eddins and Andrew Cross of IBM for their helpful discussions and insight that lead to the use of echo pulses in our protocol. We thank Dr. V.A. Bagryansky (V.V. Voevodsky Institute of Chemical Kinetics and Combustion, Novosibirsk, Russia) for allowing use of his experimental data on quantum beats. This work was supported by the U.S. Department of Energy, Office of Science, Basic Energy Sciences, Division of Materials Sciences and Engineering under Grant No. de-sc0019469. B. R. was also supported by the National Science Foundation under Award DMR-1747426 (QISE-NET). M. V. is supported by the IBM Global University Program. We acknowledge the use of IBM Q for this work. The views expressed are those of the authors and do not reflect the official policy or position of IBM or the IBM Q team.\\
\appendix

\section{Commutation of Time Evolution and Decoherence Channels}\label{appendixC}

\par It is fairly simple to show that the generalized amplitude damping, $\mathcal{E}_x$, and dephasing, $\mathcal{E}_z$, channels commute. Note that sufficient (but not necessary) condition for two quantum channels to commute is that all of the Kraus operators from the two channels satisfy $K_i^1 K_j^2=e^{i\phi}K_j^2K_i^1$. This follows directly from Eq. \ref{eq:channel} and \ref{eq:commute} because all of the phases simply cancel, $K\rho K^\dagger=(e^{i\phi}K)\rho (e^{i\phi}K)^\dagger$. 
\par It is easy to see then that $\mathcal{E}_x$ and $\mathcal{E}_z$ commute because $K_i^x K_j^z = \pm K_j^z K_i^x$ which follows directly from anti-commutation relations of the Paulis and the form of the Kraus operators in Eqs. \ref{eq:krausX} and \ref{eq:krausZ}. Let $\mathcal{E}(\cdot)\equiv\mathcal{E}_x(\mathcal{E}_z(\cdot))$. 
\par It is less obvious that coherent time evolution and $\mathcal{E}$ commute, largely because it is untrue. However, we show that at laboratory temperatures the time evolution of \textit{only} the electronic DOFs commutes with $\mathcal{E}$. Denote $\mathcal{E}$ operating at infinite temperature to be $\mathcal{E}^\infty$.
\par We argue as follows, by tracing out the nuclear subsystem we are formally treating it as a finite bath interacting with the electronic subsystem. The interaction can be read off of Eq. \ref{eq:ham} and is given by
\begin{equation}\label{eq:interaction}
  \sum_{i,j} a_{ij}\vb{I}_{ij}\cdot\vb{S}_{i}=\sum_{i,j} \frac{a_{ij}}{2}\left(I_{ij}^{+}S_{i}^{-}+I_{ij}^{-}S_{i}^{+}+2I_{ij}^{z}S_{i}^{z}\right).
\end{equation}
The two terms involving $S^{\pm}$ can be thought of as performing amplitude damping. Because the initial nuclear state is the maximally mixed state, i.e. the infinite temperature state, this interaction applies infinite temperature amplitude damping. The term involving $S^{z}$ performs the phase damping channel. As such, the interaction between the electronic and nuclear subsystems performs $\mathcal{E}^\infty$ on the electronic subsystem. Since $\mathcal{E}^\infty$ trivially commutes with itself, it is not surprising that the coherent time evolution and decoherence channel commute as well. We now make the argument more precise, although a rigorous proof is beyond the scope of this work. 
\par The action of coherent time evolution, from an initial time $t=0$ to $t$, on the electronic subsystem only is itself represented by a quantum channel, $\mathcal{E}_U$. The Kraus operators for $\mathcal{E}_U$ are given by 
\begin{equation}
  K^U_{ij}=\sqrt{P(\ket{\vb{I}_i})}\mel{\vb{I}_i}{e^{-iHt}}{\vb{I}_j}
\end{equation}
such that
\begin{equation}\label{eq:rhoe}
  \rho_U=\mathcal{E}_U(\rho(0))=\sum_{i}\mel{\vb{I}_i}{e^{-iHt}\rho(0) e^{iHt}}{\vb{I}_i}
\end{equation}
where the sum traces out the nuclear DOF and
\begin{equation}\label{eq:mel}
  \mel{\vb{I}_i}{\hat A}{\vb{I}_i}=\sum_{j,k}\mel{\vb{I}_i,e_j}{\hat A}{\vb{I}_i,e_k}\ket{e_k}\bra{e_j}.
\end{equation}
Here $\ket{\vb{I}_i}(\ket{e_i})$ is the $i^\text{th}$ basis state for the nuclear (electronic) subsystem, $P(\ket{\vb{I}_i})$ is the probability of being in that nuclear state at $t=0$ and $\rho_U$ is the state of the electronic subsystem at time $t$. For our case the initial nuclear state is the maximally mixed state, $P(\ket{\vb{I}_i})=P=1/N_{\vb{I}}$. 
From here, a few observations may be made. For simplicity we take $\rho(0)=\ket{S}\bra{S}$ moving forward, although the argument holds for any initial state of the form $\rho(0)=e^{-i \alpha H}\ket{S}\bra{S}e^{i \alpha H}$.
\par Transitions from the singlet state $\ket{S}$ to either triplet state with $m_s\neq 0$ ($\ket{T_{+}}=\ket{\uparrow\uparrow},\ket{T_{-}}=\ket{\downarrow\downarrow}$) occur from the terms containing an $S_i^\pm$ in Eq. \ref{eq:interaction}. These terms come with a corresponding, distinct transition in the nuclear state from $I_{ij}^\pm$. Furthermore, due to symmetry in the Hamiltonian, and therefore in Eq. \ref{eq:interaction}, the probability of transition to $\ket{T_{+}}$ is equal to $\ket{T_{-}}$. Due to spin conservation, a nuclear state corresponding the electronic state being in either $\ket{T_{\pm}}$, cannot correspond to any other electronic state. From Eqs. \ref{eq:rhoe} and \ref{eq:mel} then, we see $\rho_U$ has a term of $a(\ket{T_+}\bra{T_+}+\ket{T_{-}}\bra{T_{-}})$ and no other terms involving $\ket{T_{\pm}}$. 
\par Likewise, transitions from $\ket{S}$ to $\ket{T_0}=$ $\frac{1}{\sqrt{2}}\left(\ket{\uparrow\downarrow}+\ket{\downarrow\uparrow}\right)$ mediated by nuclear interactions will contribute a term of $b\ket{T_0}\bra{T_0}$. Finally, evolution that leaves the nuclear state unchanged, i.e. evolution mediated by $\vb{B}$, coherently rotates between $\ket{S}$ and $\ket{T_0}$. This contributes a term to $\rho_U$ of the form $c (R_z(\phi)\otimes R_z(\varphi))\ket{S}\bra{S}(R_z(-\phi)\otimes R_z(-\varphi))$. Explicitly writing this out gives
\begin{equation}\label{eq:rhoMatU}
  \rho_U= \left(
\begin{array}{c|cccc}
 \hphantom{a}&\ket{\uparrow\uparrow}&\ket{\downarrow\uparrow}&\ket{\uparrow\downarrow}&\ket{\downarrow\downarrow}\\\hline
 \ket{\uparrow\uparrow}&a & 0 & 0 & 0 \\
 \ket{\downarrow\uparrow}&0 & b+c & b-c e^{-i \theta } & 0 \\
 \ket{\uparrow\downarrow}&0 & b-c e^{i \theta } & b+c & 0 \\
 \ket{\downarrow\downarrow}&0 & 0 & 0 & a \\
\end{array}
\right)
\end{equation}
where normalization implies $a=\frac{1}{2}-b-c$ and we've defined $\theta=\phi-\varphi$. It is much more straightforward to show that $\rho_D=\mathcal{E}^{\infty}(\rho(0))=$
\begin{equation}\label{eq:rhoMatD}
  \frac{1}{4}\left(
\begin{array}{cccc}
 1-\bar p_x^2 & 0 & 0 & 0 \\
 0 & \bar p_x^2+1 & -2 \bar p_x \mathcal{P}_z^2 & 0 \\
 0 & -2 \bar p_x \mathcal{P}_z^{*2} & \bar p_x^2+1 & 0 \\
 0 & 0 & 0 & 1-\bar p_x^2 \\
\end{array}
\right)
\end{equation}
where $\mathcal{P}_z=1-2 p_z$. If we then modify this slightly by adding an $R_z(\phi/2)$ gate to act on one of the qubits before acting with $\mathcal{E}^\infty$, we get the same result as in Eq. \ref{eq:rhoMatD} but with $\mathcal{P}_z\mapsto e^{i\phi}\mathcal{P}_z$. A lone $R_z(\phi/2)$ itself defines a quantum channel, $\mathcal{E}^z(\rho)=R_z(\phi/2)\rho R_z^\dagger(\phi/2)$.
\par It is now explicitly obvious that this new $\mathcal{E}^\infty \!\! \circ \! \mathcal{E}^z$ map implements the same transformation as $\mathcal{E}_U$, i.e. $\mathcal{E}_U(\rho(0))=\mathcal{E}^\infty(\mathcal{E}^z(\rho(0)))$. By the same logic that showed $\mathcal{E}_x$ and $\mathcal{E}_z$ commute, it is easy to see that $\mathcal{E}^z$ and $\mathcal{E}$ commute. Furthermore, since $\mathcal{E}$ commutes with itself, we conclude that $\mathcal{E}^\infty$ and $\mathcal{E}_U$ must themselves commute. Note that for $T\neq\infty$ or for a different Hamiltonian, e.g. one that contains direct interactions between the two species, the above no longer holds.

\section{Noise in Experimental Data}\label{appendix}

% \begin{figure}[htp]
% 	\includegraphics[width = \columnwidth]{DPSappend}
% 	\caption{Reconstructed $\tilde{M}(t)$ from the adding independent Gaussian noise ($\mu=0$ and $\sigma=700$) to $\tilde S_0(t)$ and $\tilde S_B(t)$ at each time point vs. experimental data for DPS/PTP. We see that such a simple model reproduces the data well and explains the main effect behind increasing noise in the TR MFE vs. time.}
% 	\label{fig:append2}
% \end{figure}

% \begin{figure}[htp]
% 	\includegraphics[width = \columnwidth]{AppendixPlot}
% 	\caption{Reconstructed $\tilde{M}(t)$ from the adding independent Gaussian noise ($\mu=0$ and $\sigma=75$) to $\tilde S_0(t)$ and $\tilde S_B(t)$ at each time point vs. experimental data for TMP/PTP. We see that such a simple model reproduces the data well and explains the main effect behind increasing noise in the TR MFE vs. time.}
% 	\label{fig:append}
% \end{figure}

The TR MFE is calculated as per Eq. \ref{eq:MFE} in order to cancel $F(t)$ from \ref{eq:intense}. This is sound as $F(t)$ has no $B$ dependence, however $F(t)$ is a rapidly decaying function representing lifetime distribution of the radical pairs. So relative noise in the experiment grows for long times as $F(t)$ damps the signal while the noise stays approximately constant.
\par More technically, $F(t)$ for radical pairs is well approximated by\cite{veselov1987induction, Bagryansky2007}
\begin{equation}\label{eq:Ft}
  F(t)\approx A e^{-t/t_1}+B\left(1+\frac{t}{t_2}\right)^{-\alpha}
\end{equation}
with best fit parameters $A=3.12\times 10^6$, $B=2.21\times 10^5$, $t_1=3.47$ns, $t_2=123$ns, $\alpha =6.11$ for TMP/PTP as determined by non-linear regression. And $A=1.317\times 10^6$, $B=6.658\times 10^5$, $t_1=2.1432$ns, $t_2=5.1549$ns, $\alpha =1.223$ for DPS/PTP\cite{veselov1987induction,BagryanskyPrivate}.

\begin{figure}[htp]
\begin{subfigure}[t]{0.49\columnwidth}
	\includegraphics[width = \columnwidth]{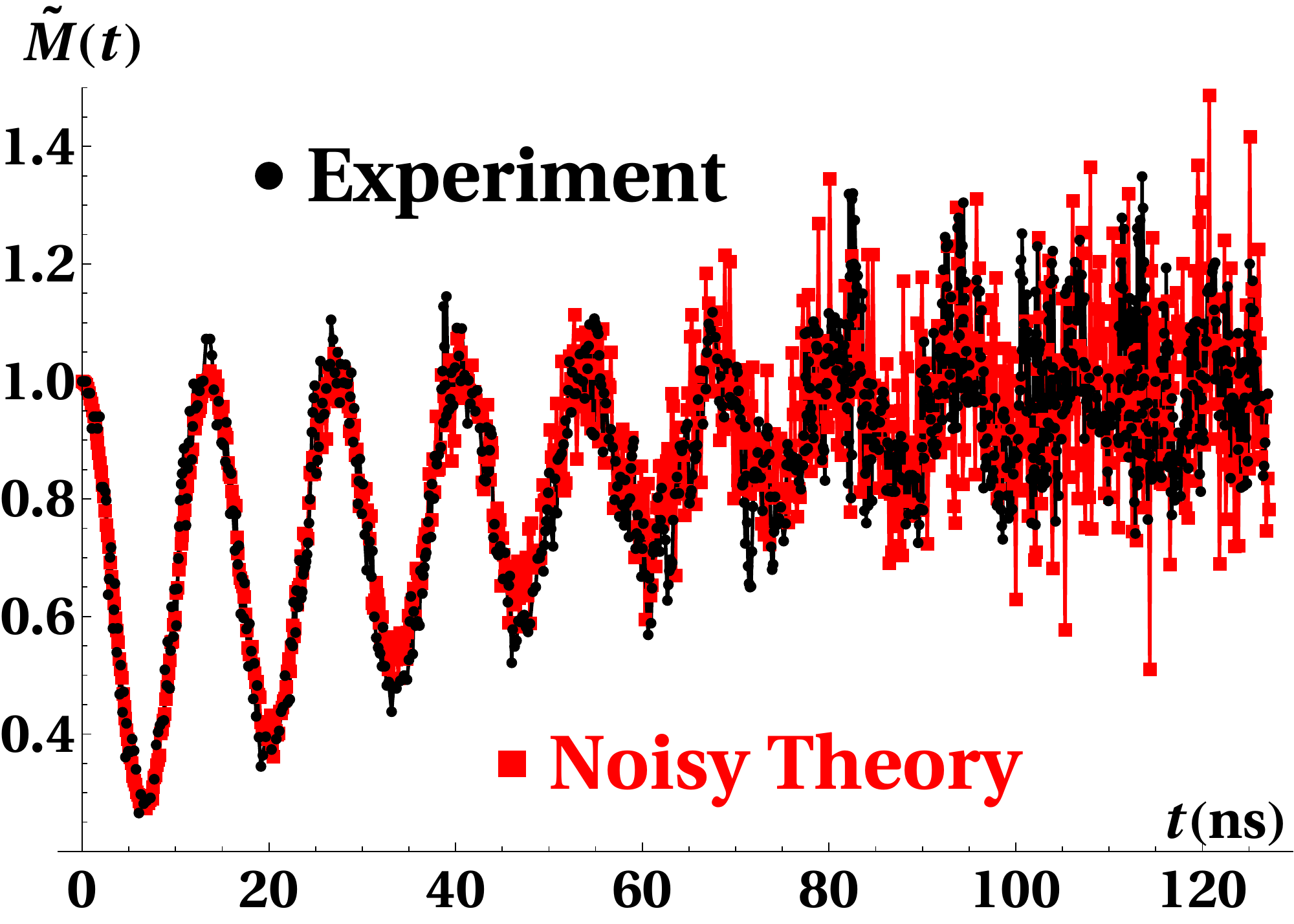}
	\caption{DPS/PTP with $\mu=0$ and $\sigma=700$.}
	\label{fig:append2}
	\end{subfigure}%
	~
\begin{subfigure}[t]{0.49\columnwidth}
	\includegraphics[width = \columnwidth]{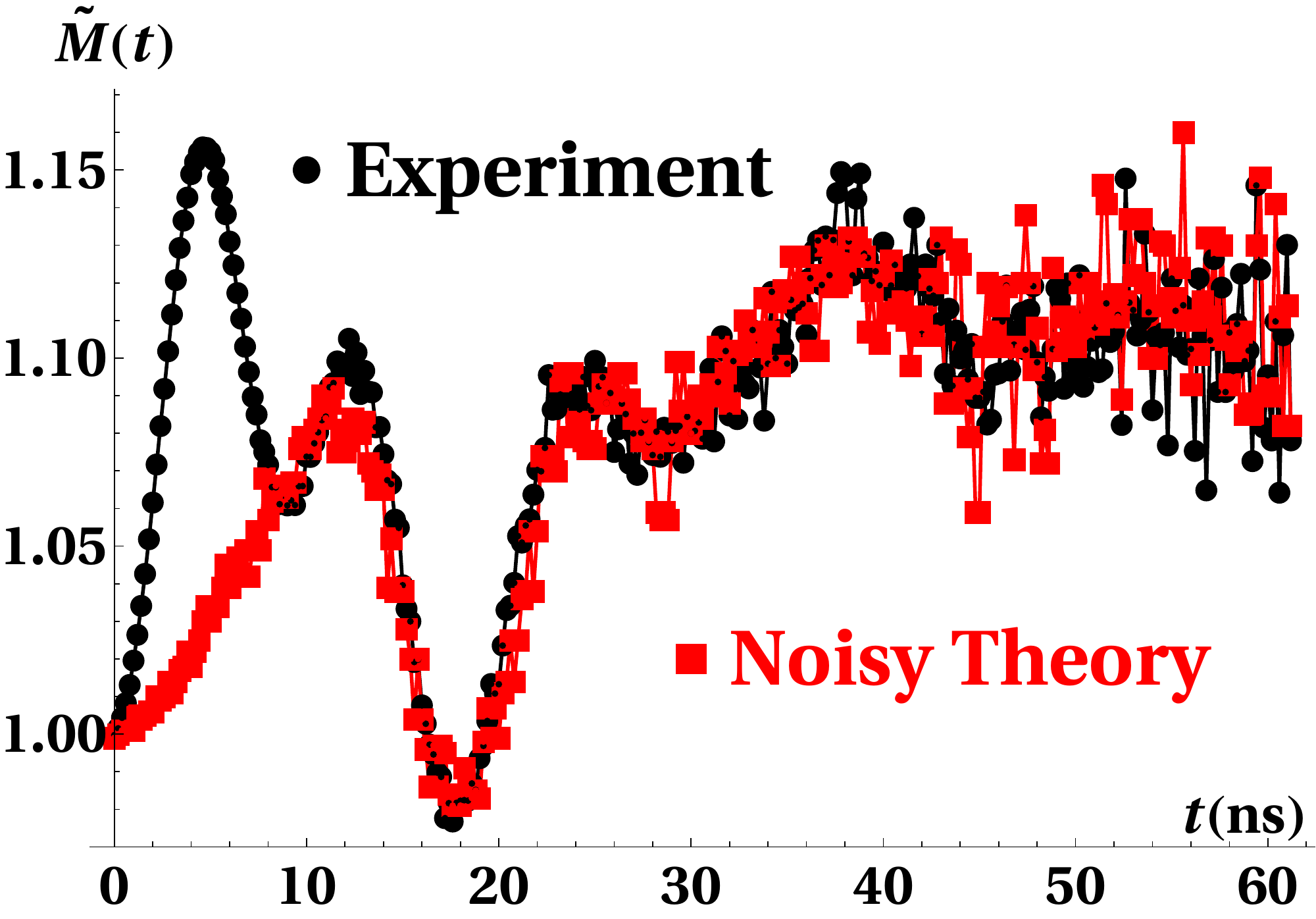}
	\caption{TMP/PTP with $\mu=0$ and $\sigma=75$.}
	\label{fig:append}
	\end{subfigure}
	\caption{Comparisons of noise seen in experimental data vs. noise from assuming Gaussian background noise on the detectors in an otherwise ideal experiment. $\tilde{M}(t)$ is constructed by adding independent Gaussian noise to $\tilde S_0(t)$ and $\tilde S_B(t)$ at each time point. We see that such a simple model reproduces the data well and explains the main effect behind increasing noise in the TR MFE vs. time.}
	\label{fig:InherRes}
\end{figure}

\par We assumed that the PMT detector is a primary source of noise in the experimental data. We approximated the PMT noise with white noise following a normal distribution and applied it to both the high-field and low-field theoretical recombination fluorescence functions, $I(t)$ (Eq. \ref{eq:intense}) constructed from Eqs \ref{eq:Ft}, \ref{eq:stilde}. We then reconstruct the TR MFE (Eq. \ref{eq:MFE}) using these noisy data. The result is shown in Fig. \ref{fig:append}.

We find a normal distribution of $\mu=0$ and $\sigma=75$ reproduces the noise behavior from the experiment very well. Such errors are an inherent challenge to producing good experimental data out to long times. In this regard, performing a quantum simulation opens the potential to investigate the behavior of such systems in finer detail and to longer times than is possible in experiments - once Hamiltonian simulation becomes feasible on these machines.

\bibliography{citations}

\begin{thebibliography}{63}%
\makeatletter
\providecommand \@ifxundefined [1]{%
 \@ifx{#1\undefined}
}%
\providecommand \@ifnum [1]{%
 \ifnum #1\expandafter \@firstoftwo
 \else \expandafter \@secondoftwo
 \fi
}%
\providecommand \@ifx [1]{%
 \ifx #1\expandafter \@firstoftwo
 \else \expandafter \@secondoftwo
 \fi
}%
\providecommand \natexlab [1]{#1}%
\providecommand \enquote  [1]{``#1''}%
\providecommand \bibnamefont  [1]{#1}%
\providecommand \bibfnamefont [1]{#1}%
\providecommand \citenamefont [1]{#1}%
\providecommand \href@noop [0]{\@secondoftwo}%
\providecommand \href [0]{\begingroup \@sanitize@url \@href}%
\providecommand \@href[1]{\@@startlink{#1}\@@href}%
\providecommand \@@href[1]{\endgroup#1\@@endlink}%
\providecommand \@sanitize@url [0]{\catcode `\\12\catcode `\$12\catcode
  `\&12\catcode `\#12\catcode `\^12\catcode `\_12\catcode `\%12\relax}%
\providecommand \@@startlink[1]{}%
\providecommand \@@endlink[0]{}%
\providecommand \url  [0]{\begingroup\@sanitize@url \@url }%
\providecommand \@url [1]{\endgroup\@href {#1}{\urlprefix }}%
\providecommand \urlprefix  [0]{URL }%
\providecommand \Eprint [0]{\href }%
\providecommand \doibase [0]{http://dx.doi.org/}%
\providecommand \selectlanguage [0]{\@gobble}%
\providecommand \bibinfo  [0]{\@secondoftwo}%
\providecommand \bibfield  [0]{\@secondoftwo}%
\providecommand \translation [1]{[#1]}%
\providecommand \BibitemOpen [0]{}%
\providecommand \bibitemStop [0]{}%
\providecommand \bibitemNoStop [0]{.\EOS\space}%
\providecommand \EOS [0]{\spacefactor3000\relax}%
\providecommand \BibitemShut  [1]{\csname bibitem#1\endcsname}%
\let\auto@bib@innerbib\@empty
%</preamble>
\bibitem [{\citenamefont {Matsuoka}\ and\ \citenamefont
  {Schiemann}(2016)}]{Matsuoka_2016}%
  \BibitemOpen
  \bibfield  {author} {\bibinfo {author} {\bibfnamefont {H.}~\bibnamefont
  {Matsuoka}}\ and\ \bibinfo {author} {\bibfnamefont {O.}~\bibnamefont
  {Schiemann}},\ }in\ \href {\doibase 10.1007/978-1-4939-3658-8_3} {\emph
  {\bibinfo {booktitle} {Electron Spin Resonance ({ESR}) Based Quantum
  Computing}}}\ (\bibinfo  {publisher} {Springer New York},\ \bibinfo {year}
  {2016})\ pp.\ \bibinfo {pages} {51--77}\BibitemShut {NoStop}%
\bibitem [{\citenamefont {Atzori}\ and\ \citenamefont
  {Sessoli}(2019)}]{Atzori_2019}%
  \BibitemOpen
  \bibfield  {author} {\bibinfo {author} {\bibfnamefont {M.}~\bibnamefont
  {Atzori}}\ and\ \bibinfo {author} {\bibfnamefont {R.}~\bibnamefont
  {Sessoli}},\ }\href {\doibase 10.1021/jacs.9b00984} {\bibfield  {journal}
  {\bibinfo  {journal} {Journal of the American Chemical Society}\ }\textbf
  {\bibinfo {volume} {141}},\ \bibinfo {pages} {11339} (\bibinfo {year}
  {2019})}\BibitemShut {NoStop}%
\bibitem [{\citenamefont {Forbes}(2019)}]{Forbes_2019}%
  \BibitemOpen
  \bibfield  {author} {\bibinfo {author} {\bibfnamefont {M.~D.~E.}\
  \bibnamefont {Forbes}},\ }\href {\doibase 10.1002/cptc.201900181} {\bibfield
  {journal} {\bibinfo  {journal} {{ChemPhotoChem}}\ }\textbf {\bibinfo {volume}
  {3}},\ \bibinfo {pages} {971} (\bibinfo {year} {2019})}\BibitemShut {NoStop}%
\bibitem [{\citenamefont {Hore}\ \emph {et~al.}(2020)\citenamefont {Hore},
  \citenamefont {Ivanov},\ and\ \citenamefont {Wasielewski}}]{Hore_2020}%
  \BibitemOpen
  \bibfield  {author} {\bibinfo {author} {\bibfnamefont {P.~J.}\ \bibnamefont
  {Hore}}, \bibinfo {author} {\bibfnamefont {K.~L.}\ \bibnamefont {Ivanov}}, \
  and\ \bibinfo {author} {\bibfnamefont {M.~R.}\ \bibnamefont {Wasielewski}},\
  }\href {\doibase 10.1063/5.0006547} {\bibfield  {journal} {\bibinfo
  {journal} {The Journal of Chemical Physics}\ }\textbf {\bibinfo {volume}
  {152}},\ \bibinfo {pages} {120401} (\bibinfo {year} {2020})}\BibitemShut
  {NoStop}%
\bibitem [{\citenamefont {Nelson}\ \emph {et~al.}(2017)\citenamefont {Nelson},
  \citenamefont {Krzyaniak}, \citenamefont {Horwitz}, \citenamefont {Rugg},
  \citenamefont {Phelan},\ and\ \citenamefont {Wasielewski}}]{Nelson_2017}%
  \BibitemOpen
  \bibfield  {author} {\bibinfo {author} {\bibfnamefont {J.~N.}\ \bibnamefont
  {Nelson}}, \bibinfo {author} {\bibfnamefont {M.~D.}\ \bibnamefont
  {Krzyaniak}}, \bibinfo {author} {\bibfnamefont {N.~E.}\ \bibnamefont
  {Horwitz}}, \bibinfo {author} {\bibfnamefont {B.~K.}\ \bibnamefont {Rugg}},
  \bibinfo {author} {\bibfnamefont {B.~T.}\ \bibnamefont {Phelan}}, \ and\
  \bibinfo {author} {\bibfnamefont {M.~R.}\ \bibnamefont {Wasielewski}},\
  }\href {\doibase 10.1021/acs.jpca.7b00587} {\bibfield  {journal} {\bibinfo
  {journal} {The Journal of Physical Chemistry A}\ }\textbf {\bibinfo {volume}
  {121}},\ \bibinfo {pages} {2241} (\bibinfo {year} {2017})}\BibitemShut
  {NoStop}%
\bibitem [{\citenamefont {Wu}\ \emph {et~al.}(2018)\citenamefont {Wu},
  \citenamefont {Zhou}, \citenamefont {Nelson}, \citenamefont {Young},
  \citenamefont {Krzyaniak},\ and\ \citenamefont {Wasielewski}}]{Wu_2018}%
  \BibitemOpen
  \bibfield  {author} {\bibinfo {author} {\bibfnamefont {Y.}~\bibnamefont
  {Wu}}, \bibinfo {author} {\bibfnamefont {J.}~\bibnamefont {Zhou}}, \bibinfo
  {author} {\bibfnamefont {J.~N.}\ \bibnamefont {Nelson}}, \bibinfo {author}
  {\bibfnamefont {R.~M.}\ \bibnamefont {Young}}, \bibinfo {author}
  {\bibfnamefont {M.~D.}\ \bibnamefont {Krzyaniak}}, \ and\ \bibinfo {author}
  {\bibfnamefont {M.~R.}\ \bibnamefont {Wasielewski}},\ }\href {\doibase
  10.1021/jacs.8b08105} {\bibfield  {journal} {\bibinfo  {journal} {Journal of
  the American Chemical Society}\ }\textbf {\bibinfo {volume} {140}},\ \bibinfo
  {pages} {13011} (\bibinfo {year} {2018})}\BibitemShut {NoStop}%
\bibitem [{\citenamefont {Nelson}\ \emph {et~al.}(2020)\citenamefont {Nelson},
  \citenamefont {Zhang}, \citenamefont {Zhou}, \citenamefont {Rugg},
  \citenamefont {Krzyaniak},\ and\ \citenamefont {Wasielewski}}]{Nelson_2020}%
  \BibitemOpen
  \bibfield  {author} {\bibinfo {author} {\bibfnamefont {J.~N.}\ \bibnamefont
  {Nelson}}, \bibinfo {author} {\bibfnamefont {J.}~\bibnamefont {Zhang}},
  \bibinfo {author} {\bibfnamefont {J.}~\bibnamefont {Zhou}}, \bibinfo {author}
  {\bibfnamefont {B.~K.}\ \bibnamefont {Rugg}}, \bibinfo {author}
  {\bibfnamefont {M.~D.}\ \bibnamefont {Krzyaniak}}, \ and\ \bibinfo {author}
  {\bibfnamefont {M.~R.}\ \bibnamefont {Wasielewski}},\ }\href {\doibase
  10.1063/1.5128132} {\bibfield  {journal} {\bibinfo  {journal} {The Journal of
  Chemical Physics}\ }\textbf {\bibinfo {volume} {152}},\ \bibinfo {pages}
  {014503} (\bibinfo {year} {2020})}\BibitemShut {NoStop}%
\bibitem [{\citenamefont {Salikhov}(2006)}]{Salikhov_2006}%
  \BibitemOpen
  \bibfield  {author} {\bibinfo {author} {\bibfnamefont {K.~M.}\ \bibnamefont
  {Salikhov}},\ }\href {\doibase 10.3367/ufnr.0176.200606i.0664} {\bibfield
  {journal} {\bibinfo  {journal} {Uspekhi Fizicheskih Nauk}\ }\textbf {\bibinfo
  {volume} {176}},\ \bibinfo {pages} {664} (\bibinfo {year}
  {2006})}\BibitemShut {NoStop}%
\bibitem [{\citenamefont {Volkov}\ and\ \citenamefont
  {Salikhov}(2011)}]{Volkov_2011}%
  \BibitemOpen
  \bibfield  {author} {\bibinfo {author} {\bibfnamefont {M.~Y.}\ \bibnamefont
  {Volkov}}\ and\ \bibinfo {author} {\bibfnamefont {K.~M.}\ \bibnamefont
  {Salikhov}},\ }\href {\doibase 10.1007/s00723-011-0297-2} {\bibfield
  {journal} {\bibinfo  {journal} {Applied Magnetic Resonance}\ }\textbf
  {\bibinfo {volume} {41}},\ \bibinfo {pages} {145} (\bibinfo {year}
  {2011})}\BibitemShut {NoStop}%
\bibitem [{\citenamefont {Kominis}(2009)}]{Kominis_2009}%
  \BibitemOpen
  \bibfield  {author} {\bibinfo {author} {\bibfnamefont {I.~K.}\ \bibnamefont
  {Kominis}},\ }\href {\doibase 10.1103/physreve.80.056115} {\bibfield
  {journal} {\bibinfo  {journal} {Physical Review E}\ }\textbf {\bibinfo
  {volume} {80}} (\bibinfo {year} {2009}),\
  10.1103/physreve.80.056115}\BibitemShut {NoStop}%
\bibitem [{\citenamefont {Cai}\ \emph {et~al.}(2010)\citenamefont {Cai},
  \citenamefont {Guerreschi},\ and\ \citenamefont {Briegel}}]{Cai_2010}%
  \BibitemOpen
  \bibfield  {author} {\bibinfo {author} {\bibfnamefont {J.}~\bibnamefont
  {Cai}}, \bibinfo {author} {\bibfnamefont {G.~G.}\ \bibnamefont {Guerreschi}},
  \ and\ \bibinfo {author} {\bibfnamefont {H.~J.}\ \bibnamefont {Briegel}},\
  }\href {\doibase 10.1103/physrevlett.104.220502} {\bibfield  {journal}
  {\bibinfo  {journal} {Physical Review Letters}\ }\textbf {\bibinfo {volume}
  {104}} (\bibinfo {year} {2010}),\ 10.1103/physrevlett.104.220502}\BibitemShut
  {NoStop}%
\bibitem [{\citenamefont {Gauger}\ \emph {et~al.}(2011)\citenamefont {Gauger},
  \citenamefont {Rieper}, \citenamefont {Morton}, \citenamefont {Benjamin},\
  and\ \citenamefont {Vedral}}]{Gauger_2011}%
  \BibitemOpen
  \bibfield  {author} {\bibinfo {author} {\bibfnamefont {E.~M.}\ \bibnamefont
  {Gauger}}, \bibinfo {author} {\bibfnamefont {E.}~\bibnamefont {Rieper}},
  \bibinfo {author} {\bibfnamefont {J.~J.~L.}\ \bibnamefont {Morton}}, \bibinfo
  {author} {\bibfnamefont {S.~C.}\ \bibnamefont {Benjamin}}, \ and\ \bibinfo
  {author} {\bibfnamefont {V.}~\bibnamefont {Vedral}},\ }\href {\doibase
  10.1103/physrevlett.106.040503} {\bibfield  {journal} {\bibinfo  {journal}
  {Physical Review Letters}\ }\textbf {\bibinfo {volume} {106}} (\bibinfo
  {year} {2011}),\ 10.1103/physrevlett.106.040503}\BibitemShut {NoStop}%
\bibitem [{\citenamefont {Kominis}(2011{\natexlab{a}})}]{Kominis_2011}%
  \BibitemOpen
  \bibfield  {author} {\bibinfo {author} {\bibfnamefont {I.~K.}\ \bibnamefont
  {Kominis}},\ }\href {\doibase 10.1103/physreve.83.056118} {\bibfield
  {journal} {\bibinfo  {journal} {Physical Review E}\ }\textbf {\bibinfo
  {volume} {83}} (\bibinfo {year} {2011}{\natexlab{a}}),\
  10.1103/physreve.83.056118}\BibitemShut {NoStop}%
\bibitem [{\citenamefont {Tiersch}\ and\ \citenamefont
  {Briegel}(2012)}]{Tiersch_2012_Deco}%
  \BibitemOpen
  \bibfield  {author} {\bibinfo {author} {\bibfnamefont {M.}~\bibnamefont
  {Tiersch}}\ and\ \bibinfo {author} {\bibfnamefont {H.~J.}\ \bibnamefont
  {Briegel}},\ }\href {\doibase 10.1098/rsta.2011.0488} {\bibfield  {journal}
  {\bibinfo  {journal} {Philosophical Transactions of the Royal Society A:
  Mathematical, Physical and Engineering Sciences}\ }\textbf {\bibinfo {volume}
  {370}},\ \bibinfo {pages} {4517} (\bibinfo {year} {2012})}\BibitemShut
  {NoStop}%
\bibitem [{\citenamefont {Hogben}\ \emph {et~al.}(2012)\citenamefont {Hogben},
  \citenamefont {Biskup},\ and\ \citenamefont {Hore}}]{Hogben_2012}%
  \BibitemOpen
  \bibfield  {author} {\bibinfo {author} {\bibfnamefont {H.~J.}\ \bibnamefont
  {Hogben}}, \bibinfo {author} {\bibfnamefont {T.}~\bibnamefont {Biskup}}, \
  and\ \bibinfo {author} {\bibfnamefont {P.~J.}\ \bibnamefont {Hore}},\ }\href
  {\doibase 10.1103/physrevlett.109.220501} {\bibfield  {journal} {\bibinfo
  {journal} {Physical Review Letters}\ }\textbf {\bibinfo {volume} {109}}
  (\bibinfo {year} {2012}),\ 10.1103/physrevlett.109.220501}\BibitemShut
  {NoStop}%
\bibitem [{\citenamefont {Pauls}\ \emph {et~al.}(2013)\citenamefont {Pauls},
  \citenamefont {Zhang}, \citenamefont {Berman},\ and\ \citenamefont
  {Kais}}]{Pauls_2013}%
  \BibitemOpen
  \bibfield  {author} {\bibinfo {author} {\bibfnamefont {J.~A.}\ \bibnamefont
  {Pauls}}, \bibinfo {author} {\bibfnamefont {Y.}~\bibnamefont {Zhang}},
  \bibinfo {author} {\bibfnamefont {G.~P.}\ \bibnamefont {Berman}}, \ and\
  \bibinfo {author} {\bibfnamefont {S.}~\bibnamefont {Kais}},\ }\href {\doibase
  10.1103/physreve.87.062704} {\bibfield  {journal} {\bibinfo  {journal}
  {Physical Review E}\ }\textbf {\bibinfo {volume} {87}} (\bibinfo {year}
  {2013}),\ 10.1103/physreve.87.062704}\BibitemShut {NoStop}%
\bibitem [{\citenamefont {Kritsotakis}\ and\ \citenamefont
  {Kominis}(2014)}]{Kritsotakis_2014}%
  \BibitemOpen
  \bibfield  {author} {\bibinfo {author} {\bibfnamefont {M.}~\bibnamefont
  {Kritsotakis}}\ and\ \bibinfo {author} {\bibfnamefont {I.~K.}\ \bibnamefont
  {Kominis}},\ }\href {\doibase 10.1103/physreve.90.042719} {\bibfield
  {journal} {\bibinfo  {journal} {Physical Review E}\ }\textbf {\bibinfo
  {volume} {90}} (\bibinfo {year} {2014}),\
  10.1103/physreve.90.042719}\BibitemShut {NoStop}%
\bibitem [{\citenamefont {Zhang}\ \emph {et~al.}(2015)\citenamefont {Zhang},
  \citenamefont {Berman},\ and\ \citenamefont {Kais}}]{Zhang_2015}%
  \BibitemOpen
  \bibfield  {author} {\bibinfo {author} {\bibfnamefont {Y.}~\bibnamefont
  {Zhang}}, \bibinfo {author} {\bibfnamefont {G.~P.}\ \bibnamefont {Berman}}, \
  and\ \bibinfo {author} {\bibfnamefont {S.}~\bibnamefont {Kais}},\ }\href
  {\doibase 10.1002/qua.24943} {\bibfield  {journal} {\bibinfo  {journal}
  {International Journal of Quantum Chemistry}\ }\textbf {\bibinfo {volume}
  {115}},\ \bibinfo {pages} {1327} (\bibinfo {year} {2015})}\BibitemShut
  {NoStop}%
\bibitem [{\citenamefont {Guo}\ \emph {et~al.}(2017)\citenamefont {Guo},
  \citenamefont {Xu}, \citenamefont {Zou},\ and\ \citenamefont
  {Shao}}]{Guo_2017}%
  \BibitemOpen
  \bibfield  {author} {\bibinfo {author} {\bibfnamefont {L.-S.}\ \bibnamefont
  {Guo}}, \bibinfo {author} {\bibfnamefont {B.-M.}\ \bibnamefont {Xu}},
  \bibinfo {author} {\bibfnamefont {J.}~\bibnamefont {Zou}}, \ and\ \bibinfo
  {author} {\bibfnamefont {B.}~\bibnamefont {Shao}},\ }\href {\doibase
  10.1038/s41598-017-06187-y} {\bibfield  {journal} {\bibinfo  {journal}
  {Scientific Reports}\ }\textbf {\bibinfo {volume} {7}} (\bibinfo {year}
  {2017}),\ 10.1038/s41598-017-06187-y}\BibitemShut {NoStop}%
\bibitem [{\citenamefont {Vitalis}\ and\ \citenamefont
  {Kominis}(2017)}]{Vitalis_2017}%
  \BibitemOpen
  \bibfield  {author} {\bibinfo {author} {\bibfnamefont {K.~M.}\ \bibnamefont
  {Vitalis}}\ and\ \bibinfo {author} {\bibfnamefont {I.~K.}\ \bibnamefont
  {Kominis}},\ }\href {\doibase 10.1103/physreva.95.032129} {\bibfield
  {journal} {\bibinfo  {journal} {Physical Review A}\ }\textbf {\bibinfo
  {volume} {95}} (\bibinfo {year} {2017}),\
  10.1103/physreva.95.032129}\BibitemShut {NoStop}%
\bibitem [{\citenamefont {Mouloudakis}\ and\ \citenamefont
  {Kominis}(2017)}]{Mouloudakis_2017}%
  \BibitemOpen
  \bibfield  {author} {\bibinfo {author} {\bibfnamefont {K.}~\bibnamefont
  {Mouloudakis}}\ and\ \bibinfo {author} {\bibfnamefont {I.~K.}\ \bibnamefont
  {Kominis}},\ }\href {\doibase 10.1103/physreve.95.022413} {\bibfield
  {journal} {\bibinfo  {journal} {Physical Review E}\ }\textbf {\bibinfo
  {volume} {95}} (\bibinfo {year} {2017}),\
  10.1103/physreve.95.022413}\BibitemShut {NoStop}%
\bibitem [{\citenamefont {Kominis}(2020)}]{Kominis_2020}%
  \BibitemOpen
  \bibfield  {author} {\bibinfo {author} {\bibfnamefont {I.~K.}\ \bibnamefont
  {Kominis}},\ }\href {\doibase 10.1103/physrevresearch.2.023206} {\bibfield
  {journal} {\bibinfo  {journal} {Physical Review Research}\ }\textbf {\bibinfo
  {volume} {2}} (\bibinfo {year} {2020}),\
  10.1103/physrevresearch.2.023206}\BibitemShut {NoStop}%
\bibitem [{\citenamefont {Fay}\ \emph {et~al.}(2020)\citenamefont {Fay},
  \citenamefont {Lindoy}, \citenamefont {Manolopoulos},\ and\ \citenamefont
  {Hore}}]{Fay_2020}%
  \BibitemOpen
  \bibfield  {author} {\bibinfo {author} {\bibfnamefont {T.~P.}\ \bibnamefont
  {Fay}}, \bibinfo {author} {\bibfnamefont {L.~P.}\ \bibnamefont {Lindoy}},
  \bibinfo {author} {\bibfnamefont {D.~E.}\ \bibnamefont {Manolopoulos}}, \
  and\ \bibinfo {author} {\bibfnamefont {P.~J.}\ \bibnamefont {Hore}},\ }\href
  {\doibase 10.1039/c9fd00049f} {\bibfield  {journal} {\bibinfo  {journal}
  {Faraday Discussions}\ }\textbf {\bibinfo {volume} {221}},\ \bibinfo {pages}
  {77} (\bibinfo {year} {2020})}\BibitemShut {NoStop}%
\bibitem [{\citenamefont {Bagryansky}\ \emph {et~al.}(2007)\citenamefont
  {Bagryansky}, \citenamefont {Borovkov},\ and\ \citenamefont
  {Molin}}]{Bagryansky2007}%
  \BibitemOpen
  \bibfield  {author} {\bibinfo {author} {\bibfnamefont {V.~A.}\ \bibnamefont
  {Bagryansky}}, \bibinfo {author} {\bibfnamefont {V.~I.}\ \bibnamefont
  {Borovkov}}, \ and\ \bibinfo {author} {\bibfnamefont {Y.~N.}\ \bibnamefont
  {Molin}},\ }\href@noop {} {\bibfield  {journal} {\bibinfo  {journal} {Russian
  Chemical Reviews}\ }\textbf {\bibinfo {volume} {76}},\ \bibinfo {pages} {493}
  (\bibinfo {year} {2007})}\BibitemShut {NoStop}%
\bibitem [{\citenamefont {Molin}(2004)}]{molin2004}%
  \BibitemOpen
  \bibfield  {author} {\bibinfo {author} {\bibfnamefont {Y.~N.}\ \bibnamefont
  {Molin}},\ }\href@noop {} {\bibfield  {journal} {\bibinfo  {journal}
  {Mendeleev Communications}\ }\textbf {\bibinfo {volume} {14}},\ \bibinfo
  {pages} {85} (\bibinfo {year} {2004})}\BibitemShut {NoStop}%
\bibitem [{\citenamefont {Yu}(1999)}]{yu1999quantum}%
  \BibitemOpen
  \bibfield  {author} {\bibinfo {author} {\bibfnamefont {N.~M.}\ \bibnamefont
  {Yu}},\ }\href@noop {} {\bibfield  {journal} {\bibinfo  {journal} {Bulletin
  of the Korean Chemical Society}\ }\textbf {\bibinfo {volume} {20}},\ \bibinfo
  {pages} {7} (\bibinfo {year} {1999})}\BibitemShut {NoStop}%
\bibitem [{\citenamefont {Brocklehurst}(2002)}]{brocklehurst2002magnetic}%
  \BibitemOpen
  \bibfield  {author} {\bibinfo {author} {\bibfnamefont {B.}~\bibnamefont
  {Brocklehurst}},\ }\href@noop {} {\bibfield  {journal} {\bibinfo  {journal}
  {Chemical Society Reviews}\ }\textbf {\bibinfo {volume} {31}},\ \bibinfo
  {pages} {301} (\bibinfo {year} {2002})}\BibitemShut {NoStop}%
\bibitem [{\citenamefont {Bagryansky}\ \emph {et~al.}(2005)\citenamefont
  {Bagryansky}, \citenamefont {Ivanov}, \citenamefont {Borovkov}, \citenamefont
  {Lukzen},\ and\ \citenamefont {Molin}}]{bagryansky2005spin}%
  \BibitemOpen
  \bibfield  {author} {\bibinfo {author} {\bibfnamefont {V.}~\bibnamefont
  {Bagryansky}}, \bibinfo {author} {\bibfnamefont {K.}~\bibnamefont {Ivanov}},
  \bibinfo {author} {\bibfnamefont {V.}~\bibnamefont {Borovkov}}, \bibinfo
  {author} {\bibfnamefont {N.}~\bibnamefont {Lukzen}}, \ and\ \bibinfo {author}
  {\bibfnamefont {Y.~N.}\ \bibnamefont {Molin}},\ }\href@noop {} {\bibfield
  {journal} {\bibinfo  {journal} {The Journal of chemical physics}\ }\textbf
  {\bibinfo {volume} {122}},\ \bibinfo {pages} {224503} (\bibinfo {year}
  {2005})}\BibitemShut {NoStop}%
\bibitem [{\citenamefont {Jones}\ and\ \citenamefont
  {Hore}(2010)}]{Jones_2010}%
  \BibitemOpen
  \bibfield  {author} {\bibinfo {author} {\bibfnamefont {J.}~\bibnamefont
  {Jones}}\ and\ \bibinfo {author} {\bibfnamefont {P.}~\bibnamefont {Hore}},\
  }\href {\doibase 10.1016/j.cplett.2010.01.063} {\bibfield  {journal}
  {\bibinfo  {journal} {Chemical Physics Letters}\ }\textbf {\bibinfo {volume}
  {488}},\ \bibinfo {pages} {90} (\bibinfo {year} {2010})}\BibitemShut
  {NoStop}%
\bibitem [{\citenamefont {Shushin}(2010)}]{Shushin_2010}%
  \BibitemOpen
  \bibfield  {author} {\bibinfo {author} {\bibfnamefont {A.~I.}\ \bibnamefont
  {Shushin}},\ }\href {\doibase 10.1063/1.3461133} {\bibfield  {journal}
  {\bibinfo  {journal} {The Journal of Chemical Physics}\ }\textbf {\bibinfo
  {volume} {133}},\ \bibinfo {pages} {044505} (\bibinfo {year}
  {2010})}\BibitemShut {NoStop}%
\bibitem [{\citenamefont {Il'ichov}\ and\ \citenamefont
  {Anishchik}(2010)}]{il2010should}%
  \BibitemOpen
  \bibfield  {author} {\bibinfo {author} {\bibfnamefont {L.~V.}\ \bibnamefont
  {Il'ichov}}\ and\ \bibinfo {author} {\bibfnamefont {S.~V.}\ \bibnamefont
  {Anishchik}},\ }\href@noop {} {\bibfield  {journal} {\bibinfo  {journal}
  {arXiv preprint arXiv:1003.1793}\ } (\bibinfo {year} {2010})}\BibitemShut
  {NoStop}%
\bibitem [{\citenamefont {Ivanov}\ \emph {et~al.}(2010)\citenamefont {Ivanov},
  \citenamefont {Petrova}, \citenamefont {Lukzen},\ and\ \citenamefont
  {Maeda}}]{Ivanov_2010}%
  \BibitemOpen
  \bibfield  {author} {\bibinfo {author} {\bibfnamefont {K.~L.}\ \bibnamefont
  {Ivanov}}, \bibinfo {author} {\bibfnamefont {M.~V.}\ \bibnamefont {Petrova}},
  \bibinfo {author} {\bibfnamefont {N.~N.}\ \bibnamefont {Lukzen}}, \ and\
  \bibinfo {author} {\bibfnamefont {K.}~\bibnamefont {Maeda}},\ }\href
  {\doibase 10.1021/jp1048265} {\bibfield  {journal} {\bibinfo  {journal} {The
  Journal of Physical Chemistry A}\ }\textbf {\bibinfo {volume} {114}},\
  \bibinfo {pages} {9447} (\bibinfo {year} {2010})}\BibitemShut {NoStop}%
\bibitem [{\citenamefont {Purtov}(2010)}]{Purtov_2010}%
  \BibitemOpen
  \bibfield  {author} {\bibinfo {author} {\bibfnamefont {P.}~\bibnamefont
  {Purtov}},\ }\href {\doibase 10.1016/j.cplett.2010.07.006} {\bibfield
  {journal} {\bibinfo  {journal} {Chemical Physics Letters}\ }\textbf {\bibinfo
  {volume} {496}},\ \bibinfo {pages} {335} (\bibinfo {year}
  {2010})}\BibitemShut {NoStop}%
\bibitem [{\citenamefont {Jones}\ \emph
  {et~al.}(2011{\natexlab{a}})\citenamefont {Jones}, \citenamefont {Maeda},\
  and\ \citenamefont {Hore}}]{Jones_2011}%
  \BibitemOpen
  \bibfield  {author} {\bibinfo {author} {\bibfnamefont {J.}~\bibnamefont
  {Jones}}, \bibinfo {author} {\bibfnamefont {K.}~\bibnamefont {Maeda}}, \ and\
  \bibinfo {author} {\bibfnamefont {P.}~\bibnamefont {Hore}},\ }\href {\doibase
  10.1016/j.cplett.2011.03.082} {\bibfield  {journal} {\bibinfo  {journal}
  {Chemical Physics Letters}\ }\textbf {\bibinfo {volume} {507}},\ \bibinfo
  {pages} {269} (\bibinfo {year} {2011}{\natexlab{a}})}\BibitemShut {NoStop}%
\bibitem [{\citenamefont {Kominis}(2011{\natexlab{b}})}]{Kominis_2011_Com}%
  \BibitemOpen
  \bibfield  {author} {\bibinfo {author} {\bibfnamefont {I.~K.}\ \bibnamefont
  {Kominis}},\ }\href {\doibase 10.1016/j.cplett.2011.04.026} {\bibfield
  {journal} {\bibinfo  {journal} {Chemical Physics Letters}\ }\textbf {\bibinfo
  {volume} {508}},\ \bibinfo {pages} {182} (\bibinfo {year}
  {2011}{\natexlab{b}})}\BibitemShut {NoStop}%
\bibitem [{\citenamefont {Jones}\ \emph
  {et~al.}(2011{\natexlab{b}})\citenamefont {Jones}, \citenamefont {Maeda},
  \citenamefont {Steiner},\ and\ \citenamefont {Hore}}]{Jones_2011_Com}%
  \BibitemOpen
  \bibfield  {author} {\bibinfo {author} {\bibfnamefont {J.}~\bibnamefont
  {Jones}}, \bibinfo {author} {\bibfnamefont {K.}~\bibnamefont {Maeda}},
  \bibinfo {author} {\bibfnamefont {U.}~\bibnamefont {Steiner}}, \ and\
  \bibinfo {author} {\bibfnamefont {P.}~\bibnamefont {Hore}},\ }\href {\doibase
  10.1016/j.cplett.2011.04.022} {\bibfield  {journal} {\bibinfo  {journal}
  {Chemical Physics Letters}\ }\textbf {\bibinfo {volume} {508}},\ \bibinfo
  {pages} {184} (\bibinfo {year} {2011}{\natexlab{b}})}\BibitemShut {NoStop}%
\bibitem [{\citenamefont {Tiersch}\ \emph {et~al.}(2012)\citenamefont
  {Tiersch}, \citenamefont {Steiner}, \citenamefont {Popescu},\ and\
  \citenamefont {Briegel}}]{Tiersch_2012_Op}%
  \BibitemOpen
  \bibfield  {author} {\bibinfo {author} {\bibfnamefont {M.}~\bibnamefont
  {Tiersch}}, \bibinfo {author} {\bibfnamefont {U.~E.}\ \bibnamefont
  {Steiner}}, \bibinfo {author} {\bibfnamefont {S.}~\bibnamefont {Popescu}}, \
  and\ \bibinfo {author} {\bibfnamefont {H.~J.}\ \bibnamefont {Briegel}},\
  }\href {\doibase 10.1021/jp209196a} {\bibfield  {journal} {\bibinfo
  {journal} {The Journal of Physical Chemistry A}\ }\textbf {\bibinfo {volume}
  {116}},\ \bibinfo {pages} {4020} (\bibinfo {year} {2012})}\BibitemShut
  {NoStop}%
\bibitem [{\citenamefont {Dellis}\ and\ \citenamefont
  {Kominis}(2012)}]{Dellis_2012}%
  \BibitemOpen
  \bibfield  {author} {\bibinfo {author} {\bibfnamefont {A.}~\bibnamefont
  {Dellis}}\ and\ \bibinfo {author} {\bibfnamefont {I.}~\bibnamefont
  {Kominis}},\ }\href {\doibase 10.1016/j.cplett.2012.06.023} {\bibfield
  {journal} {\bibinfo  {journal} {Chemical Physics Letters}\ }\textbf {\bibinfo
  {volume} {543}},\ \bibinfo {pages} {170} (\bibinfo {year}
  {2012})}\BibitemShut {NoStop}%
\bibitem [{\citenamefont {Bagryansky}\ \emph {et~al.}(2013)\citenamefont
  {Bagryansky}, \citenamefont {Borovkov},\ and\ \citenamefont
  {Molin}}]{Bagryansky_2013}%
  \BibitemOpen
  \bibfield  {author} {\bibinfo {author} {\bibfnamefont {V.}~\bibnamefont
  {Bagryansky}}, \bibinfo {author} {\bibfnamefont {V.}~\bibnamefont
  {Borovkov}}, \ and\ \bibinfo {author} {\bibfnamefont {Y.}~\bibnamefont
  {Molin}},\ }\href {\doibase 10.1016/j.cplett.2013.03.047} {\bibfield
  {journal} {\bibinfo  {journal} {Chemical Physics Letters}\ }\textbf {\bibinfo
  {volume} {570}},\ \bibinfo {pages} {141} (\bibinfo {year}
  {2013})}\BibitemShut {NoStop}%
\bibitem [{ibm(2019)}]{ibm}%
  \BibitemOpen
  \href@noop {} {} (\bibinfo {year} {2019}),\ \bibinfo {note} {20-qubit
  backend: IBM Q team, ``IBM Q 20 Almaden backend specification V1.3.1,"
  (2019)}\BibitemShut {NoStop}%
\bibitem [{\citenamefont {Vyushkova}\ \emph {et~al.}(2008)\citenamefont
  {Vyushkova}, \citenamefont {Borovkov}, \citenamefont {Shchegoleva},
  \citenamefont {Beregovaya}, \citenamefont {Bagryansky},\ and\ \citenamefont
  {Molin}}]{Mariya1}%
  \BibitemOpen
  \bibfield  {author} {\bibinfo {author} {\bibfnamefont {M.~M.}\ \bibnamefont
  {Vyushkova}}, \bibinfo {author} {\bibfnamefont {V.~I.}\ \bibnamefont
  {Borovkov}}, \bibinfo {author} {\bibfnamefont {L.~N.}\ \bibnamefont
  {Shchegoleva}}, \bibinfo {author} {\bibfnamefont {I.~V.}\ \bibnamefont
  {Beregovaya}}, \bibinfo {author} {\bibfnamefont {V.~A.}\ \bibnamefont
  {Bagryansky}}, \ and\ \bibinfo {author} {\bibfnamefont {Y.~N.}\ \bibnamefont
  {Molin}},\ }\href@noop {} {\bibfield  {journal} {\bibinfo  {journal} {Doklady
  Physical Chemistry}\ }\textbf {\bibinfo {volume} {420}},\ \bibinfo {pages}
  {125} (\bibinfo {year} {2008})}\BibitemShut {NoStop}%
\bibitem [{\citenamefont {Brocklehurst}(1976)}]{Brocklehurst_1976}%
  \BibitemOpen
  \bibfield  {author} {\bibinfo {author} {\bibfnamefont {B.}~\bibnamefont
  {Brocklehurst}},\ }\href {\doibase 10.1039/F29767201869} {\bibfield
  {journal} {\bibinfo  {journal} {J. Chem. Soc.{,} Faraday Trans. 2}\ }\textbf
  {\bibinfo {volume} {72}},\ \bibinfo {pages} {1869} (\bibinfo {year}
  {1976})}\BibitemShut {NoStop}%
\bibitem [{\citenamefont {Bagryansky}\ \emph {et~al.}(1997)\citenamefont
  {Bagryansky}, \citenamefont {Usov}, \citenamefont {Lukzen},\ and\
  \citenamefont {Molin}}]{bagryansky1997spin}%
  \BibitemOpen
  \bibfield  {author} {\bibinfo {author} {\bibfnamefont {V.}~\bibnamefont
  {Bagryansky}}, \bibinfo {author} {\bibfnamefont {O.}~\bibnamefont {Usov}},
  \bibinfo {author} {\bibfnamefont {N.}~\bibnamefont {Lukzen}}, \ and\ \bibinfo
  {author} {\bibfnamefont {Y.~N.}\ \bibnamefont {Molin}},\ }\href@noop {}
  {\bibfield  {journal} {\bibinfo  {journal} {Applied Magnetic Resonance}\
  }\textbf {\bibinfo {volume} {12}},\ \bibinfo {pages} {505} (\bibinfo {year}
  {1997})}\BibitemShut {NoStop}%
\bibitem [{\citenamefont {Steiner}\ and\ \citenamefont
  {Ulrich}(1989)}]{steiner1989magnetic}%
  \BibitemOpen
  \bibfield  {author} {\bibinfo {author} {\bibfnamefont {U.~E.}\ \bibnamefont
  {Steiner}}\ and\ \bibinfo {author} {\bibfnamefont {T.}~\bibnamefont
  {Ulrich}},\ }\href@noop {} {\bibfield  {journal} {\bibinfo  {journal}
  {Chemical Reviews}\ }\textbf {\bibinfo {volume} {89}},\ \bibinfo {pages} {51}
  (\bibinfo {year} {1989})}\BibitemShut {NoStop}%
\bibitem [{\citenamefont {Preskill}(1998)}]{Preskill}%
  \BibitemOpen
  \bibfield  {author} {\bibinfo {author} {\bibfnamefont {J.}~\bibnamefont
  {Preskill}},\ }\href@noop {} {\bibfield  {journal} {\bibinfo  {journal}
  {California Institute of Technology}\ }\textbf {\bibinfo {volume} {16}}
  (\bibinfo {year} {1998})}\BibitemShut {NoStop}%
\bibitem [{\citenamefont {Fujiwara}(2004)}]{fujiwara2004ampdamp}%
  \BibitemOpen
  \bibfield  {author} {\bibinfo {author} {\bibfnamefont {A.}~\bibnamefont
  {Fujiwara}},\ }\href {\doibase 10.1103/PhysRevA.70.012317} {\bibfield
  {journal} {\bibinfo  {journal} {Physical Review A}\ }\textbf {\bibinfo
  {volume} {89}} (\bibinfo {year} {2004}),\
  10.1103/PhysRevA.70.012317}\BibitemShut {NoStop}%
\bibitem [{\citenamefont {Low}\ and\ \citenamefont
  {Chuang}(2017)}]{low2017optimal}%
  \BibitemOpen
  \bibfield  {author} {\bibinfo {author} {\bibfnamefont {G.~H.}\ \bibnamefont
  {Low}}\ and\ \bibinfo {author} {\bibfnamefont {I.~L.}\ \bibnamefont
  {Chuang}},\ }\href@noop {} {\bibfield  {journal} {\bibinfo  {journal}
  {Physical review letters}\ }\textbf {\bibinfo {volume} {118}},\ \bibinfo
  {pages} {010501} (\bibinfo {year} {2017})}\BibitemShut {NoStop}%
\bibitem [{\citenamefont {Berry}\ \emph {et~al.}(2015)\citenamefont {Berry},
  \citenamefont {Childs}, \citenamefont {Cleve}, \citenamefont {Kothari},\ and\
  \citenamefont {Somma}}]{berry2015simulating}%
  \BibitemOpen
  \bibfield  {author} {\bibinfo {author} {\bibfnamefont {D.~W.}\ \bibnamefont
  {Berry}}, \bibinfo {author} {\bibfnamefont {A.~M.}\ \bibnamefont {Childs}},
  \bibinfo {author} {\bibfnamefont {R.}~\bibnamefont {Cleve}}, \bibinfo
  {author} {\bibfnamefont {R.}~\bibnamefont {Kothari}}, \ and\ \bibinfo
  {author} {\bibfnamefont {R.~D.}\ \bibnamefont {Somma}},\ }\href@noop {}
  {\bibfield  {journal} {\bibinfo  {journal} {Physical review letters}\
  }\textbf {\bibinfo {volume} {114}},\ \bibinfo {pages} {090502} (\bibinfo
  {year} {2015})}\BibitemShut {NoStop}%
\bibitem [{\citenamefont {Whitfield}\ \emph {et~al.}(2011)\citenamefont
  {Whitfield}, \citenamefont {Biamonte},\ and\ \citenamefont
  {Aspuru-Guzik}}]{whitfield2011simulation}%
  \BibitemOpen
  \bibfield  {author} {\bibinfo {author} {\bibfnamefont {J.~D.}\ \bibnamefont
  {Whitfield}}, \bibinfo {author} {\bibfnamefont {J.}~\bibnamefont {Biamonte}},
  \ and\ \bibinfo {author} {\bibfnamefont {A.}~\bibnamefont {Aspuru-Guzik}},\
  }\href@noop {} {\bibfield  {journal} {\bibinfo  {journal} {Molecular
  Physics}\ }\textbf {\bibinfo {volume} {109}},\ \bibinfo {pages} {735}
  (\bibinfo {year} {2011})}\BibitemShut {NoStop}%
\bibitem [{\citenamefont {Babbush}\ \emph {et~al.}(2018)\citenamefont
  {Babbush}, \citenamefont {Wiebe}, \citenamefont {McClean}, \citenamefont
  {McClain}, \citenamefont {Neven},\ and\ \citenamefont
  {Chan}}]{babbush2018low}%
  \BibitemOpen
  \bibfield  {author} {\bibinfo {author} {\bibfnamefont {R.}~\bibnamefont
  {Babbush}}, \bibinfo {author} {\bibfnamefont {N.}~\bibnamefont {Wiebe}},
  \bibinfo {author} {\bibfnamefont {J.}~\bibnamefont {McClean}}, \bibinfo
  {author} {\bibfnamefont {J.}~\bibnamefont {McClain}}, \bibinfo {author}
  {\bibfnamefont {H.}~\bibnamefont {Neven}}, \ and\ \bibinfo {author}
  {\bibfnamefont {G.~K.-L.}\ \bibnamefont {Chan}},\ }\href@noop {} {\bibfield
  {journal} {\bibinfo  {journal} {Physical Review X}\ }\textbf {\bibinfo
  {volume} {8}},\ \bibinfo {pages} {011044} (\bibinfo {year}
  {2018})}\BibitemShut {NoStop}%
\bibitem [{\citenamefont {Barreiro}\ \emph {et~al.}(2011)\citenamefont
  {Barreiro}, \citenamefont {M{\"u}ller}, \citenamefont {Schindler},
  \citenamefont {Nigg}, \citenamefont {Monz}, \citenamefont {Chwalla},
  \citenamefont {Hennrich}, \citenamefont {Roos}, \citenamefont {Zoller},\ and\
  \citenamefont {Blatt}}]{Barreiro2011}%
  \BibitemOpen
  \bibfield  {author} {\bibinfo {author} {\bibfnamefont {J.~T.}\ \bibnamefont
  {Barreiro}}, \bibinfo {author} {\bibfnamefont {M.}~\bibnamefont
  {M{\"u}ller}}, \bibinfo {author} {\bibfnamefont {P.}~\bibnamefont
  {Schindler}}, \bibinfo {author} {\bibfnamefont {D.}~\bibnamefont {Nigg}},
  \bibinfo {author} {\bibfnamefont {T.}~\bibnamefont {Monz}}, \bibinfo {author}
  {\bibfnamefont {M.}~\bibnamefont {Chwalla}}, \bibinfo {author} {\bibfnamefont
  {M.}~\bibnamefont {Hennrich}}, \bibinfo {author} {\bibfnamefont {C.~F.}\
  \bibnamefont {Roos}}, \bibinfo {author} {\bibfnamefont {P.}~\bibnamefont
  {Zoller}}, \ and\ \bibinfo {author} {\bibfnamefont {R.}~\bibnamefont
  {Blatt}},\ }\href {https://doi.org/10.1038/nature09801} {\bibfield  {journal}
  {\bibinfo  {journal} {Nature}\ }\textbf {\bibinfo {volume} {470}},\ \bibinfo
  {pages} {486 EP } (\bibinfo {year} {2011})},\ \bibinfo {note}
  {article}\BibitemShut {NoStop}%
\bibitem [{\citenamefont {Del~Re}\ \emph {et~al.}(2020)\citenamefont {Del~Re},
  \citenamefont {Rost}, \citenamefont {Kemper},\ and\ \citenamefont
  {Freericks}}]{rost2020}%
  \BibitemOpen
  \bibfield  {author} {\bibinfo {author} {\bibfnamefont {L.}~\bibnamefont
  {Del~Re}}, \bibinfo {author} {\bibfnamefont {B.}~\bibnamefont {Rost}},
  \bibinfo {author} {\bibfnamefont {A.~F.}\ \bibnamefont {Kemper}}, \ and\
  \bibinfo {author} {\bibfnamefont {J.~K.}\ \bibnamefont {Freericks}},\ }\href
  {\doibase 10.1103/PhysRevB.102.125112} {\bibfield  {journal} {\bibinfo
  {journal} {Phys. Rev. B}\ }\textbf {\bibinfo {volume} {102}},\ \bibinfo
  {pages} {125112} (\bibinfo {year} {2020})}\BibitemShut {NoStop}%
\bibitem [{\citenamefont {McKay}\ \emph {et~al.}(2017)\citenamefont {McKay},
  \citenamefont {Wood}, \citenamefont {Sheldon}, \citenamefont {Chow},\ and\
  \citenamefont {Gambetta}}]{mckay2017efficient}%
  \BibitemOpen
  \bibfield  {author} {\bibinfo {author} {\bibfnamefont {D.~C.}\ \bibnamefont
  {McKay}}, \bibinfo {author} {\bibfnamefont {C.~J.}\ \bibnamefont {Wood}},
  \bibinfo {author} {\bibfnamefont {S.}~\bibnamefont {Sheldon}}, \bibinfo
  {author} {\bibfnamefont {J.~M.}\ \bibnamefont {Chow}}, \ and\ \bibinfo
  {author} {\bibfnamefont {J.~M.}\ \bibnamefont {Gambetta}},\ }\href@noop {}
  {\bibfield  {journal} {\bibinfo  {journal} {Physical Review A}\ }\textbf
  {\bibinfo {volume} {96}},\ \bibinfo {pages} {022330} (\bibinfo {year}
  {2017})}\BibitemShut {NoStop}%
\bibitem [{\citenamefont {Abraham}\ \emph {et~al.}(2019)\citenamefont
  {Abraham}, \citenamefont {Akhalwaya}, \citenamefont {Aleksandrowicz},
  \citenamefont {Alexander}, \citenamefont {Alexandrowics}, \citenamefont
  {Arbel}, \citenamefont {Asfaw}, \citenamefont {Azaustre}, \citenamefont
  {Barkoutsos}, \citenamefont {Barron}, \citenamefont {Bello}, \citenamefont
  {Ben-Haim}, \citenamefont {Bevenius}, \citenamefont {Bishop}, \citenamefont
  {Bosch}, \citenamefont {Bucher}, \citenamefont {CZ}, \citenamefont {Cabrera},
  \citenamefont {Calpin}, \citenamefont {Capelluto}, \citenamefont {Carballo},
  \citenamefont {Carrascal}, \citenamefont {Chen}, \citenamefont {Chen},
  \citenamefont {Chen}, \citenamefont {Chow}, \citenamefont {Claus},
  \citenamefont {Clauss}, \citenamefont {Cross}, \citenamefont {Cross},
  \citenamefont {Cruz-Benito}, \citenamefont {Cryoris}, \citenamefont {Culver},
  \citenamefont {C{\'o}rcoles-Gonzales}, \citenamefont {Dague}, \citenamefont
  {Dartiailh}, \citenamefont {Davila}, \citenamefont {Ding}, \citenamefont
  {Dumitrescu}, \citenamefont {Dumon}, \citenamefont {Duran}, \citenamefont
  {Eendebak}, \citenamefont {Egger}, \citenamefont {Everitt}, \citenamefont
  {Fern{\'a}ndez}, \citenamefont {Frisch}, \citenamefont {Fuhrer},
  \citenamefont {GOULD}, \citenamefont {Gacon}, \citenamefont {Gadi},
  \citenamefont {Gago}, \citenamefont {Gambetta}, \citenamefont {Garcia},
  \citenamefont {Garion}, \citenamefont {Gawel-Kus}, \citenamefont
  {Gomez-Mosquera}, \citenamefont {de~la Puente~Gonz{\'a}lez}, \citenamefont
  {Greenberg}, \citenamefont {Gunnels}, \citenamefont {Haide}, \citenamefont
  {Hamamura}, \citenamefont {Havlicek}, \citenamefont {Hellmers}, \citenamefont
  {Herok}, \citenamefont {Horii}, \citenamefont {Howington}, \citenamefont
  {Hu}, \citenamefont {Hu}, \citenamefont {Imai}, \citenamefont {Imamichi},
  \citenamefont {Iten}, \citenamefont {Itoko}, \citenamefont {Javadi-Abhari},
  \citenamefont {Jessica}, \citenamefont {Johns}, \citenamefont {Kanazawa},
  \citenamefont {Karazeev}, \citenamefont {Kassebaum}, \citenamefont
  {Kovyrshin}, \citenamefont {Krishnan}, \citenamefont {Krsulich},
  \citenamefont {Kus}, \citenamefont {LaRose}, \citenamefont {Lambert},
  \citenamefont {Latone}, \citenamefont {Lawrence}, \citenamefont {Liu},
  \citenamefont {Liu}, \citenamefont {Mac}, \citenamefont {Maeng},
  \citenamefont {Malyshev}, \citenamefont {Marecek}, \citenamefont {Marques},
  \citenamefont {Mathews}, \citenamefont {Matsuo}, \citenamefont {McClure},
  \citenamefont {McGarry}, \citenamefont {McKay}, \citenamefont {Meesala},
  \citenamefont {Mezzacapo}, \citenamefont {Midha}, \citenamefont {Minev},
  \citenamefont {Mooring}, \citenamefont {Morales}, \citenamefont {Moran},
  \citenamefont {Murali}, \citenamefont {M{\"u}ggenburg}, \citenamefont
  {Nadlinger}, \citenamefont {Nannicini}, \citenamefont {Nation}, \citenamefont
  {Naveh}, \citenamefont {Nick-Singstock}, \citenamefont {Niroula},
  \citenamefont {Norlen}, \citenamefont {O'Riordan}, \citenamefont
  {Ollitrault}, \citenamefont {Oud}, \citenamefont {Padilha}, \citenamefont
  {Paik}, \citenamefont {Perriello}, \citenamefont {Phan}, \citenamefont
  {Pistoia}, \citenamefont {Pozas-iKerstjens}, \citenamefont {Prutyanov},
  \citenamefont {P{\'e}rez}, \citenamefont {Quintiii}, \citenamefont {Raymond},
  \citenamefont {Redondo}, \citenamefont {Reuter}, \citenamefont
  {Rodr{\'\i}guez}, \citenamefont {Ryu}, \citenamefont {Sandberg},
  \citenamefont {Sathaye}, \citenamefont {Schmitt}, \citenamefont {Schnabel},
  \citenamefont {Scholten}, \citenamefont {Schoute}, \citenamefont {Sertage},
  \citenamefont {Shammah}, \citenamefont {Shi}, \citenamefont {Silva},
  \citenamefont {Siraichi}, \citenamefont {Sivarajah}, \citenamefont {Smolin},
  \citenamefont {Soeken}, \citenamefont {Steenken}, \citenamefont
  {Stypulkoski}, \citenamefont {Takahashi}, \citenamefont {Taylor},
  \citenamefont {Taylour}, \citenamefont {Thomas}, \citenamefont {Tillet},
  \citenamefont {Tod}, \citenamefont {de~la Torre}, \citenamefont {Trabing},
  \citenamefont {Treinish}, \citenamefont {TrishaPe}, \citenamefont {Turner},
  \citenamefont {Vaknin}, \citenamefont {Valcarce}, \citenamefont {Varchon},
  \citenamefont {Vogt-Lee}, \citenamefont {Vuillot}, \citenamefont {Weaver},
  \citenamefont {Wieczorek}, \citenamefont {Wildstrom}, \citenamefont {Wille},
  \citenamefont {Winston}, \citenamefont {Woehr}, \citenamefont {Woerner},
  \citenamefont {Woo}, \citenamefont {Wood}, \citenamefont {Wood},
  \citenamefont {Wood}, \citenamefont {Wootton}, \citenamefont {Yeralin},
  \citenamefont {Yu}, \citenamefont {Zdanski}, \citenamefont {Zoufalc},
  \citenamefont {anedumla}, \citenamefont {azulehner}, \citenamefont
  {bcamorrison}, \citenamefont {brandhsn}, \citenamefont {dennis-liu 1},
  \citenamefont {drholmie}, \citenamefont {elfrocampeador}, \citenamefont
  {fanizzamarco}, \citenamefont {gruu}, \citenamefont {kanejess}, \citenamefont
  {klinvill}, \citenamefont {lerongil}, \citenamefont {ma5x}, \citenamefont
  {merav aharoni}, \citenamefont {mrossinek}, \citenamefont {ordmoj},
  \citenamefont {strickroman}, \citenamefont {tigerjack}, \citenamefont
  {yang.luh},\ and\ \citenamefont {yotamvakninibm}}]{Qiskit}%
  \BibitemOpen
  \bibfield  {author} {\bibinfo {author} {\bibfnamefont {H.}~\bibnamefont
  {Abraham}}, \bibinfo {author} {\bibfnamefont {I.~Y.}\ \bibnamefont
  {Akhalwaya}}, \bibinfo {author} {\bibfnamefont {G.}~\bibnamefont
  {Aleksandrowicz}}, \bibinfo {author} {\bibfnamefont {T.}~\bibnamefont
  {Alexander}}, \bibinfo {author} {\bibfnamefont {G.}~\bibnamefont
  {Alexandrowics}}, \bibinfo {author} {\bibfnamefont {E.}~\bibnamefont
  {Arbel}}, \bibinfo {author} {\bibfnamefont {A.}~\bibnamefont {Asfaw}},
  \bibinfo {author} {\bibfnamefont {C.}~\bibnamefont {Azaustre}}, \bibinfo
  {author} {\bibfnamefont {P.}~\bibnamefont {Barkoutsos}}, \bibinfo {author}
  {\bibfnamefont {G.}~\bibnamefont {Barron}}, \bibinfo {author} {\bibfnamefont
  {L.}~\bibnamefont {Bello}}, \bibinfo {author} {\bibfnamefont
  {Y.}~\bibnamefont {Ben-Haim}}, \bibinfo {author} {\bibfnamefont
  {D.}~\bibnamefont {Bevenius}}, \bibinfo {author} {\bibfnamefont {L.~S.}\
  \bibnamefont {Bishop}}, \bibinfo {author} {\bibfnamefont {S.}~\bibnamefont
  {Bosch}}, \bibinfo {author} {\bibfnamefont {D.}~\bibnamefont {Bucher}},
  \bibinfo {author} {\bibnamefont {CZ}}, \bibinfo {author} {\bibfnamefont
  {F.}~\bibnamefont {Cabrera}}, \bibinfo {author} {\bibfnamefont
  {P.}~\bibnamefont {Calpin}}, \bibinfo {author} {\bibfnamefont
  {L.}~\bibnamefont {Capelluto}}, \bibinfo {author} {\bibfnamefont
  {J.}~\bibnamefont {Carballo}}, \bibinfo {author} {\bibfnamefont
  {G.}~\bibnamefont {Carrascal}}, \bibinfo {author} {\bibfnamefont
  {A.}~\bibnamefont {Chen}}, \bibinfo {author} {\bibfnamefont {C.-F.}\
  \bibnamefont {Chen}}, \bibinfo {author} {\bibfnamefont {R.}~\bibnamefont
  {Chen}}, \bibinfo {author} {\bibfnamefont {J.~M.}\ \bibnamefont {Chow}},
  \bibinfo {author} {\bibfnamefont {C.}~\bibnamefont {Claus}}, \bibinfo
  {author} {\bibfnamefont {C.}~\bibnamefont {Clauss}}, \bibinfo {author}
  {\bibfnamefont {A.~J.}\ \bibnamefont {Cross}}, \bibinfo {author}
  {\bibfnamefont {A.~W.}\ \bibnamefont {Cross}}, \bibinfo {author}
  {\bibfnamefont {J.}~\bibnamefont {Cruz-Benito}}, \bibinfo {author}
  {\bibnamefont {Cryoris}}, \bibinfo {author} {\bibfnamefont {C.}~\bibnamefont
  {Culver}}, \bibinfo {author} {\bibfnamefont {A.~D.}\ \bibnamefont
  {C{\'o}rcoles-Gonzales}}, \bibinfo {author} {\bibfnamefont {S.}~\bibnamefont
  {Dague}}, \bibinfo {author} {\bibfnamefont {M.}~\bibnamefont {Dartiailh}},
  \bibinfo {author} {\bibfnamefont {A.~R.}\ \bibnamefont {Davila}}, \bibinfo
  {author} {\bibfnamefont {D.}~\bibnamefont {Ding}}, \bibinfo {author}
  {\bibfnamefont {E.}~\bibnamefont {Dumitrescu}}, \bibinfo {author}
  {\bibfnamefont {K.}~\bibnamefont {Dumon}}, \bibinfo {author} {\bibfnamefont
  {I.}~\bibnamefont {Duran}}, \bibinfo {author} {\bibfnamefont
  {P.}~\bibnamefont {Eendebak}}, \bibinfo {author} {\bibfnamefont
  {D.}~\bibnamefont {Egger}}, \bibinfo {author} {\bibfnamefont
  {M.}~\bibnamefont {Everitt}}, \bibinfo {author} {\bibfnamefont {P.~M.}\
  \bibnamefont {Fern{\'a}ndez}}, \bibinfo {author} {\bibfnamefont
  {A.}~\bibnamefont {Frisch}}, \bibinfo {author} {\bibfnamefont
  {A.}~\bibnamefont {Fuhrer}}, \bibinfo {author} {\bibfnamefont
  {I.}~\bibnamefont {GOULD}}, \bibinfo {author} {\bibfnamefont
  {J.}~\bibnamefont {Gacon}}, \bibinfo {author} {\bibnamefont {Gadi}}, \bibinfo
  {author} {\bibfnamefont {B.~G.}\ \bibnamefont {Gago}}, \bibinfo {author}
  {\bibfnamefont {J.~M.}\ \bibnamefont {Gambetta}}, \bibinfo {author}
  {\bibfnamefont {L.}~\bibnamefont {Garcia}}, \bibinfo {author} {\bibfnamefont
  {S.}~\bibnamefont {Garion}}, \bibinfo {author} {\bibnamefont {Gawel-Kus}},
  \bibinfo {author} {\bibfnamefont {J.}~\bibnamefont {Gomez-Mosquera}},
  \bibinfo {author} {\bibfnamefont {S.}~\bibnamefont {de~la
  Puente~Gonz{\'a}lez}}, \bibinfo {author} {\bibfnamefont {D.}~\bibnamefont
  {Greenberg}}, \bibinfo {author} {\bibfnamefont {J.~A.}\ \bibnamefont
  {Gunnels}}, \bibinfo {author} {\bibfnamefont {I.}~\bibnamefont {Haide}},
  \bibinfo {author} {\bibfnamefont {I.}~\bibnamefont {Hamamura}}, \bibinfo
  {author} {\bibfnamefont {V.}~\bibnamefont {Havlicek}}, \bibinfo {author}
  {\bibfnamefont {J.}~\bibnamefont {Hellmers}}, \bibinfo {author}
  {\bibfnamefont {{\L}.}~\bibnamefont {Herok}}, \bibinfo {author}
  {\bibfnamefont {H.}~\bibnamefont {Horii}}, \bibinfo {author} {\bibfnamefont
  {C.}~\bibnamefont {Howington}}, \bibinfo {author} {\bibfnamefont
  {S.}~\bibnamefont {Hu}}, \bibinfo {author} {\bibfnamefont {W.}~\bibnamefont
  {Hu}}, \bibinfo {author} {\bibfnamefont {H.}~\bibnamefont {Imai}}, \bibinfo
  {author} {\bibfnamefont {T.}~\bibnamefont {Imamichi}}, \bibinfo {author}
  {\bibfnamefont {R.}~\bibnamefont {Iten}}, \bibinfo {author} {\bibfnamefont
  {T.}~\bibnamefont {Itoko}}, \bibinfo {author} {\bibfnamefont
  {A.}~\bibnamefont {Javadi-Abhari}}, \bibinfo {author} {\bibnamefont
  {Jessica}}, \bibinfo {author} {\bibfnamefont {K.}~\bibnamefont {Johns}},
  \bibinfo {author} {\bibfnamefont {N.}~\bibnamefont {Kanazawa}}, \bibinfo
  {author} {\bibfnamefont {A.}~\bibnamefont {Karazeev}}, \bibinfo {author}
  {\bibfnamefont {P.}~\bibnamefont {Kassebaum}}, \bibinfo {author}
  {\bibfnamefont {A.}~\bibnamefont {Kovyrshin}}, \bibinfo {author}
  {\bibfnamefont {V.}~\bibnamefont {Krishnan}}, \bibinfo {author}
  {\bibfnamefont {K.}~\bibnamefont {Krsulich}}, \bibinfo {author}
  {\bibfnamefont {G.}~\bibnamefont {Kus}}, \bibinfo {author} {\bibfnamefont
  {R.}~\bibnamefont {LaRose}}, \bibinfo {author} {\bibfnamefont
  {R.}~\bibnamefont {Lambert}}, \bibinfo {author} {\bibfnamefont
  {J.}~\bibnamefont {Latone}}, \bibinfo {author} {\bibfnamefont
  {S.}~\bibnamefont {Lawrence}}, \bibinfo {author} {\bibfnamefont
  {D.}~\bibnamefont {Liu}}, \bibinfo {author} {\bibfnamefont {P.}~\bibnamefont
  {Liu}}, \bibinfo {author} {\bibfnamefont {P.~B.~Z.}\ \bibnamefont {Mac}},
  \bibinfo {author} {\bibfnamefont {Y.}~\bibnamefont {Maeng}}, \bibinfo
  {author} {\bibfnamefont {A.}~\bibnamefont {Malyshev}}, \bibinfo {author}
  {\bibfnamefont {J.}~\bibnamefont {Marecek}}, \bibinfo {author} {\bibfnamefont
  {M.}~\bibnamefont {Marques}}, \bibinfo {author} {\bibfnamefont
  {D.}~\bibnamefont {Mathews}}, \bibinfo {author} {\bibfnamefont
  {A.}~\bibnamefont {Matsuo}}, \bibinfo {author} {\bibfnamefont {D.~T.}\
  \bibnamefont {McClure}}, \bibinfo {author} {\bibfnamefont {C.}~\bibnamefont
  {McGarry}}, \bibinfo {author} {\bibfnamefont {D.}~\bibnamefont {McKay}},
  \bibinfo {author} {\bibfnamefont {S.}~\bibnamefont {Meesala}}, \bibinfo
  {author} {\bibfnamefont {A.}~\bibnamefont {Mezzacapo}}, \bibinfo {author}
  {\bibfnamefont {R.}~\bibnamefont {Midha}}, \bibinfo {author} {\bibfnamefont
  {Z.}~\bibnamefont {Minev}}, \bibinfo {author} {\bibfnamefont {M.~D.}\
  \bibnamefont {Mooring}}, \bibinfo {author} {\bibfnamefont {R.}~\bibnamefont
  {Morales}}, \bibinfo {author} {\bibfnamefont {N.}~\bibnamefont {Moran}},
  \bibinfo {author} {\bibfnamefont {P.}~\bibnamefont {Murali}}, \bibinfo
  {author} {\bibfnamefont {J.}~\bibnamefont {M{\"u}ggenburg}}, \bibinfo
  {author} {\bibfnamefont {D.}~\bibnamefont {Nadlinger}}, \bibinfo {author}
  {\bibfnamefont {G.}~\bibnamefont {Nannicini}}, \bibinfo {author}
  {\bibfnamefont {P.}~\bibnamefont {Nation}}, \bibinfo {author} {\bibfnamefont
  {Y.}~\bibnamefont {Naveh}}, \bibinfo {author} {\bibnamefont
  {Nick-Singstock}}, \bibinfo {author} {\bibfnamefont {P.}~\bibnamefont
  {Niroula}}, \bibinfo {author} {\bibfnamefont {H.}~\bibnamefont {Norlen}},
  \bibinfo {author} {\bibfnamefont {L.~J.}\ \bibnamefont {O'Riordan}}, \bibinfo
  {author} {\bibfnamefont {P.}~\bibnamefont {Ollitrault}}, \bibinfo {author}
  {\bibfnamefont {S.}~\bibnamefont {Oud}}, \bibinfo {author} {\bibfnamefont
  {D.}~\bibnamefont {Padilha}}, \bibinfo {author} {\bibfnamefont
  {H.}~\bibnamefont {Paik}}, \bibinfo {author} {\bibfnamefont {S.}~\bibnamefont
  {Perriello}}, \bibinfo {author} {\bibfnamefont {A.}~\bibnamefont {Phan}},
  \bibinfo {author} {\bibfnamefont {M.}~\bibnamefont {Pistoia}}, \bibinfo
  {author} {\bibfnamefont {A.}~\bibnamefont {Pozas-iKerstjens}}, \bibinfo
  {author} {\bibfnamefont {V.}~\bibnamefont {Prutyanov}}, \bibinfo {author}
  {\bibfnamefont {J.}~\bibnamefont {P{\'e}rez}}, \bibinfo {author}
  {\bibnamefont {Quintiii}}, \bibinfo {author} {\bibfnamefont {R.}~\bibnamefont
  {Raymond}}, \bibinfo {author} {\bibfnamefont {R.~M.-C.}\ \bibnamefont
  {Redondo}}, \bibinfo {author} {\bibfnamefont {M.}~\bibnamefont {Reuter}},
  \bibinfo {author} {\bibfnamefont {D.~M.}\ \bibnamefont {Rodr{\'\i}guez}},
  \bibinfo {author} {\bibfnamefont {M.}~\bibnamefont {Ryu}}, \bibinfo {author}
  {\bibfnamefont {M.}~\bibnamefont {Sandberg}}, \bibinfo {author}
  {\bibfnamefont {N.}~\bibnamefont {Sathaye}}, \bibinfo {author} {\bibfnamefont
  {B.}~\bibnamefont {Schmitt}}, \bibinfo {author} {\bibfnamefont
  {C.}~\bibnamefont {Schnabel}}, \bibinfo {author} {\bibfnamefont {T.~L.}\
  \bibnamefont {Scholten}}, \bibinfo {author} {\bibfnamefont {E.}~\bibnamefont
  {Schoute}}, \bibinfo {author} {\bibfnamefont {I.~F.}\ \bibnamefont
  {Sertage}}, \bibinfo {author} {\bibfnamefont {N.}~\bibnamefont {Shammah}},
  \bibinfo {author} {\bibfnamefont {Y.}~\bibnamefont {Shi}}, \bibinfo {author}
  {\bibfnamefont {A.}~\bibnamefont {Silva}}, \bibinfo {author} {\bibfnamefont
  {Y.}~\bibnamefont {Siraichi}}, \bibinfo {author} {\bibfnamefont
  {S.}~\bibnamefont {Sivarajah}}, \bibinfo {author} {\bibfnamefont {J.~A.}\
  \bibnamefont {Smolin}}, \bibinfo {author} {\bibfnamefont {M.}~\bibnamefont
  {Soeken}}, \bibinfo {author} {\bibfnamefont {D.}~\bibnamefont {Steenken}},
  \bibinfo {author} {\bibfnamefont {M.}~\bibnamefont {Stypulkoski}}, \bibinfo
  {author} {\bibfnamefont {H.}~\bibnamefont {Takahashi}}, \bibinfo {author}
  {\bibfnamefont {C.}~\bibnamefont {Taylor}}, \bibinfo {author} {\bibfnamefont
  {P.}~\bibnamefont {Taylour}}, \bibinfo {author} {\bibfnamefont
  {S.}~\bibnamefont {Thomas}}, \bibinfo {author} {\bibfnamefont
  {M.}~\bibnamefont {Tillet}}, \bibinfo {author} {\bibfnamefont
  {M.}~\bibnamefont {Tod}}, \bibinfo {author} {\bibfnamefont {E.}~\bibnamefont
  {de~la Torre}}, \bibinfo {author} {\bibfnamefont {K.}~\bibnamefont
  {Trabing}}, \bibinfo {author} {\bibfnamefont {M.}~\bibnamefont {Treinish}},
  \bibinfo {author} {\bibnamefont {TrishaPe}}, \bibinfo {author} {\bibfnamefont
  {W.}~\bibnamefont {Turner}}, \bibinfo {author} {\bibfnamefont
  {Y.}~\bibnamefont {Vaknin}}, \bibinfo {author} {\bibfnamefont {C.~R.}\
  \bibnamefont {Valcarce}}, \bibinfo {author} {\bibfnamefont {F.}~\bibnamefont
  {Varchon}}, \bibinfo {author} {\bibfnamefont {D.}~\bibnamefont {Vogt-Lee}},
  \bibinfo {author} {\bibfnamefont {C.}~\bibnamefont {Vuillot}}, \bibinfo
  {author} {\bibfnamefont {J.}~\bibnamefont {Weaver}}, \bibinfo {author}
  {\bibfnamefont {R.}~\bibnamefont {Wieczorek}}, \bibinfo {author}
  {\bibfnamefont {J.~A.}\ \bibnamefont {Wildstrom}}, \bibinfo {author}
  {\bibfnamefont {R.}~\bibnamefont {Wille}}, \bibinfo {author} {\bibfnamefont
  {E.}~\bibnamefont {Winston}}, \bibinfo {author} {\bibfnamefont {J.~J.}\
  \bibnamefont {Woehr}}, \bibinfo {author} {\bibfnamefont {S.}~\bibnamefont
  {Woerner}}, \bibinfo {author} {\bibfnamefont {R.}~\bibnamefont {Woo}},
  \bibinfo {author} {\bibfnamefont {C.~J.}\ \bibnamefont {Wood}}, \bibinfo
  {author} {\bibfnamefont {R.}~\bibnamefont {Wood}}, \bibinfo {author}
  {\bibfnamefont {S.}~\bibnamefont {Wood}}, \bibinfo {author} {\bibfnamefont
  {J.}~\bibnamefont {Wootton}}, \bibinfo {author} {\bibfnamefont
  {D.}~\bibnamefont {Yeralin}}, \bibinfo {author} {\bibfnamefont
  {J.}~\bibnamefont {Yu}}, \bibinfo {author} {\bibfnamefont {L.}~\bibnamefont
  {Zdanski}}, \bibinfo {author} {\bibnamefont {Zoufalc}}, \bibinfo {author}
  {\bibnamefont {anedumla}}, \bibinfo {author} {\bibnamefont {azulehner}},
  \bibinfo {author} {\bibnamefont {bcamorrison}}, \bibinfo {author}
  {\bibnamefont {brandhsn}}, \bibinfo {author} {\bibnamefont {dennis-liu 1}},
  \bibinfo {author} {\bibnamefont {drholmie}}, \bibinfo {author} {\bibnamefont
  {elfrocampeador}}, \bibinfo {author} {\bibnamefont {fanizzamarco}}, \bibinfo
  {author} {\bibnamefont {gruu}}, \bibinfo {author} {\bibnamefont {kanejess}},
  \bibinfo {author} {\bibnamefont {klinvill}}, \bibinfo {author} {\bibnamefont
  {lerongil}}, \bibinfo {author} {\bibnamefont {ma5x}}, \bibinfo {author}
  {\bibnamefont {merav aharoni}}, \bibinfo {author} {\bibnamefont {mrossinek}},
  \bibinfo {author} {\bibnamefont {ordmoj}}, \bibinfo {author} {\bibnamefont
  {strickroman}}, \bibinfo {author} {\bibnamefont {tigerjack}}, \bibinfo
  {author} {\bibnamefont {yang.luh}}, \ and\ \bibinfo {author} {\bibnamefont
  {yotamvakninibm}},\ }\href {\doibase 10.5281/zenodo.2562110} {\enquote
  {\bibinfo {title} {Qiskit: An open-source framework for quantum computing},}\
  } (\bibinfo {year} {2019})\BibitemShut {NoStop}%
\bibitem [{\citenamefont {Jurcevic}\ \emph {et~al.}(2020)\citenamefont
  {Jurcevic}, \citenamefont {Javadi-Abhari}, \citenamefont {Bishop},
  \citenamefont {Lauer}, \citenamefont {Bogorin}, \citenamefont {Brink},
  \citenamefont {Capelluto}, \citenamefont {G{\"u}nl{\"u}k}, \citenamefont
  {Itoko}, \citenamefont {Kanazawa} \emph
  {et~al.}}]{jurcevic2020demonstration}%
  \BibitemOpen
  \bibfield  {author} {\bibinfo {author} {\bibfnamefont {P.}~\bibnamefont
  {Jurcevic}}, \bibinfo {author} {\bibfnamefont {A.}~\bibnamefont
  {Javadi-Abhari}}, \bibinfo {author} {\bibfnamefont {L.~S.}\ \bibnamefont
  {Bishop}}, \bibinfo {author} {\bibfnamefont {I.}~\bibnamefont {Lauer}},
  \bibinfo {author} {\bibfnamefont {D.~F.}\ \bibnamefont {Bogorin}}, \bibinfo
  {author} {\bibfnamefont {M.}~\bibnamefont {Brink}}, \bibinfo {author}
  {\bibfnamefont {L.}~\bibnamefont {Capelluto}}, \bibinfo {author}
  {\bibfnamefont {O.}~\bibnamefont {G{\"u}nl{\"u}k}}, \bibinfo {author}
  {\bibfnamefont {T.}~\bibnamefont {Itoko}}, \bibinfo {author} {\bibfnamefont
  {N.}~\bibnamefont {Kanazawa}},  \emph {et~al.},\ }\href@noop {} {\bibfield
  {journal} {\bibinfo  {journal} {arXiv preprint arXiv:2008.08571}\ } (\bibinfo
  {year} {2020})}\BibitemShut {NoStop}%
\bibitem [{\citenamefont {Sundaresan}\ \emph {et~al.}(2020)\citenamefont
  {Sundaresan}, \citenamefont {Lauer}, \citenamefont {Pritchett}, \citenamefont
  {Magesan}, \citenamefont {Jurcevic},\ and\ \citenamefont
  {Gambetta}}]{sundaresan2020reducing}%
  \BibitemOpen
  \bibfield  {author} {\bibinfo {author} {\bibfnamefont {N.}~\bibnamefont
  {Sundaresan}}, \bibinfo {author} {\bibfnamefont {I.}~\bibnamefont {Lauer}},
  \bibinfo {author} {\bibfnamefont {E.}~\bibnamefont {Pritchett}}, \bibinfo
  {author} {\bibfnamefont {E.}~\bibnamefont {Magesan}}, \bibinfo {author}
  {\bibfnamefont {P.}~\bibnamefont {Jurcevic}}, \ and\ \bibinfo {author}
  {\bibfnamefont {J.~M.}\ \bibnamefont {Gambetta}},\ }\href@noop {} {\bibfield
  {journal} {\bibinfo  {journal} {arXiv preprint arXiv:2007.02925}\ } (\bibinfo
  {year} {2020})}\BibitemShut {NoStop}%
\bibitem [{\citenamefont {Wang}\ \emph {et~al.}(2017)\citenamefont {Wang},
  \citenamefont {Um}, \citenamefont {Zhang}, \citenamefont {An}, \citenamefont
  {Lyu}, \citenamefont {Zhang}, \citenamefont {Duan}, \citenamefont {Yum},\
  and\ \citenamefont {Kim}}]{wang2017single}%
  \BibitemOpen
  \bibfield  {author} {\bibinfo {author} {\bibfnamefont {Y.}~\bibnamefont
  {Wang}}, \bibinfo {author} {\bibfnamefont {M.}~\bibnamefont {Um}}, \bibinfo
  {author} {\bibfnamefont {J.}~\bibnamefont {Zhang}}, \bibinfo {author}
  {\bibfnamefont {S.}~\bibnamefont {An}}, \bibinfo {author} {\bibfnamefont
  {M.}~\bibnamefont {Lyu}}, \bibinfo {author} {\bibfnamefont {J.-N.}\
  \bibnamefont {Zhang}}, \bibinfo {author} {\bibfnamefont {L.-M.}\ \bibnamefont
  {Duan}}, \bibinfo {author} {\bibfnamefont {D.}~\bibnamefont {Yum}}, \ and\
  \bibinfo {author} {\bibfnamefont {K.}~\bibnamefont {Kim}},\ }\href@noop {}
  {\bibfield  {journal} {\bibinfo  {journal} {Nature Photonics}\ }\textbf
  {\bibinfo {volume} {11}},\ \bibinfo {pages} {646} (\bibinfo {year}
  {2017})}\BibitemShut {NoStop}%
\bibitem [{\citenamefont {Kandala}\ \emph {et~al.}(2019)\citenamefont
  {Kandala}, \citenamefont {Temme}, \citenamefont {C{\'o}rcoles}, \citenamefont
  {Mezzacapo}, \citenamefont {Chow},\ and\ \citenamefont
  {Gambetta}}]{kandala2019error}%
  \BibitemOpen
  \bibfield  {author} {\bibinfo {author} {\bibfnamefont {A.}~\bibnamefont
  {Kandala}}, \bibinfo {author} {\bibfnamefont {K.}~\bibnamefont {Temme}},
  \bibinfo {author} {\bibfnamefont {A.~D.}\ \bibnamefont {C{\'o}rcoles}},
  \bibinfo {author} {\bibfnamefont {A.}~\bibnamefont {Mezzacapo}}, \bibinfo
  {author} {\bibfnamefont {J.~M.}\ \bibnamefont {Chow}}, \ and\ \bibinfo
  {author} {\bibfnamefont {J.~M.}\ \bibnamefont {Gambetta}},\ }\href@noop {}
  {\bibfield  {journal} {\bibinfo  {journal} {Nature}\ }\textbf {\bibinfo
  {volume} {567}},\ \bibinfo {pages} {491} (\bibinfo {year}
  {2019})}\BibitemShut {NoStop}%
\bibitem [{\citenamefont {Molin}\ and\ \citenamefont
  {Salikhov}(1993)}]{Molin1993}%
  \BibitemOpen
  \bibfield  {author} {\bibinfo {author} {\bibfnamefont {Y.}~\bibnamefont
  {Molin}}\ and\ \bibinfo {author} {\bibfnamefont {K.}~\bibnamefont
  {Salikhov}},\ }\href@noop {} {\bibfield  {journal} {\bibinfo  {journal}
  {Chemical Physics Letters}\ }\textbf {\bibinfo {volume} {211}},\ \bibinfo
  {pages} {484} (\bibinfo {year} {1993})}\BibitemShut {NoStop}%
\bibitem [{\citenamefont {Krantz}\ \emph {et~al.}(2019)\citenamefont {Krantz},
  \citenamefont {Kjaergaard}, \citenamefont {Yan}, \citenamefont {Orlando},
  \citenamefont {Gustavsson},\ and\ \citenamefont
  {Oliver}}]{krantz2019quantum}%
  \BibitemOpen
  \bibfield  {author} {\bibinfo {author} {\bibfnamefont {P.}~\bibnamefont
  {Krantz}}, \bibinfo {author} {\bibfnamefont {M.}~\bibnamefont {Kjaergaard}},
  \bibinfo {author} {\bibfnamefont {F.}~\bibnamefont {Yan}}, \bibinfo {author}
  {\bibfnamefont {T.~P.}\ \bibnamefont {Orlando}}, \bibinfo {author}
  {\bibfnamefont {S.}~\bibnamefont {Gustavsson}}, \ and\ \bibinfo {author}
  {\bibfnamefont {W.~D.}\ \bibnamefont {Oliver}},\ }\href@noop {} {\bibfield
  {journal} {\bibinfo  {journal} {Applied Physics Reviews}\ }\textbf {\bibinfo
  {volume} {6}},\ \bibinfo {pages} {021318} (\bibinfo {year}
  {2019})}\BibitemShut {NoStop}%
\bibitem [{\citenamefont {Usov}\ \emph {et~al.}(1997)\citenamefont {Usov},
  \citenamefont {Stass}, \citenamefont {Tadjikov},\ and\ \citenamefont
  {Molin}}]{usov1997highly}%
  \BibitemOpen
  \bibfield  {author} {\bibinfo {author} {\bibfnamefont {O.~M.}\ \bibnamefont
  {Usov}}, \bibinfo {author} {\bibfnamefont {D.~V.}\ \bibnamefont {Stass}},
  \bibinfo {author} {\bibfnamefont {B.~M.}\ \bibnamefont {Tadjikov}}, \ and\
  \bibinfo {author} {\bibfnamefont {Y.~N.}\ \bibnamefont {Molin}},\ }\href@noop
  {} {\bibfield  {journal} {\bibinfo  {journal} {The Journal of Physical
  Chemistry A}\ }\textbf {\bibinfo {volume} {101}},\ \bibinfo {pages} {7711}
  (\bibinfo {year} {1997})}\BibitemShut {NoStop}%
\bibitem [{\citenamefont {Veselov}\ \emph {et~al.}(1987)\citenamefont
  {Veselov}, \citenamefont {Melekhov}, \citenamefont {Anisimov},\ and\
  \citenamefont {Molin}}]{veselov1987induction}%
  \BibitemOpen
  \bibfield  {author} {\bibinfo {author} {\bibfnamefont {A.}~\bibnamefont
  {Veselov}}, \bibinfo {author} {\bibfnamefont {V.}~\bibnamefont {Melekhov}},
  \bibinfo {author} {\bibfnamefont {O.}~\bibnamefont {Anisimov}}, \ and\
  \bibinfo {author} {\bibfnamefont {Y.~N.}\ \bibnamefont {Molin}},\ }\href@noop
  {} {\bibfield  {journal} {\bibinfo  {journal} {CPL}\ }\textbf {\bibinfo
  {volume} {136}},\ \bibinfo {pages} {263} (\bibinfo {year}
  {1987})}\BibitemShut {NoStop}%
\bibitem [{\citenamefont {Bagryansky}(2020)}]{BagryanskyPrivate}%
  \BibitemOpen
  \bibfield  {author} {\bibinfo {author} {\bibfnamefont {V.}~\bibnamefont
  {Bagryansky}},\ }\href@noop {} {\enquote {\bibinfo {title} {Private
  communication},}\ } (\bibinfo {year} {2020})\BibitemShut {NoStop}%
\end{thebibliography}%

\end{document}